\documentclass[12pt,twoside]{article}
\usepackage{amsmath}

\input BoxedEPS
\SetRokickiEPSFSpecial  
\HideDisplacementBoxes

\newcommand{\rd}[1]{\mathop{\mathrm{d}#1}}
\newcommand{\tr}{\mathop{\mathrm{tr}}}
\newcommand{\im}{\mathop{\mathrm{Im}}}
\newcommand{\fract}[2]{{\textstyle\frac{#1}{#2}}}
\newcommand{\grad}{\vec\nabla}
\newcommand{\dtr}{\rd{\vec r}}
\newcommand{\nA}{non-Abelian}
\newcommand{\CS}{Chern-Simons}
\newcommand{\Cpr}{Clebsch pa\-ra\-me\-ter\-iza\-tion}
\newcommand{\pr}{parameterization}
\newcommand{\BI}{\mathrm{Born\mbox{\scriptsize -}Infeld}}
\newcommand{\balpha}{\mbox{\boldmath$\alpha$}}
\newcommand{\bcP}{\mbox{\boldmath$\cal P$}}

\newcommand{\numeq}[2]{\begin{equation}
#2
\label{#1}
\end{equation}}
\newcommand{\refeq}[1]{(\ref{#1})}

\newtheorem{problem}{Problem} 

\let\vec\boldsymbol
\let\eps\varepsilon
\let\epsilon\varepsilon
\let\phi\varphi

\begin{document}
 
\title{(A Particle Field Theorist's)\\Lectures on\\(Supersymmetric,
Non-Abelian)\\Fluid Mechanics (and d-Branes)}
\author{R. Jackiw\\
\small\it Center for Theoretical Physics\\[-1ex]
\small\it Massachusetts Institute of Technology\\[-1ex]
\small\it Cambridge, MA 02139-4307}
\date{\small Text transcribed by M.-A.~Lewis, typeset in \LaTeX\ by M. Stock\\
MIT-CTP\#3000
}
\maketitle

\abstract{\noindent
This monograph is derived from a series of six lectures I gave at
the Centre de Recherches Math\'ematiques in Montr\'eal, in March and June
2000, while titulary of the Aisenstadt Chair.}

\pagestyle{myheadings}
\markboth{\small {\it R. Jackiw} --- \footnotesize\rm(A Particle Field Theorist's)
}{\footnotesize\rm  Lectures on (Supersymmetric, Non-Abelian) Fluid Mechanics
(and d-Branes)}
\thispagestyle{empty}

\newpage
\thispagestyle{empty}
\null

\newpage
\thispagestyle{empty}

\tableofcontents
 \thispagestyle{empty}

\newpage

{\def\Cpr{Clebsch para\-meter\-iza\-tion}
\def\CS{Chern-Simons}
\def\CSt{Chern-Simons term}
\def\Chg{Chaplygin gas}
\def\BI{Born-Infeld}
\def\BIm{Born-Infeld model}
\def\NG{Nambu-Goto}
\def\NGa{Nambu-Goto action}
\def\pr{para\-me\-ter\-iza\-tion}

\section*{Pr\'ecis}
\addcontentsline{toc}{section}{\kern1.4em Pr\'ecis}

\noindent
During the March 2000 meeting of the Workshop on Strings, Duality, and
Geometry in Montr\'eal, Canada,  I delivered three lectures on topics in fluid
mechanics, while titulary of the Aisenstadt Chair.  Three more lectures were
presented in June 2000, during the Montr\'eal Workshop on Integrable  Models in
Condensed Matter and Non-Equilibrium Physics.  Here are brief descriptive
remarks on the content of the lectures.
\begin{enumerate}

\item	 Introduction -- The motivation for the research is explained.

\item		Classical Equations -- The classical theory is reviewed, but in a
    manner different from textbook discussions.

\begin{enumerate}
				
			\item   Equations of  motion -- Summary of conservation and Euler
equations.

    	\item   A word on canonical formulations -- An advertisement of the 
method for finding the canonical structure for the above (developed with
L.D.~Faddeev).

		  \item   The irrotational case -- C.~Eckart's Lagrangian  and a
relativistic generalization for vortex-free motion. 

		  \item  Nonvanishing vorticity and the \Cpr\ -- In
the presence of vorticity, the velocity \CSt\ (kinetic helicity)
provides an obstruction to the construction of a Lagrangian for the motion.
C.C.~Lin's method  overcomes the obstruction, and  leads
to the \Cpr\ for the velocity vector. 

 \item   Some further remarks on the \Cpr\ -- Properties and
peculiarities of this presentation for a 3-vector.
\end{enumerate}

\item Specific Models -- Nonrelativistic and relativistic fluid mechanics in
spatial dimensions greater than one. 

\begin{enumerate}
\item  Galileo-invariant nonrelativistic model -- The Chaplygin gas
[negative pressure, inversely proportional to density] is studied, selected
solutions are presented, unexpected symmetries are identified.

\item  Lorentz-invariant relativistic model -- The scalar Born-Infeld
model is found to be the relativistic generalization of the Chaplygin gas, and
shares with it unexpected symmetries.

\item  Some remarks on relativistic fluid mechanics -- Dynamics for
isentropic relativistic fluids is given a Lagrangian formulation, and the
Born-Infeld model is fitted into that framework.
\end{enumerate}

\item Common Ancestry: The \NG\ Action -- Both the \Chg\ and the
\BIm\ devolve from the \pr-invariant \NGa, when specific \pr\ is made. 

\begin{enumerate}
\item  Light-cone \pr\ --   \Chg\ is derived.

\item  Cartesian \pr\ --  \BIm\ is derived.

\item  Hodographic transformation --  \Chg\ is derived (again).

\item  Interrelations -- The \Chg\ and \BI\ are related because (1)~the
former is the nonrelativistic limit of the latter; (2)~both descend from the
same \NGa.
\end{enumerate}

\item Supersymmetric Generalization -- Fluid
mechanics enhanced by supersymmetry.
\begin{enumerate}
\item  \Chg\ with Grassmann variables -- Vorticity is parameterized by
Grassmann variables, which act like Gaussian potentials of the \Cpr. 

\item  Supersymmetry -- Supercharges,
transformations generated by\linebreak[4] them, and their algebra.

\item  Supermembrane connection -- Supermembrane Lagrangian in
three spatial dimensions.

\item  Hodographic transformation -- Supersymmetric \Chg\ in
two spatial dimensions is derived.

\item  Light-cone \pr\ -- Supersymmetric \Chg\ in two spatial
dimensions is derived (again).

\item  Further consequences of the supermembrane connection -- Hidden
symmetries of the supersymmetric model. 
\end{enumerate}

\item One-dimensional Case -- The previous models in
one spatial dimension  are completely integrable. 
\begin{enumerate}
\item  Solutions for the \Chg\ on a line -- Some special solutions are
presented; infinite number of constants of motion is identified; Riemann
coordinates are introduced and the fluid equations as well as constants of
motion are expressed in terms of them.

\item  Aside on the integrability of the cubic potential in one dimension
-- The one-dimen\-sional problem with pressure $\propto
(\hbox{density})^3$ possesses the SO(2,1) ``Schr\"odinger symmetry'' and the
equations of motion, in Riemann form, become free. 

\item  General solution of the \Chg\ on a line -- Solution  obtained by
linearization.

\item  \BIm\ on a line -- When formulated in terms of its Riemann
coordinates,  it becomes trivially equivalent to the \Chg.

\item  General solution of the \NG\ theory for a ($d=1$)-brane (string) in
two spatial dimensions (on a plane) -- The explicit string solution is
transformed by a hodographic transformation to the \Chg\ solution, and a
relation is established between this solution and the one found by
linearization. 
\end{enumerate}

\item Towards a  Non-Abelian Fluid Mechanics -- Motivation for
this theory is given. 
\begin{enumerate}

\item  Proposal for non-Abelian fluid mechanics -- A Lagrangian is
proposed; it involves a non-Abelian auxiliary field whose \CS\ density
should be a total derivative. 

\item  Non-Abelian \Cpr\ (or, casting the non-Abelian \CS\ density
into total derivative form) -- Total derivative form for the non-Abelian  
\CS\ density is found,  thereby generalizing the Abelian \Cpr, which achieves a
total derivative form for the  Abelian density.

\item  Proposal for non-Abelian magnetohydrodynamics -- Our proposal,
which generalizes the one in Section~7.1 to include a dynamical non-Abelian
gauge field, reduces in the Abelian limit to conventional
magnetohydrodynamics. 
\end{enumerate}

\end{enumerate}
} 
%
\newpage 
\section{Introduction}

Field theory, as developed by particle physicists in the last quarter century,
has enjoyed a tremendous expansion in concepts and calculational
possibilities.  

We learned about higher and unexpected symmetries, and discovered evidence
for  partial or complete integrability facilitated by these
symmetries. We appreciated the relevance of topological ideas and structures,
like solitons and instantons, and introduced new dynamical quantities, like the
Chern-Simons terms in odd-dimensional gauge theories. We enlarged and
unified numerous degrees of freedom by introducing organizing
principles such as non-Abelian symmetries and supersymmetries.  Indeed,
application of field theory to particle physics has now been replaced by the
study of fundamentally extended structures like strings and membranes, which
bring with them new mathematically intricate ideas.

Thinking about research possibilities, I decided to investigate 
whether the novelties that we have introduced into particle physics field theory
can be used in a different, non-particle physics, yet still  field-theoretic
context.  In these Aisenstadt lectures I shall describe an approach to fluid
mechanics, which is an ancient field theory, but which can be enhanced by the
ideas that we gleaned from particle physics. 

As an introduction, I shall begin with a review of the classical theory.  Mostly, I 
duplicate what can be found in textbooks, but perhaps the emphasis will be new
and different.  After this I shall describe how some instances of the classical
theory are related to d-branes and how this relation explains some integrability
properties of various models.  I shall then show how the degrees of freedom can
be enlarged to accomodate supersymmetry and non-Abelian structures in fluid
mechanics. 
A few problems are scattered throughout;  solutions are given at the end
of the text, before the references.  

New work that I shall describe here was done in
collaboration with D. Bazeia, V.P.~Nair, S.-Y.~Pi, and A.P.~Polychronakos.
Textbooks for the classical theory, which I recommend, are  by Landau and
 Lifschitz~\cite{LanLif} as well as by Arnold and Khesin~\cite{ArnKhe}.

\newpage
\section{Classical Equations}
\subsection{Equations of motion}
We begin with nonrelativistic equations that govern a matter density field
$\rho(t,\vec r)$ and a velocity field vector $\vec v(t,\vec r)$, taken in any number
of dimensions.  The equations of motion comprise a continuity equation,
\numeq{conteq}{
\frac{\partial}{\partial t} \rho(t,\vec r) + \grad \cdot \bigl(\rho(t,\vec r)\vec
v(t,\vec r)\bigl)=0
}
 which ensures matter conservation, that is,
 time independence, of
$N=\int
\rd r \rho$, and Euler's equation, which is the expression of a nonrelativistic
force law.
\begin{equation}
\frac{\partial}{\partial t} \vec v(t,\vec r) + \vec v(t,\vec r)\cdot\grad
\vec v(t,\vec r) = \vec f(t,\vec r)
\label{Euler}
\end{equation}
 Here $\rho\vec v$ is the current $\vec j$ and $\vec f$ is the force.  We shall deal
with an isentropic fluid, that is,
 entropy is constant and does not appear in our theory.  Also we ignore
dissipation and take the force to be given by the pressure $P$: $\vec f =
-\frac{1}{\rho}\grad P$. 
 For isentropic motion  $P$ is a function only of~$\rho$, so  $\vec f$ can also be
written as
$-\grad V'(\rho)$:
\begin{equation}
\vec f = -\frac{1}{\rho}\grad P = -\grad V'(\rho) \label{force}
\end{equation}
 with  the dash (also known as ``prime'') designating the derivative with respect
to argument.
$V'(\rho)$ is the enthalpy, $\rho V'(\rho) - V(\rho) = P(\rho)$, and
$\sqrt{P'(\rho)}\equiv s$ is the speed of sound.
(Those familiar with the subject will recognize that I am using an Eulerian rather
than a Lagrangian description of a fluid~\cite{Salm}.)

The dynamics summarized in \refeq{conteq} and \refeq{Euler} and the
definition \refeq{force} may be presented as continuity equations for an
energy momentum tensor. The energy density ${\cal E} = T^{oo}$
\begin{subequations}\label{conteqs}
\numeq{conteqsa}{
{\cal E} = \fract12 \rho v^2 + V(\rho) = T^{oo}
}
together with the energy flux
\numeq{conteqsb}{
T^{jo} = \rho v^j (\fract12 v^2 +  V')
}
obey
\numeq{conteqsc}{
\frac{\partial}{\partial t} T^{oo} + \partial_j T^{jo} = 0 . 
}
\end{subequations}
Similarly the momentum density, which in the nonrelativistic theory coincides with
the current,
\begin{subequations}\label{conteqsII}
\numeq{conteqsIIa}{
{\cal P}^i = \rho v^i  = T^{oi}
}
and the stress tensor $T^{ij}$
\numeq{conteqsIIb}{
T^{ij} = \delta^{ij} (\rho V' - V) + \rho v^i v^j = \delta^{ij} P +  \rho v^i v^j
}
satisfy
\numeq{conteqsIIc}{
\frac{\partial}{\partial t} T^{oi} + \partial_j T^{ji} = 0 . 
}
\end{subequations}
Note that $T^{oi} \neq T^{io}$ because the theory is not Lorentz invariant, but
$T^{ij}=T^{ji}$ because it is invariant against spatial rotations. [Thus $T^{\mu\nu}$ is
not, properly speaking, a ``tensor'', but an energy-momentum ``complex''.]

A simplification occurs for the irrotational case  when the vorticity
\begin{equation}
\omega_{ij} \equiv \partial_i v^j - \partial_j v^i \label{vort}
\end{equation}
 vanishes.  For then the velocity can be given in terms of a velocity
potential
$\theta$,
\begin{equation}
\vec v = \grad \theta
\label{potential}
\end{equation}
 and equation (\ref{Euler}) can be replaced by Bernoulli's equation.
\begin{equation}
\frac{\partial\theta}{\partial t} + \frac{v^2}{2} = -V'(\rho)
\label{Bernoulli}
\end{equation}
 The gradient of (\ref{Bernoulli}) gives (\ref{Euler}), with help of
(\ref{force}) and (\ref{potential}).
\smallskip

\centerline{\hfill\hbox to 2in{\hrulefill} \hfill}

\begin{problem}\label{prob:1}
\emph{In the free Schr\"odinger equation for a unit-mass particle, $i\hbar
\frac{\partial\psi}{\partial t} = -\frac{\hbar^2}2 \nabla^2 \psi$, set $\psi =
\rho^{1/2} e^{i\theta/\hbar}$, and separate real and imaginary parts. Show that
the resulting equations are like those of fluid mechanics. What is the velocity? Is
vorticity supported? What is the force~$\vec f$?}

\centerline{\hfill\hbox to 2in{\hrulefill} \hfill}
\end{problem}

Our equations can be presented in any dimensionality, but we shall mostly
consider the cases of three, two, and one spatial dimensions.  In the first case, the
vorticity is a (pseudo-)\linebreak[1] vector
\begin{equation} 
\vec \omega = \grad \times \vec v  
\end{equation}
 in the second, it is a (pseudo-)scalar
\begin{equation}
\omega=\epsilon^{ij}\partial_i v^j  
\end{equation}
 while the last, lineal case is always simple because there is no vorticity
and the velocity can always be written as the derivative (with respect to the
single spatial variable) of a potential.

Dynamics of any particular system is most economically
presented when a canonical/action formulation is available.  To this end we
note that the above equations of motion can be obtained by (Poisson) bracketing
with the Hamiltonian
\begin{gather} H=\int \rd r \left(\fract{1}{2}\rho v^2 + V(\rho)\right)
=\int \rd r {\cal E}
\label{Hamilt}\\
\frac{\partial\rho}{\partial t} = \{H,\rho\}\\
\frac{\partial\vec v}{\partial t} = \{H,\vec v\}
\end{gather}
 provided the nonvanishing brackets of the fundamental $(\rho,\vec v)$
variables are taken to be
\begin{align}
\{v^i(\vec r),\rho(\vec r')\}&=\partial_i\delta(\vec r-\vec r')\nonumber\\
\{v^i(\vec r),v^j(\vec r')\}&=-\frac{\omega_{ij}(\vec r)}{\rho(\vec r)}\delta(\vec
r-\vec r').\label{algebra}
\end{align} 
(The fields in curly brackets are at equal times, hence the time
argument is suppressed.)  An equivalent, more transparent version of the
algebra (\ref{algebra}) is satisfied by the field momentum density,
\numeq{fimode}{
{\cal P}=\rho\vec v\ .
}
As a consequence of (\ref{algebra}) we have
\begin{align}
\{{\cal P}^i(\vec r),\rho(\vec r')\} &= \rho(\vec r)\partial_i \delta(\vec
r-\vec r')\nonumber\\
\{{\cal P}^i(\vec r),{\cal P}^j(\vec r')\} &= {\cal P}^j(\vec r) \partial_i
\delta(\vec r-\vec r') + {\cal P}^i(\vec r')\partial_j \delta(\vec r-\vec
r').\label{algebrap}
\end{align}
This is the familiar algebra of momentum densities. One verifies that the Jacobi
identity is satisfied~\cite{MorGre}.

Naturally one asks whether there exists a Lagrangian whose canonical variables
lead to the Poisson brackets~(\ref{algebra}) or~(\ref{algebrap}) and to the
Hamiltonian~(\ref{Hamilt}). In  more mathematical language, we seek a canonical
1-form and a symplectic 2-form that lead to the algebra~(\ref{algebra})
or~(\ref{algebrap}).

\centerline{\hfill\hbox to 2in{\hrulefill} \hfill}

\begin{problem}\label{prob:2}
\emph{Second-quantized Schr\"odinger fields satisfy equal-time commutation
(anticommutation) relations, when describing bosons (fermions):
$
\lceil \psi(\vec r), \psi^* (\vec r')\rfloor_{\pm} = \delta(\vec r -
\vec r')
$.
Show that the algebra \refeq{algebrap} is reproduced (apart from factors of
$i\hbar$) when
$\rho=\psi^*\psi$,
$\vec{\cal P} = \im \hbar \psi^*\grad\psi$. Since in the nonrelativistic theory
$\vec{\cal P}=\vec j$, find $\vec j$ in terms of $\rho$ and $\vec v$,
with  $\vec v$ as determined in Problem~\ref{prob:1}. }
\end{problem}
\vspace*{-\bigskipamount}

\centerline{\hfill\hbox to 2in{\hrulefill} \hfill}

\subsection{A word on canonical formulations}

I shall now describe an approach to canonical formulations of dynamics,
publicized by Faddeev and me~\cite{FadJac}, which circumvents and simplifies
the more elaborate approach of Dirac. 

We begin with a Lagrangian that is first order in time.  This entails no loss of
generality because all second-order Lagrangians can be converted to first
order by the familiar Legendre transformation, which produces a Hamiltonian:
$H(p,q)=p\dot q-L(\dot q,q)$, where $p\equiv  \partial L/\partial\dot q$
 (the over-dot designates
the time derivative). The 
equations of motion gotten by taking the Euler-Lagrange derivative with respect to
$p$ and $q$ of the
 Lagrangian
$L(\dot p,p;\dot q,q)\equiv p\dot q - H(p,q)$ coincide with  the ``usual''
equations of motion 
 obtained by taking the $q$ Euler-Lagrange derivative of $L(\dot q,q)$.  
[In fact $L(\dot p,p;\dot q,q)$ does not depend on~$\dot p$.] Moreover, some
Lagrangians possess only a first-order formulation (for example, Lagrangians for
the Schr\"odinger or Dirac fields; also the Klein-Gordon Lagrangian in light-cone
coordinates is first order in the light-cone ``time'' derivative).

Denoting all variables by the generic symbol $\xi^i$, the most general first-order
Lagrangian is
\begin{equation} L=a_i(\xi)\dot\xi^i-H(\xi).
\label{lag}
\end{equation}
  Note that  although we shall ultimately be interested in fields defined on
space-time, for present didactic purposes  it  suffices to consider
variables~$\xi^i(t)$ that are functions only of time.  The Euler-Lagrange equation 
that is implied by (\ref{lag}) reads
\begin{equation} f_{ij}(\xi)\dot\xi^j = \frac{\partial H(\xi)}{\partial
\xi^i}\label{E-L}
\end{equation}
 where
\begin{equation} f_{ij}(\xi)=  \frac{\partial a_j(\xi)}{\partial \xi^i}- \frac{\partial
a_i(\xi)}{\partial
\xi^j}.\label{fij}
\end{equation}
The first term in \refeq{lag} determines the canonical 1-form: $a_i(\xi)\dot\xi^i
\rd t = a_i(\xi)\rd{\xi^i}$, while $f_{ij}$ gives the symplectic 2-form: 
$\rd{a_i(\xi)\rd{\xi^i}} = \fract12 f_{ij} (\xi) \rd{\xi^i}\rd{\xi^j}$.

 To set up a canonical formalism, we proceed directly.  We \emph{do not} make
the frequently heard statement that ``the canonical momenta
$ \partial L/\partial\dot\xi^i = a_i(\xi)$ are constrained to depend on the
coordinates $\xi$'', and we \emph{do not}  embark on Dirac's method for
constrained systems.

In fact, if the matrix $f_{ij}$ possesses the inverse $f^{ij}$  there are no
constraints.  Then (\ref{E-L}) implies
\begin{equation}
\dot\xi^i = f^{ij}(\xi)  \frac{\partial H(\xi)}{\partial \xi^j}.\label{eqm}
\end{equation}
 When one wants to express this equation of motion by bracketing with
the Hamiltonian
\begin{equation}
\dot\xi^i=\{H(\xi),\xi^i\} = \{\xi^j,\xi^i\}\frac{\partial H(\xi)}{\partial \xi^j} 
\end{equation}
 one is directly led to postulating the fundamental bracket as
\begin{equation}
\{\xi^i,\xi^j\}=-f^{ij}(\xi).\label{fundbrack}
\end{equation}
 The Poisson bracket between functions of $\xi$ is then defined by
\begin{equation}
\{F_1(\xi),F_2(\xi)\} = -\frac{\partial F_1(\xi)}{\partial\xi^i} f^{ij}
\frac{\partial F_2(\xi)}{\partial\xi^j}.
\end{equation}
 One verifies that (\ref{fundbrack}) satisfies Jacobi identity by
virtue of (\ref{fij}).  

When $f_{ij}$ is singular and has no inverse, constraints
do arise, and the development becomes more complicated (see~\cite{FadJac}).  

Our problem in connection with (\ref{algebra}) and (\ref{algebrap}) is in fact the
inverse of what I have here summarized.  From (\ref{algebra}) and
(\ref{algebrap}), we know the form of
$f^{ij}$ and that the Jacobi identity holds.  We then wish to determine the
inverse $f_{ij}$, and also
$a_i$ from (\ref{fij}). Since we  know the Hamiltonian from~(\ref{Hamilt}),
construction of  the Lagrangian~(\ref{lag}) should follow immediately.

However, an obstacle may arise: If there exists a quantity $C(\xi)$ whose
Poisson bracket with all the $\xi^i$ vanishes, then
\begin{equation} 0 = \{\xi^i,C(\xi)\} = -f^{ij}\frac{\partial}{\partial \xi^j}C(\xi)\ . 
\end{equation}
That is, $f^{ij}$ has the zero mode $\frac{\partial}{\partial \xi^j} C(\xi)$, and the  
inverse to $f^{ij}$, namely the symplectic 2-form
$f_{ij}$, does not exist.  In that case, something more has to be done, and we
shall come back to this problem.

Totally commuting quantities like $C(\xi)$ are called ``Casimir invariants''. Since
they Poisson-commute with \emph{all} the dynamical variables, they commute
with the Hamiltonian, and are constants of motion. But these constants do not
reflect any symmetry of the specific Hamiltonian, nor do they generate any
infinitesimal transformation on $\xi^i$, since the $\{ C(\xi), \xi^i\}$ bracket
vanishes. 

As will be demonstrated below, the algebra \refeq{algebra}, \refeq{algebrap}
admits Casimir invariants, which create an obstruction to the construction of a
canonical formalism for fluid mechanics; this obstruction must be overcome to
make progress. 
(In the Lagrangian formulation of fluid mechanics these Casimirs
are related to a parameterization-invariance of that formalism~\cite{Salm}.) 

\subsection{The irrotational case}

We now return to the specific issue of determining the fluid dynamical
Lagrangian.  The problem of constructing a Lagrangian which leads to
(\ref{algebra}) and (\ref{algebrap}) can be solved by inspection for the
irrotational case, with vanishing vorticity [see (\ref{vort})].  For then the velocity
commutator in (\ref{algebra}) vanishes and (\ref{potential}) shows that the first
equation in (\ref{algebra}) can be satisfied by taking $\rho$ and $\theta$ to be
canonically conjugate.
\begin{equation}
\{\theta(\vec r),\rho(\vec r')\}=\delta(\vec r-\vec r')
\label{canconj}
\end{equation}
Thus the Lagrangian  reads
\begin{equation} 
 L_{\mathrm{irrotational}} = \int \rd r \bigl(\theta\dot\rho -
H_{\vec v=\grad\theta}\bigr)\label{lagirr}
\end{equation}
 where $H$ is given by (\ref{Hamilt}) with $\vec v$ taken as in
(\ref{potential}).  
The form of this Lagrangian can be understood by the following
argument, due to C.~Eckart~\cite{Eck}.

Consider the Lagragian for $N$ point-particles in free nonrelativistic motion. 
With the mass
$m$ set to unity, the Galileo-invariant, free Lagrangian is just the kinetic energy.
\begin{equation} L_0 =\fract{1}{2}\sum_{n=1}^N v_n^2(t)
\end{equation}
 In a continuum description, the particle-counting index $n$ becomes the
continuous variable $\vec r$, and the particles are distributed with density
$\rho$, so that $\sum_{n=1}^N v_n^2(t)$ becomes $\int \rd r \rho(t,\vec r)
v^2(t,\vec r)$.  But we also need to link the density with the current $\vec
j=\rho\vec v$, so that the continuity equation holds.  This can be enforced
with the help of a Lagrange multiplier~$\theta$.  We thus arrive at the free,
continuum Lagrangian.
\begin{equation} \bar L_0^{\mathrm{Galileo}} = \int \rd r \Bigl(\fract{1}{2}\rho
v^2 +
\theta\bigl(\dot\rho + \grad\cdot(\rho\vec v)\bigr)\Bigr)
\label{laggal}
\end{equation}
 Since $\bar L_0^{\mathrm{Galileo}}$ is first order in time and the canonical
1-form
$\int \rd r \theta\dot \rho$ does not contain $\vec v$, the latter may be varied,
evaluated, and eliminated~\cite{FadJac}.  Doing this, we find
\begin{equation}
\rho\vec v - \rho\grad\theta = 0
\end{equation}
 and we conclude that $\grad\theta$ is the velocity or, more precisely, that
$\grad\theta$ is the $\vec v$ derivative of the kinetic energy, that is,
 the momentum
$\vec p$, which in this nonrelativistic setting coincides with~$\vec v$.
\numeq{coincvecv}{
\grad \theta = \frac\partial{\partial \vec v} \fract12 v^2 \equiv \vec p = \vec v 
}
 Substituting this in (\ref{laggal}), we obtain
\begin{equation}\label{eq:31}
 L_0^{\mathrm{Galileo}} = \int \rd r   \left(\theta\dot \rho -
\fract{1}{2}\rho(\grad\theta)^2\right) 
\end{equation}
 which reproduces (\ref{lagirr}) with the interaction  $V(\rho)$  in \refeq{Hamilt}
set to zero, and leads to the free version of the Bernoulli equation of motion
(\ref{Bernoulli}).
\begin{equation}
\dot\theta  + \frac{(\grad \theta)^2}{2} =0
\end{equation}
 Taking the gradient gives
\numeq{takgrad}{
\dot{\vec v} +\vec v \cdot \grad \vec v=0.
}

\centerline{\hfill\hbox to 2in{\hrulefill} \hfill}
\begin{problem}\label{prob:3}
\emph{The Lagrange density for the unit-mass Schr\"odinger equation can be
taken as 
$
{\cal L}_{\textrm{Schr\"odinger}} = i\hbar \psi^* \frac\partial{\partial t}\psi -
\frac{\hbar^2}{2m} \grad \psi^* \cdot \grad \psi 
$.
What form does this take after $\psi$ is represented by $\rho^{1/2}
e^{i\theta/\hbar}$? Compare with \refeq{lagirr}.}

\centerline{\hfill\hbox to 2in{\hrulefill} \hfill}
\end{problem}

 Remarkably, the same equation \refeq{takgrad} emerges for a kinetic energy
$T(\vec v)$ that is an arbitrary function of $\vec v$.  This will be useful for us 
when we  study a relativistic generalization of the theory.  If we replace
(\ref{laggal}) by
\begin{equation}\label{eq:34}
 \bar L_0=\int \rd r \Bigl(\rho T(\vec v) + \theta\bigl(\dot\rho +
\grad\cdot (\rho\vec v)\bigr)\Bigr)
\end{equation}
 and  vary $\vec v$ to eliminate it, we get in a generalization of \refeq{coincvecv}
\begin{equation}
\grad \theta = \frac{\partial T}{\partial\vec v} \equiv \vec p.\label{thetarel}
\end{equation}
So in the general case, it is the momentum -- the $\vec v$ derivative of
$T(\vec v)$ -- that is irrotational.
 The Lagrange density becomes
\begin{equation}\label{eq:36}
 L_0=\int \rd r \bigl(\theta\dot\rho - \rho h(\vec p)_{\vec p =\grad
\theta}\bigr) 
\end{equation}
 where $h(\vec p)$ is the Legendre transform of $T(\vec v)$.
\begin{equation}\label{eq:37}
 h(\vec p) = \vec v\cdot\vec p - T(\vec v)  \qquad \frac{\partial
h}{\partial
\vec p} = \vec v 
\end{equation}
 Again varying $\theta$ in \refeq{eq:36}  gives the continuity equation
\begin{align} 
0=\frac{\delta L_0}{\delta \theta} &= \dot\rho -\int \rd r
\rho\frac{\partial h(\vec p)}{\partial\vec p}\cdot \frac{\delta \vec p}{\delta
\theta}\nonumber\\ 
&= \dot\rho - \int \rd r \rho \vec v\cdot \frac{\delta}{\delta \theta}\grad
\theta \nonumber\\ 
 &= \dot \rho +\grad \cdot(\rho\vec v).\label{contrel}
\end{align} 
Varying $\rho$ gives
\begin{equation} 0=\frac{\delta L_0}{\delta \rho} = -\dot\theta - h(\vec p).
\end{equation}
 Taking the gradient, this implies with the help of~(\ref{thetarel})
\begin{align}
\partial_i\dot\theta &= -\vec v\cdot\frac\partial{\partial r^i}\vec p \nonumber\\ 
&= -v^j\frac\partial{\partial r^j} p^i \nonumber\\ 
&= -v^j\frac{\partial p^i}{\partial v^k}\frac\partial{\partial r^j} v^k.
\end{align} On the other hand, (\ref{thetarel}) implies that
\begin{equation}
\partial_i \dot\theta = \frac{\partial p^i}{\partial v^k} \dot v^k.
\end{equation}
 The two are consistent, provided the free Euler equation holds, that is,
\begin{equation}
\dot v^k + v^j\partial_j v^k = 0  \label{freeEuler}
\end{equation}
 (as long as ${\partial p^i}/{\partial v^k} = {\partial^2 T}/{\partial
v^i\partial v^k}$ has an inverse).

Let me observe that free motion is here governed by a Lagrangian that  is
not quadratic and the free equations are not linear.  Nevertheless, the equations
of motion (\ref{contrel}) and (\ref{freeEuler}) can be solved in terms of initial
data.
\begin{align}
\rho(t=0,\vec r)&\equiv\rho_0(\vec r)\label{initdata1}\\
\vec v(t=0,\vec r) &\equiv \vec v_0(\vec r) \label{initdata}
\end{align}
 Upon determining the retarded position $\vec q(t,\vec r)$  from the equation
\begin{equation}
\vec q + t\vec v_0(\vec q) = \vec r \label{initdata2}
\end{equation}
 one verifies that the solution to the free equations reads
\begin{align}
\vec v(t,\vec r) &= \vec v_0(\vec q)\label{initdata3}\\
\rho(t,\vec r) &= \rho_0(\vec q) \bigl|\det \frac{\partial q^i}{\partial
r^j}\bigr|.\label{initdata4}
\end{align}

A final remark: Note that the free Bernoulli equation \refeq{Bernoulli} coincides
with the free Hamilton-Jacobi equation for the action.

\subsection{Nonvanishing vorticity and the \Cpr}\label{sec:2.4}

We now return to our original Galileo-invariant problem and enquire about the
Lagrangian for velocity fields that are not irrotational, that is,
 whose
vorticity is nonvanishing.  Here we specify the spatial dimensionality to be~$3$,
and observe that the algebra (\ref{algebra}) possesses a zero mode, since the
quantity
\begin{equation} C(\vec v)\equiv  \int \rd r \epsilon^{ijk} v^i\partial_j v^k = \int
\rd r
\,\vec v\cdot \vec\omega
\end{equation}
(Poisson)  commutes with both $\rho$ and $\vec v$.  So the symplectic
$2$-form does not exist: in the language developed above, $f^{ij}$ has no
inverse. (Notice that $C$, also called the ``fluid helicity'', coincides with the
Abelian Chern-Simons term for~$\vec v$~\cite{DesJac}.) (In the irrotational case
with vanishing
$\vec\omega$, the obstacle obviously is absent.)

To make progress, one must neuteralize the obstruction.  This is achieved in the
following manner, as was shown by C.C. Lin~\cite{Lin}.

We use the Clebsch parameterization for the vector field $\vec v$.  Any
three-dimensional vector, which involves three functions, can be presented as
\begin{equation}
\vec v = \grad \theta+\alpha\grad \beta \label{Cpar}
\end{equation}
 with three suitably chosen scalar functions $\theta,\alpha$, and $\beta$. 
This is called the ``Clebsch parameterization'', and $(\alpha,\beta)$ are called
``Gaussian potentials''~\cite{Cle}.  In this parameterization, the vorticity reads
\begin{equation}
\vec \omega = \grad\alpha \times \grad \beta 
\end{equation}
 and the Lagrangian is taken as
\begin{equation}L =  -\int \dtr \rho(\dot\theta+\alpha\dot\beta) - H_{\vec
v=\grad\theta+\alpha\grad\beta} \label{takenas}
\end{equation}
 with $\vec v$ in $H$ expressed as in (\ref{Cpar}).  Thus $(\rho,\theta)$
remain canonically conjugate but another canonical pair appears:
$(\rho\alpha,\beta)$.  The phase space is $4$-dimensional, corresponding to
the four observables $\rho$ and $\vec v$, and a straightforward calculation
shows that the Poisson brackets (\ref{algebra}) are reproduced, with $\vec v$
constructed by (\ref{Cpar}).  

But how has the obstacle been overcome?  Let us
observe that in the Clebsch parameterization~$C$ is given by
\begin{equation}
 C =  \int \rd r \epsilon^{ijk}\,\partial_i\theta
\,\partial_j\alpha\,\partial_k\beta \label{cpars}
\end{equation}
 which is just a surface integral
\begin{equation} 
C= \int \rd {\vec S}\cdot (\theta\vec \omega).
\end{equation}
 In this form, $C$ has no bulk contribution, and presents no obstacle to
constructing a symplectic $2$-form and a canonical $1$-form in terms of
$\theta,\alpha$ and $\beta$, which are defined in the bulk, that is, for all
finite~$\vec r$. 

Lin gave an Eckart-type derivation of (\ref{takenas}):  Return to
$\bar L_0^{\mathrm{Galileo}}$ in (\ref{laggal}) and add a further constraint,
beyond the one enforcing current conservation~\cite{Lin}.
\begin{equation}
\bar L_0^{\mathrm{Galileo}} = \int \rd r \left(\fract{1}{2}\rho v^2 +
\theta\bigl(\dot\rho +
\grad \cdot (\rho\vec v)\bigr)- \rho\alpha (\dot\beta+\vec v\cdot\grad
\beta)\right) \label{enfcur}
\end{equation}
Setting the variation (with respect to $\vec v$) to zero evaluates $\vec v$ as in
\refeq{Cpar}; 
 eliminating $\vec v$ from \refeq{enfcur} gives rise to \refeq{takenas}.  

This procedure works in any number of dimensions, producing the same
canonical 1-form in any dimension.  This means that in two spatial dimensions,
on the plane, where the $(\rho, \vec v)$ space is three-dimensional, the
four-dimensional phase space $(\rho, \theta; \rho \alpha, \beta)$  is larger. 
Moreover, the analog to~$C$ in two spatial dimensions, that is, the obstruction to
constructing a symplectic 2-form, is not a single quantity: an infinite number of
objects (Poisson) commute with~$\rho$ and~$\vec v$.  These are 
\numeq{Pcommute}{
C_n = \int \rd r \rho \Bigl(\frac\omega\rho\Bigr)^n\qquad n=0,\pm1,
\pm2,\dots.
}
(Of course, the $C_n$ vanish in the irrotational case where there is no obstruction.)
One can understand why there is an infinite number of obstructions by
observing that phase space must be even dimensional, but $(\rho, \vec v)$
comprise three quantities on the plane.  So a nonsingular symplectic form can
be constructed either by increasing the number of canonical variables to four,
or decreasing to two.  The Lin/Clebsch method increases the variables.  On the
other hand, decreasing to two entails suppressing of one continuous and local
degree of freedom, and evidently this is equivalent to neutralizing the infinite
number of global obstructions, namely, the~$C_n$.   But I do not know how to
effect such a suppression, so I remain with (\ref{takenas}). 

Note that $\bar
L_0^{\mathrm{Galileo}}$ in \refeq{enfcur}, apart from a total time derivative, can
also be written in any number of dimensions as
\begin{align}
\bar L_0^{\mathrm{Galileo}}&=\int \rd r \bigl(\rho T(\vec v) - \rho(\dot\theta +
\alpha\dot \beta) -\rho\vec v\cdot(\grad \theta+\alpha\grad\beta)\bigr)
\nonumber\\
&= \int \rd r \bigl(\rho T(\vec v) - j^{\mu} (\partial_\mu
\theta+\alpha\partial_\mu
\beta)\bigr)\label{notethat}
\end{align} 
where we have introduced the   current four-vector
\begin{equation}
 j^\mu = (c\rho,\rho\vec v) ,
\label{curr4vec}
\end{equation}
 employed the four-vector gradient $\partial_\mu = (\frac1c
\frac\partial{\partial t},\grad)$, and denoted the kinetic energy by~$T(\vec v)$.
These expressions will form our starting point for a relativistic generalization of
the theory as well as a non-Abelian generalization. (That is why we have
introduced the velocity of light in the above definitions; of course $c$ disappears
in the Galilean theory, as it has no role there.)

Finally we observed that in one spatial dimension, where~$v$ can always be
written as~$\theta'$ and the vorticity vanishes, the irrotational canonical 1-form
$\int \rd x \theta\dot\rho$ is generally applicable and can equivalently be
written as $-\fract12 \int \rd x \rd y\linebreak[2] \rho(x) \eps(x-y) \dot v(y)$,
where~$\eps$ is a $\pm1$-step function, determined by the sign of its argument. 
Evidently this leads to a spatially nonlocal, but otherwise completely satisfactory
canonical formulation of fluid motion on a line.  

\subsection{Some further remarks on the   \Cpr}\label{sec:2.5}

Let me elaborate on the \Cpr\ for a vector field, which was
presented for the velocity vector in~(\ref{Cpar}). Here I shall use the notation
of electromagnetism and discuss the \Cpr\ of a vector potential $\vec A$, which
also leads to the magnetic field $\vec B =\grad \times \vec A$. (Of course the
same observations apply when the vector in question is the velocity field $\vec
v$, with $\grad\times\vec v$ giving the vorticity.) 

The familiar \pr\ of a three-component vector employs a scalar
function~$\theta$  (the ``gauge'' or ``longitudinal'' part) and a two-component
transverse vector
$\vec A_T$: $\vec A=
\grad\theta + \vec A_T$, $\grad\cdot\vec A_T = 0$. This decomposition is
unique and invertible (on a space with simple topology). In contrast, the \Cpr\
uses three scalar functions, $\theta$, $\alpha$, and~$\beta$,
\begin{equation}
\vec A = \grad\theta + \alpha \grad \beta
\label{threescalar}
\end{equation}
which are not uniquely determined by $\vec A$ (see below). The
associated magnetic field reads 
\begin{equation}
\vec B=\grad\alpha\times\grad\beta.
\label{assmag}
\end{equation}
Repeating the above in form notation, the 1-form $A=A_i \rd r^i$ is presented as 
\begin{equation}
A= \rd \theta + \alpha \rd \beta
\label{1-form}
\end{equation}
and the 2-form is
\begin{equation}
\rd A= \rd \alpha \rd \beta. 
\label{2-form}
\end{equation}

Darboux's theorem ensures that the \Cpr\ is attainable locally in space [in the
form (\ref{1-form})]~\cite{Darb}. Additionally, an explicit construction of $\alpha$,
$\beta$, and $\theta$ can be given by the following~\cite{Lam}. 

Solve the equations
\begin{subequations}\label{solve}
\begin{equation}
\frac{\rd x}{B_x} =\frac{\rd y}{B_y} =\frac{\rd z}{B_z}
\label{solvea}  
\end{equation}
which may also be presented as
\begin{equation}
\eps^{ijk} \rd {r^j} B^k = 0.
\label{solveb}  
\end{equation}
\end{subequations}
Solutions of these relations define two surfaces, called ``magnetic surfaces'', that
are given by equations of the form
\begin{equation}
S_n(\vec r) = \mathrm{constant}\qquad (n=1,2).
\label{magsurfs}
\end{equation}
It follows from \refeq{solve} that these also satisfy
\numeq{follows}{
\vec B\cdot \grad S_n = 0 \qquad (n=1,2)
}
that is, the normals to $S_n$ are orthogonal to $\vec B$, or $\vec B$ is
parallel to the tangent of~$S_n$.  The intersection of the two surfaces forms the
so-called ``magnetic lines'', that is, loci that solve the dynamical system
\numeq{loci}{
\frac{\dtr(\tau)}{\rd \tau} = \vec B\bigl(\vec r(\tau)\bigr)
}
where $\tau$ is an evolution parameter. Finally, the Gaussian potentials $\alpha$
and~$\beta$ are constructed as functions of~$\vec r$ only through a dependence
on the magnetic surfaces,
\begin{align}
\alpha(\vec r) &= \alpha \bigl( S_n (\vec r)\bigr)\nonumber\\
\beta(\vec r) &= \beta \bigl( S_n (\vec r)\bigr)
\label{depmags}
\end{align}
so that
\numeq{depmags2}{
\grad \alpha \times \grad\beta = (\grad S_1 \times \grad S_2) \eps^{mn}
\frac{\partial \alpha}{\partial S_m}\frac{\partial \beta}{\partial S_n}.
}
Evidently as a consequence of \refeq{follows}, $\grad
\alpha\times\grad\beta$ in~\refeq{depmags2} is parallel to~$\vec B$, and
because
$\vec B$ is divergence-free $\alpha$ and
$\beta$ can be adjusted so that the  norm of $\grad \alpha \times \grad\beta$ 
coincides with $|\vec B|$. Once $\alpha$ and
$\beta$ have been fixed in this way, $\theta$ can easily be computed from $\vec
A-\alpha\grad\beta$. 

Neither the individual magnetic surfaces nor the Gauss potentials are unique.
[By viewing $A$ as a canonical 1-form, it is clear that the expression
\refeq{1-form} retains its form after a canonical transformation of
$\alpha,\beta$.] One may therefore require that the Gaussian potentials $\alpha$
and $\beta$ simply coincide with the two magnetic surfaces: $\alpha=S_1$,
$\beta=S_2$. Nevertheless, for a given
$\vec A$ and $\vec B$ it may not be possible to solve \refeq{solve} explicitly.

The \CS\ integrand $\vec A\cdot \vec B$ becomes in the \Cpr\
\numeq{CSi}{
\vec A\cdot \vec B = \grad\theta\cdot (\grad \alpha\times\grad\beta)
= \grad\cdot(\theta\vec B) = \vec B\cdot \grad\theta.
}
Thus having identified the Gauss potentials $\alpha$ and $\beta$ with the two
magnetic surfaces, we deduce from \refeq{follows} and \refeq{CSi} three
equations for the three functions ($\theta$, $\alpha$, $\beta$) that comprise
the \Cpr.
\begin{align}
\vec B\cdot\grad \alpha &= \vec B\cdot\grad\beta = 0 \nonumber\\
\vec B\cdot\grad \theta &= \mbox{\CS\ density } \vec A\cdot \vec
B\label{follows2}
\end{align}

Eq.~\refeq{CSi} also shows that in the \Cpr\ the \CS\ density becomes a total
derivative.
\numeq{totderiv}{
\vec A\cdot \vec B = \grad\cdot(\theta\vec B)
}
This does \emph{not} mean that the \Cpr\ is unavailable when the \CS\ integral 
over all space is nonzero. Rather for a nonvanishing integral and well-behaved
$\vec B$ field, one must conclude that the Clebsch function~$\theta$ is singular
either in the finite volume of the integration region or on the surface at infinity
bounding the integration domain. Then the \CS\ volume integral over ($\Omega$)
becomes a surface integral on the surfaces ($\partial\Omega$) bounding the
singularities.
\numeq{boundsing}{
\int_\Omega \rd r \vec A\cdot \vec B = \int_{\partial\Omega} \rd {\vec
S}\cdot\, (\theta\vec B) 
}
Eq.~\refeq{boundsing} shows that the \CS\ integral measures the magnetic flux,
modulated by~$\theta$ and passing through the surfaces that surround the
singularities of~$\theta$. 

The following  explicit example  illustrates the above points. 

 Consider the vector potential whose spherical components are
given by 
\begin{align}
A_r &= (\cos\Theta) a'(r)\nonumber\\
A_\Theta &= -(\sin\Theta)\frac1r \sin a(r)\nonumber\\
A_\Phi &= -(\sin\Theta)\frac1r \bigl(1-\cos a(r)\bigr).
\label{vecpot}
\end{align}
($r$, and $\Theta$, $\Phi$ denote the conventional radial coordinate and the
polar, azimuthal angles.)  The function $a(r)$ is taken to vanish at the origin, and
to behave as
$2\pi\nu$ at infinity ($\nu$~integer or half-integer). The corresponding
magnetic field reads 
\begin{align}
B_r &= -2(\cos\Theta)\frac1{r^2} \bigl(1-\cos a(r)\bigr)\nonumber\\
B_\Theta &=  (\sin\Theta)\frac1r a'(r)\sin a(r)\nonumber\\
B_\Phi &= (\sin\Theta)\frac1r a'(r) \bigl(1-\cos a(r)\bigr)
\label{corrmag}
\end{align}
and the \CS\ integral -- also called the ``magnetic helicity'' in the
electrodynamical context --  is quantized (in multiples of
$16\pi^2$) by the behavior of
$a(r)$ at infinity
\begin{align}
\int \rd r \vec A\cdot \vec B &= 
-8\pi\int_0^\infty \rd r \frac {\rd\null}{\rd r} \bigl(a(r) -\sin a(r)\bigr) 
\nonumber\\
&= -16\pi^2\nu.
\label{atinf}
\end{align}

In spite of the nonvanishing magnetic helicity, a \Cpr\ for \refeq{vecpot} is readily
constructed. In form notation, it reads 
\numeq{constr}{
A=\rd{(-2\Phi)} +2\Bigl(1-\bigl(\sin^2 \frac a2\bigr) \sin^2\Theta\Bigr) 
\rd{\!\left(\Phi + \tan^{-1}\Bigl[ \bigl(\tan\frac a2\bigr)\cos\Theta\Bigr]\right)}
}
The
magnetic surfaces can be taken from  formula~\refeq{constr} to coincide with
the Gauss potentials. 
\begin{align}
S_1 &= 2\Bigl(1- \bigl(\sin^2 \frac a2\bigr)\sin^2\Theta\Bigr) =
\mbox{constant}\nonumber\\
S_2 &= \Phi + \tan^{-1} \Bigl[\bigl(\tan \frac a2\bigr)\cos\Theta\Bigr]=
\mbox{constant}
\label{gausspot}
\end{align}
The magnetic lines are determined by the intersection of $S_1$ and $S_2$. 
\begin{align}
\cos \frac a2 &= \eps \cos(\Phi - \phi_0)\nonumber\\
\sin\Theta &= \sqrt{\frac{1-\eps^2}{1-\eps^2\cos^2 (\Phi-\phi_0)}}
\label{maglines}
\end{align}
where $\eps$ and $\phi_0$ are constants. The potential $\theta= -2\Phi$ is
multivalued. Consequently the ``surface'' integral determining the \CS\ term
reads
\numeq{surfaceint}{
\int \rd r \vec A\cdot \vec B = \int \rd r \grad\cdot(-2\Phi\vec B) 
= \int_0^\infty r\rd r \int_0^\pi \rd \Theta B_\Phi\Bigr|_{\Phi=2\pi} \ .
}
That is, the magnetic helicity is the flux of the toroidal magnetic field through the
positive-$x$ ($x,z$) half-plane. 
\medskip

\centerline{\hfill\hbox to 2in{\hrulefill} \hfill}
\begin{problem}\label{prob:4}
\emph{Consider a vector potential $\vec A$, whose \Cpr\ reads
$
A_i = \partial_i \Phi + \cos \Theta \partial_i h(r)
$,
where $\Theta$ and $\Phi$ are the azimuthal and polar angles of the vector~$\vec
r$, and $h$ is a nonsingular function of the magnitude of~$\vec r$. Show that the
\CS\ density (magnetic helicity density) is given by 
$
\eps^{ijk} A_i \partial_j A_k = \eps^{ijk} \partial_i \Phi \partial_j \cos\Theta
\partial_k h(r) 
$.
Consider the integral of the \CS\ density over all space. This integral may first be
evaluated over a spherical ball $\Omega$, and then the radius $R$ of the ball is
taken to infinity. When the integrand is a divergence of a vector, Gauss's theorem
casts the volume integral onto a surface integral over the sphere $\partial \Omega$
bounding the ball: 
$
I = \int_\Omega \rd{^3 r} \grad\cdot\vec V = 
\int_{\partial \Omega} \rd{\vec S}\cdot\vec V = 
\int_0^{2\pi} \rd\Phi \int_0^\pi \rd\Theta \sin \Theta  r^2 V^r \,|_{r=R}
$, 
but singularities in $\vec V$ may modify the  equality. The three
derivatives in the above \CS\ density may be extracted in three different ways.
Show that the result always is $4\pi\bigl[ h(R) - h(0)\bigr]$, but various
singularities must be carefully handled. }
\end{problem}
\begin{problem}\label{prob:5}
\emph{
Show that $\int \rd{^3 r} \vec B\cdot \delta \vec A$ vanishes (apart from surface
terms) where $\delta \vec A$ is  a  variation and $\vec A$,  $\vec B =
\grad\times\vec A$, as well as $\delta \vec A$ are presented in the
\Cpr. When the variational principle is implemented by varying the components of
$\vec A$, one finds that $\fract12 \int \rd{^3 r} B^2$ is stationary provided
$\grad\times\vec B=0$. Show that implementing the variation by varying the
scalar functions in the \Cpr\ for $\vec A$ gives the weaker condition
$\grad\times\vec B=\mu\vec B$, where $\mu$ can depend on $\vec r$. How is
this
$\vec r$-dependence constrained? How is the constraint satisfied? }
\end{problem}
\vspace*{-\medskipamount}

\centerline{\hfill\hbox to 2in{\hrulefill} \hfill}
\smallskip

There is another approach to the construction of (Abelian) vector potentials for
which the (Abelian) \CS\ density is a total derivative, and as a consequence a
\Cpr\ for these potentials is readily found. The method relies on projecting an
Abelian potential from a non-Abelian one, and it can be generalized to a
construction of non-Abelian vectors for which the non-Abelian \CS\ density is
again a total derivative. This will be useful for us when we come to discuss
non-Abelian fluid mechanics. Therefore, I shall now explain this method -- in its
Abelian realization~\cite{JacPi}. 

We consider an SU(2) group element~$g$ and a pure gauge SU(2) gauge field,
whose matrix-valued 1-form is 
\numeq{gfield}{
g^{-1} \rd g = V^a \frac{\sigma^a}{2i}
}
where $\sigma^a$ are Pauli matrices. It is known that 
\numeq{knowntotderiv}{
\tr (g^{-1} \rd g)^3 = -\fract14 \eps^{abc} V^a V^b V^c = -\fract32 V^1 V^2 V^3
}
is a total derivative; indeed its spatial integral measures the winding number of
the gauge function~$g$~\cite{TrJaZuWi}. Since $V^a$ is a pure gauge, we have
\numeq{Va}{
\rd{V^a} = -\fract12 \eps^{abc} V^b V^c
}
so that if we define an Abelian gauge potential $A$ by projecting one SU(2)
component of~\refeq{gfield} (say the third) $A=V^3$, the Abelian \CS\ density
for~$A$ is a total derivative, as is seen from the chain of equation that relies
on~\refeq{knowntotderiv} and~\refeq{Va}. 
\numeq{chaineq}{
A\rd A = V^3 \rd{V^3} = -V^1V^2V^3 = \fract23 \tr (g^{-1}\rd g)^3
}
Of course $A=V^3$ is not an Abelian pure gauge. 

Note that $g$ depends on three arbitrary functions, the three SU(2) local gauge
functions. Hence $V^3$ enjoys sufficient generality to represent the
3-dimensional vector $A$. Moreover, since  $A$'s Abelian \CS\ density is
given by $\tr(g^{-1}\rd g)^3$, which is a total derivative, a \Cpr\ for $A$ is
easily constructed. We also observe that when the SU(2) group element~$g$ has
nonvanishing winding number, the resultant Abelian vector possesses a
nonvanishing
\CS\ integral, that is, nonzero magnetic helicity.

Finally we remark that the example of a Clebsch-parameterized gauge
potential~$A$,  presented above in~\refeq{vecpot}, is gotten by a projection
onto the third isospin direction of a pure gauge  SU(2) potential, constructed
from  group element
$g=\exp
\bigl(({\sigma^a}/{2i}) \hat r^a a(r)\bigr)$~\cite{JacPi}. 

Further intricacies arise when the \Cpr\ is used in variational calculations
involving vector fields; see Ref.~\cite{DesJacPol}. 

\newpage
\section{Specific Models}

We now return to our irrotational models both relativistic and nonrelativistic, for
which we shall specify an explicit force law and discuss further properties. 

\subsection{Galileo-invariant nonrelativistic model}\label{sec:3.1}
 Recall
that the nonrelativistic Lagrangian for irrotational motion reads
\begin{equation} L^{\mathrm{Galileo}}=\int \rd r \Bigl(\theta\dot\rho - \rho
\frac{(\grad\theta)^2}{2}-V(\rho)\Bigr) 
\end{equation}
 where $\grad\theta=\vec v$.  The Hamiltonian density $\cal H$ is
composed of the last two terms beyond the canonical $1$-form $\int  \rd r
\theta\dot\rho$,
\begin{equation}
 H=\int \rd r \Bigl(\rho\frac{(\grad\theta)^2}{2}+V(\rho)\Bigr)=\int\rd r
{\cal H}.
\end{equation}
 Various expressions for $V$ appear in the literature.  $V(\rho)\propto
\rho^n$  is a popular choice, appropriate for the adiabatic equation of state. 
We shall be specifically interested in the ``Chaplygin gas''.
\numeq{ChaGas}{
V(\rho)=\frac{\lambda}{\rho}, \qquad \lambda>0 
}
 According to what we said before, the Chaplygin gas has enthalpy
$V'=-{\lambda}/{\rho^2}$, negative pressure $P=-{2\lambda}/{\rho}$, 
and speed of sound $s=\sqrt{2\lambda}/\rho$ (hence $\lambda>0$).  

 Chaplygin introduced his equation of state as a mathematical approximation to
the physically relevant adiabatic expressions with $n>0$. (Constants are arranged
so that the Chaplygin formula is tangent at one point to the adiabatic
profile~\cite{Chp}.)  Also it was realized that certain deformable solids can be
described by the Chaplygin equation of state~\cite{Stan}. These days negative
pressure is recognized as a possible physical effect: exchange forces in atoms give
rise to negative pressure; stripe states in the quantum Hall effect may be a
consequence of negative pressure; the recently discovered cosmological constant
may be exerting negative pressure on the cosmos, thereby accelerating expansion. 

For any
form of $V$, the model possesses the Galileo symmetry appropriate to
nonrelativistic dynamics.  The Galileo transformations comprise the time and
space translations, as well as space rotations.  The corresponding constants of
motion are the energy~$E$.
\begin{equation} E=\int \rd r {\cal H} = \int \rd r
\Bigl(\rho\frac{(\grad\theta)^2}{2}+V(\rho)\Bigr)\qquad\mbox{(time
translation)}
\end{equation}
 the momentum~$\vec P$ (whose density~$\vec{\cal P}$ equals the spatial
current),
\begin{equation}
\vec P=\int \rd r \vec{\cal P} = \int\rd r \vec j = \int \rd r \rho\vec v\qquad
          \mbox{(space translation)}
\end{equation}
 and the angular momentum~$M^{ij}$.
\begin{equation} M^{ij}=\int\rd r \left(r^i{\cal P}^j- r^j{\cal P}^i\right)\qquad
\mbox{(rotation)}
\end{equation}
 The action of these transformations on the fields is straightforward: the time and
space arguments are shifted or the space argument is rotated.  Slightly less
trivial is the action of Galileo boosts, which boost the spatial coordinate by a
velocity
$\vec u$
\begin{equation}
\vec r \rightarrow \vec R=\vec r -t\vec u.
\end{equation}
 The density field transforms trivially: its spatial argument is boosted,
\begin{equation}
\rho(t,\vec r) \rightarrow \rho_{\vec u}(t,\vec r)=\rho(t,\vec R)
\end{equation}
 but the velocity potential acquires an inhomogeneous term.
\numeq{inhomterm}{
\theta(t,\vec r) \rightarrow \theta_{\vec u}(t,\vec r) =\theta(t,\vec R) + \vec u
\cdot \vec r - \frac{u^2}2 t
}
 Those of you familiar with field theoretic realizations of the Galileo group
will recognize the inhomogeneous term as the well-known Galileo $1$-cocyle. 
It ensures that the velocity, the gradient of $\theta$, transforms appropriately. 
\begin{equation}
\vec v(t,\vec r) \rightarrow \vec v_{\vec u} (t,\vec r) = \vec v (t,\vec r-t\vec
u) + \vec u
\end{equation}
 The associated conserved quantity is the ``boost generator''.
\begin{equation}
\vec B = t\vec P - \int\rd r \rho\vec r\qquad
          \mbox{(Galileo boost)}
\end{equation}
 Finally, also conserved is the total number. 
\begin{equation} N=\int \rd r \rho\qquad
          \mbox{(particle number)}
\end{equation}
The corresponding transformation shifts $\theta$ by a constant.

These transformations and constants of motion fit into the general theory: the
action is invariant against the transformations; Noether's theorem can be used
to derive the constants of motion; their time independence is verified with
the help of the equations of motion -- indeed, the continuity equations
\refeq{conteq}, \refeq{conteqs}, and \refeq{conteqsII} as well as the
symmetry of $T^{ij}$ guarantee this.  Also, using the basic Poisson brackets
\refeq{canconj} for
the $(\rho,\theta)$ variables, one can check that each infinitesimal
transformation is generated by Poisson bracketing with the appropriate
constant;  Poisson bracketing the constants with each other reproduces the
Galileo Lie algebra with a central extension given by $N$, which corresponds to
the familiar Galileo $2$-cocycle.  There are a total of $\fract{1}{2}(d+1)(d+2)$
Galileo generators in $d$ space plus one time dimensions.  Together with the
central term, we have a total of 
$\fract{1}{2}(d+1)(d+2)+1$ generators.

Another useful consequence of the symmetry transformations is that they
map solutions of the equations of motion into new solutions.  Of course, ``new''
solutions produced by Galileo transformations are trivially related to the old
ones: they are simply shifted, boosted or rotated.  

[The free theory as well as the adiabatic theory with $n=1+2/d$ are also invariant
against the SO(2,1) group of time translation, dilation, and conformal
transformation~\cite{HasHor}, which together with the Galileo group form the
``Schr\"odinger group'' of nonrelativistic motion, whenever the
energy-momen\-tum ``tensor'' satisfies $2T^{oo} = T^i{}_i$~\cite{Jac72}.]

But we shall now turn to the
specific Chaplygin gas model, with
$V(\rho)= {\lambda}/{\rho}$, which possesses additional and unexpected
symmetries.

The Chaplygin gas action and consequent Bernoulli equation for the Chaplygin
gas in
$(d,1)$ space-time read
\begin{gather} I^{\mathrm{Chaplygin}}_\lambda= \int \rd t \int \rd r
\Bigl(\theta\dot\rho -
\fract{1}{2}\rho(\grad \theta)^2 -
\frac{\lambda}{\rho}\Bigr)\label{chapgasact}\\
\dot\theta + \frac{(\grad \theta)^2}{2} =
\frac{\lambda}{\rho^2}\label{conseqBern}
\end{gather}
  This model possesses further space-time symmetries beyond those of
the Galileo group~\cite{BazJac}.  First of all, there is a one-parameter~($\omega$)
time rescaling transformation
\begin{equation}\label{trt}
 t\rightarrow T=e^{\omega} t,
\end{equation}
 under which the fields transform as 
 \begin{align}
 \theta(t,\vec r) &\rightarrow \theta_\omega (t,\vec r) = e^\omega
 \theta(T,\vec r)\label{trt2}\\
 \rho(t,\vec r) &\rightarrow \rho_\omega (t,\vec r) = e^{-\omega} \rho(T,\vec
 r).\label{trt3}
 \end{align}
 Second, in $d$ spatial dimensions, there is a
vectorial, $d$-parameter~($\vec\omega$) space-time mixing transformation
\begin{align} t&\rightarrow T(t,\vec r) = t+\vec\omega \cdot \vec r +
\fract{1}{2}\omega^2
\theta(T,\vec R)\label{eq:100}\\
\vec r &\rightarrow \vec R(t,\vec r) = \vec r + \vec \omega \,\theta(T,\vec R)
\end{align}
 Note that the transformation law for the coordinates involves the
$\theta$ field itself.  Under this transformation, the fields transform according
to
\begin{align}
\theta(t,\vec r) &\rightarrow \theta_{\vec\omega}(t,\vec r) = \theta(T,\vec
R)\label{eq:102}\\
\rho(t,\vec r) &\rightarrow \rho_{\vec\omega}(t,\vec r) = \rho(T,\vec
R)\frac{1}{|J|},
\end{align}
 with $J$ the Jacobian of the transformation linking $(T,\vec R)\rightarrow
(t,\vec r)$. 
\numeq{matrlink}{
J = \det \begin{pmatrix} 
\displaystyle\frac{\partial T}{\partial t} &\displaystyle  \frac{\partial
T}{\partial{r^j}}\\[2ex]
\displaystyle\frac{\partial R^i}{\partial t} &\displaystyle \frac{\partial
R^i}{\partial r^j}
\end{pmatrix} =
\Bigl(1-\vec\omega \cdot\grad \theta(T,\vec R) - \frac{\omega^2}2\,  \dot\theta
(T,\vec R)\Bigr)^{-1} }
(The time and space derivatives in the last element are with respect to~$t$
and~$\vec r$.)
 One can again tell the complete story for these transformations:
The action is invariant; Noether's theorem gives the conserved quantities,
which for the time rescaling is
\begin{equation} 
D=tH-\int \rd r \rho\theta \qquad\mbox{(time rescaling)}
\label{eq:105}
\end{equation}
 while for the space-time mixing one finds
\begin{equation}
\vec G = \int \rd r \left(\vec r {\cal H} - \theta \vec{\cal P}\right)
\qquad\mbox{(space-time mixing).}
\label{eq:106}
\end{equation}
 The time independence of $D$ and $\vec G$  can be verified with the help of the
equations of motion (continuity and Bernoulli).  Poisson bracketing the
fields~$\theta$ and~$\rho$ with 
$D$ and $\vec G$ generates the appropriate infinitesimal transformation on the
fields.

So now the total number of generators is the sum of the previous
$\fract12(d+1)(d+2) +1$ with $1+d$ additional ones
\begin{equation}
\fract{1}{2}(d+1)(d+2) + 1 + 1 +d = \fract{1}{2}(d+2)(d+3). 
\end{equation}
 When one  computes the Poisson brackets of all  these with each other one finds 
the Poincar\'e Lie algebra in one higher spatial dimension, that is,
in $(d+1,1)$-dimensional space-time, where the Poincar\'e group possesses
$\fract12 (d+2)(d+3)$ generators~\cite{BorHop}.  Moreover, one verifies that
$(t,\theta,\vec r)$ transform  linearly as a $(d+2)$ Lorentz vector in light-cone
components, with $t$ being the ``$+$'' component and $\theta$ the ``$-$''
component~\cite{Baz}.

Presently, we shall use these additional symmetries to generate new solutions
from old ones, but, in contrast with what we saw earlier, the new solutions will
be nontrivially linked to the former ones.  Note that the additional
symmetry holds even in the free theory. 

Before proceeding, let us observe that $\rho$ may be eliminated by using the
Bernoulli equation to express it in terms of $\theta$.  In this way, one is led to
the following
$\rho$-independent action for $\theta$ in the Chaplygin gas problem:
\begin{equation} 
I^{\mathrm{Chaplygin}}_\lambda=-2\sqrt\lambda \int dt \int \rd r
\sqrt{\dot\theta+\frac{(\grad
\theta)^2}{2}}.\label{rindependent}
\end{equation}
 Although this operation is possible only in the interacting case, the
interaction strength is seen to disappear from the equations of motion.
\numeq{interstren}{
\frac\partial{\partial t} \frac1{\sqrt{\dot\theta+\frac{(\grad
\theta)^2}{2}}} +\grad \cdot \frac{\grad\theta}{\sqrt{\dot\theta+\frac{(\grad
\theta)^2}{2}}} = 0 
}
$\lambda$ merely
serves as an overall factor in the action.

The action \refeq{rindependent} looks unfamiliar; yet it is Galileo invariant.
[The combination $\dot\theta+\frac12 (\grad\theta)^2$ responds to Galileo
transformations without a 1-cocycle; see \refeq{inhomterm}.]
 Also
$I^{\mathrm{Chaplygin}}_\lambda$ possesses the additional symmetries described
above, with $\theta$ transforming according to the previously recorded equations.

Let us discuss some solutions.  For example, the free theory is solved by
\begin{equation}
\theta(t,\vec r) = \frac{r^2}{2t} 
\label{freesol}
\end{equation}
 which corresponds to the velocity
\begin{equation}
\vec v(t,\vec r) = \frac{\vec r}{t}.
\end{equation}
Galileo transforms generalize this in an obvious manner into a set of  solutions. 
(The charge density~$\rho$ is determined by its initial condition.  In the free
theory, $\rho$ is an independent quantity, and I shall not discuss it here.) 
Performing on the above formula for~$\theta$ the new transformations of
time-rescaling and space-time mixing, we find that the  solution is
invariant.

We can find a   solution similar to \refeq{freesol} in the interacting case, for
$d>1$, which we henceforth assume (the $d=1$ case will be separately
discussed later).  One verifies that a solution is
\begin{equation}
\theta(t,\vec r) = -\frac{r^2}{2(d-1)t} \qquad \rho(t,\vec r) =
\sqrt{\frac{2\lambda}{d}}(d-1)\frac{|t|}{r}
\label{disclat}
\end{equation}
\begin{equation}
\vec v(t,\vec r) =  -\frac{\vec r}{(d-1)t} \qquad \vec j(t,\vec r) = -\epsilon(t)
\sqrt{\frac{2\lambda}{d}} \hat r.
\end{equation}
Note that the speed of sound
\numeq{speedsound}{
s=\frac{\sqrt{2\lambda}}\rho = \frac{\sqrt d\,  r}{(d-1)t} = \sqrt d v
}
exceeds~$v$.  Again this solution can be translated, rotated, and boosted. 
Moreover, the solution is time-rescaling--invariant.  However, the
space-time mixing transformation produces a wholly different kind of solution. 
This is best shown graphically, where the $d=2$ case is exhibited (\emph{see
figure})~\cite{AnDet}. 
\begin{figure}[hbtp]
\begin{gather*}
\BoxedEPSF{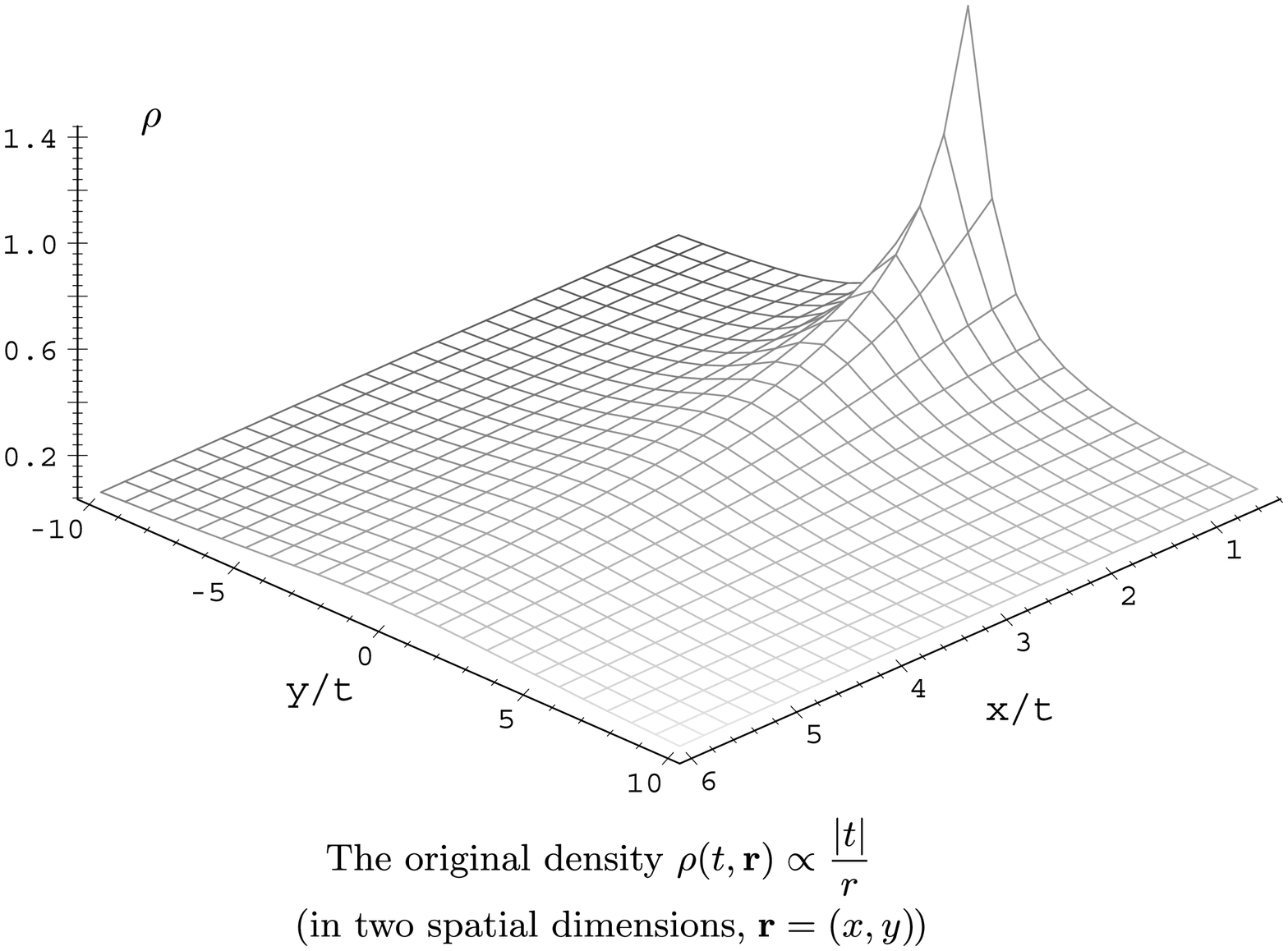 scaled 700}\\[1ex]
\BoxedEPSF{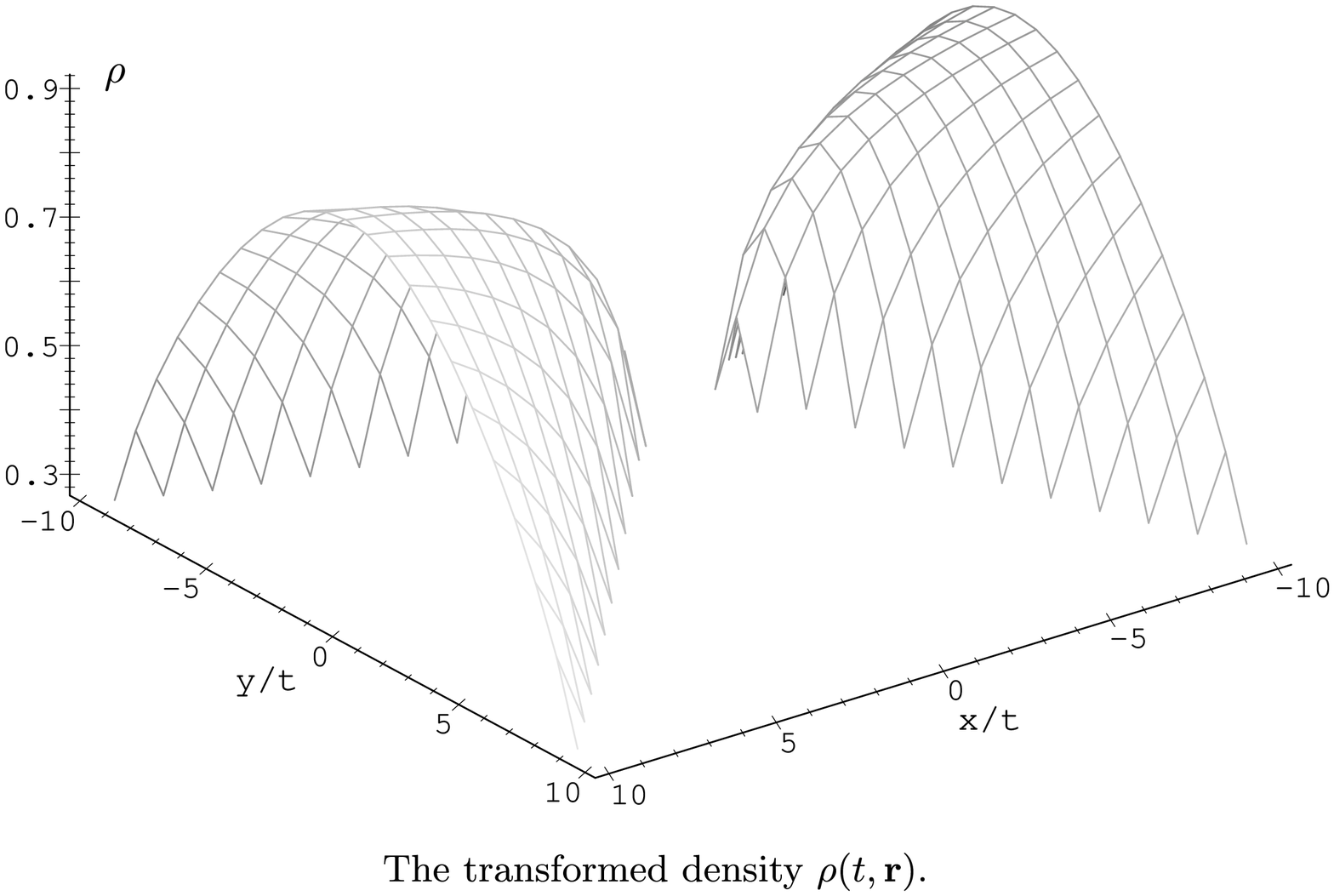 scaled 700}
\end{gather*}
\vspace*{-1pc}
\end{figure}

Another interesting solution, which is essentially one-dimensional (lineal), even
though it exists in arbitrary spatial dimension, is given by
\numeq{intersol}{
\theta(t,\vec r) = \Theta (\hat n\cdot\vec r) + \vec u \cdot \vec r - \fract12
t\bigl(u^2 - (\hat n\cdot \vec u)^2\bigr).
 }
Here $\hat n$ is a spatial unit vector,  $\vec u$ is an arbitrary vector with
dimension of velocity, while $\Theta$ is an arbitrary function with static
argument, which can be boosted by the Galileo transform~\refeq{inhomterm}. The
corresponding charge density is time-independent.
 \numeq{timeindep}{
\rho(t,\vec r) = \frac{\sqrt{2\lambda}}{\hat n\cdot\vec u + \Theta'(\hat
n\cdot\vec r)}
 }
The current is static and divergenceless.
\numeq{curstat}{
\vec j(t,\vec r) = \sqrt{2\lambda} \Bigl(\hat n + 
\frac{\vec u - \hat n(\hat n\cdot\vec u)}{\hat n\cdot\vec r +
 \Theta'(\hat n\cdot\vec r)}\Bigr)
}
The sound speed $s=\sqrt{2\lambda}/\rho = \Theta'(\hat n\cdot\vec r) + \hat
n\cdot \vec u$ is just the $\hat n$ component of the velocity $\vec v = \grad
\theta =
\hat n \Theta'(\hat n\cdot\vec r) +  \vec u$.

Finally, we record a planar static solution to \refeq{interstren}
\numeq{solinterstren}{
\theta(\vec r) = \Theta(\hat n_1\cdot\vec r / \hat n_2\cdot\vec r)
}
where $\hat n_1$ and $\hat n_2$ are two orthogonal unit vectors~\cite{Oga}. 
\smallbreak

\centerline{\hfill\hbox to 2in{\hrulefill} \hfill}
\vspace*{-\smallskipamount}

\begin{problem}\label{prob:6}
\emph{
Show that the solution for $\theta$ given in \refeq{freesol} is invariant under the
time rescaling transformation \refeq{trt}, \refeq{trt2}, and under the
space-time mixing transformation \refeq{eq:100}--\refeq{eq:102}.  }

\centerline{\hfill\hbox to 2in{\hrulefill} \hfill}
\end{problem}

\subsection{Lorentz-invariant relativistic model}\label{L-irm}

We now turn to a Lorentz-invariant generalization of our Galileo-invariant
Chaplygin model in
$(d,1)$-dimensional space-time.  We already know from
\refeq{eq:34}--\refeq{eq:37} how to construct the free Lagrangian by using
Eckart's method with a relativistic kinetic energy.
\begin{equation} T(\vec v) = -c^2\sqrt{1- {v^2}/{c^2}}
\end{equation}
 Recall that mass has been scaled to unity, and that we retain the velocity
of light $c$ to keep track of the nonrelativistic $c\rightarrow\infty$ limit. 
Evidently, the momentum is
\begin{equation}
\vec p = \frac{\partial T(\vec v)}{\partial \vec v}=\frac{\vec
v}{\sqrt{1-{v^2}/{c^2}}}.
\end{equation}
 Thus the free relativistic Lagrangian, with current conservation
enforced by the Lagrange multiplier~$\theta$, reads [compare~\refeq{eq:34}]
\begin{equation} \bar L_0^{\mathrm{Lorentz}} = \int \rd r \Bigl(-c^2\rho
\sqrt{1- {v^2}/{c^2}} +
\theta\bigl(\dot\rho+\grad\cdot(\rho\vec v)\bigr)\Bigr).\label{laglor}
\end{equation}
 This may be presented in a Lorentz-covariant form in terms of a current
four-vector
$j^{\mu}=(c\rho,\rho\vec v)$.  $\bar L_0^{\mathrm{Lorentz}}$ of equation
(\ref{laglor}) is thus equivalent to [compare~\refeq{notethat}, \refeq{curr4vec}]
\begin{equation}
\bar L_0^{\mathrm{Lorentz}} = \int \rd r \Bigl(
  - j^\mu\partial_\mu \theta -c\sqrt{j^\mu j_\mu}\Bigr).
\label{Lorcov}
\end{equation}
 Eliminating $\vec v$ in (\ref{laglor}), we find, as before, that
\begin{equation}
\vec p = \frac{\partial T}{\partial \vec v} = \frac{\vec v}{\sqrt{1-v^2}/c^2}
= \grad \theta,
\qquad
\vec v =
\frac{\grad\theta}{\sqrt{1+ {(\grad\theta)^2}/{c^2}}},
\end{equation}
 and the free Lorentz-invariant Lagrangian reads [compare~\refeq{eq:36},
\refeq{eq:37}]
\begin{equation}
 L_0^{\mathrm{Lorentz}}=\int\rd r \Bigl(\theta\dot\rho -\rho
c^2\sqrt{1+ {(\grad\theta)^2}/{c^2}}\, \Bigr).
\label{LiL}
\end{equation}

To find $L_0^{\mathrm{Galileo}}$ of~\refeq{eq:31} as the nonrelativistic limit of
$L_0^{\mathrm{Lorentz}}$ in~\refeq{LiL}, a nonrelativistic~$\theta$ variable 
must be extracted from its relativistic counterpart. Calling the former
$\theta_{\mathrm{NR}}$ and the latter, which occurs in \refeq{LiL},
$\theta_{\mathrm{R}}$, we define
\numeq{thetaRNR}{
\theta_{\mathrm{R}} \equiv -c^2 t  + \theta_{\mathrm{NR}}.
}
It then follows that apart from a total time derivative
\numeq{tottimdder}{
 L_0^{\mathrm{Lorentz}} \mathop{\hbox to 3em{\rightarrowfill}}_{c\to\infty}
L_0^{\mathrm{Galileo}}.
}

 Next, one wants to include interactions.  While there are many ways to allow for
Lorentz-invariant interactions, we seek an expression that reduces to the
Chaplygin gas in the nonrelativistic limit.  Thus, we choose~\cite{JacPol}
\begin{equation}
 L_a^{\BI}=\int\rd r
\left(\theta\dot\rho - 
\sqrt{\rho^2 c^2+a^2}\sqrt{c^2+(\grad\theta)^2}\, \right),
\label{laglorint}
\end{equation}
 where $a$ is the interaction strength.   [The reason for the nomenclature will
emerge presently.] We see that, as
$c\to\infty$,
\begin{equation} L_a^{\mathrm{\BI}}  \mathop{\hbox to
3em{\rightarrowfill}}_{c\to\infty} L_{\lambda= {a^2}/{2}}^{\mathrm{Chaplygin}}.
\end{equation}
[Again $\theta_{\mathrm{NR}}$ is extracted from $\theta_{\mathrm{R}}$ as in
\refeq{thetaRNR} and a total time derivative is ignored.]
 Although it perhaps is not obvious, (\ref{laglorint}) defines a
Poincar\'e-invariant theory, and this will be explicitly demonstrated below. 
Therefore, $L_a^{\mathrm{\BI}}$ possesses Lorentz and Poincar\'e
symmetries in
$(d,1)$ space-time, with a total of $\fract{1}{2}(d+1)(d+2)+1$ generators, where
the last ``${}+1$''  refers to the total number~$N = \int \rd r \rho$.  

When $a=0$, the
model is free and elementary. It was demonstrated previously
[eqs.~\refeq{eq:34}--\refeq{freeEuler}] that the free equations of motion are
precisely the same as in the nonrelativistic free model, so the complete solution
\refeq{initdata1}--\refeq{initdata4}
works here as well.  For
$a\ne 0$, in the presence of interactions, one can eliminate
$\rho$ as before, and one is left with a Lagrangian just for the $\theta$ field.  It
reads
\begin{equation} L_a^{\mbox{\scriptsize Born-Infeld}}=-a\int\rd r
\sqrt{c^2-(\partial_\mu
\theta)^2}.
\label{B-I}
\end{equation}
 This is a Born-Infeld-type theory for a scalar field $\theta$; its Poincar\'e
invariance is manifest, and again, the elimination of $\rho$ is only possible
with nonvanishing $a$, which however disappears from the dynamics,
serving merely to normalize the Lagrangian.

The equations of motion that follow from~\refeq{laglorint} read
\begin{align}  
\dot\rho + \grad \cdot \Biggl(
 \grad\theta  \sqrt{\frac{\rho^2
c^2+a^2}{c^2+(\grad\theta)^2}}\,\Biggr) &= 0
\label{followsfrom1}\\[1ex]
\dot\theta + \rho c^2 \sqrt{\frac{c^2+(\grad\theta)^2}{\rho^2
c^2+a^2}} &=0\ . \label{followsfrom2}\\[1ex]
\intertext{The density $\rho$ can be evaluated in terms of $\theta$ from
\refeq{followsfrom2}; then \refeq{followsfrom1} reads}
\partial^\alpha \Bigl( \frac1{\sqrt{c^2-(\partial_\mu \theta)^2}}\, 
\partial_\alpha \theta\Bigr) &= 0 \label{followsfrom}
\end{align}
 which also follows from
\refeq{B-I}.  After $\theta_{\mathrm{NR}}$ is extracted from
$\theta_{\mathrm{R}}$ as in
\refeq{thetaRNR} we see that in the
nonrelativistic limit $L_a^{\mbox{\scriptsize Born-Infeld}}$ \refeq{laglorint} or
\refeq{B-I} becomes $L_\lambda^{\mathrm{Chaplygin}}$
of~\refeq{chapgasact} or \refeq{rindependent},
\numeq{B-ItoC}{
L_a^{\mbox{\scriptsize Born-Infeld}}  \mathop{\hbox to
3em{\rightarrowfill}}_{c\to\infty} L_{\lambda=a^2/2}^{\mathrm{Chaplygin}}
 }
and the equations of motion \refeq{followsfrom1}--\refeq{followsfrom} reduce  to
\refeq{conteq}, \refeq{conseqBern}, and \refeq{interstren}.

In view of all the similarities to the nonrelativistic Chaplygin gas, it comes as
no surprise that the relativistic Born-Infeld  theory possesses additional
symmetries. These additional symmetry transformations, which leave
(\ref{laglorint}) or (\ref{B-I}) invariant, involve a one-parameter ($\omega$) 
 reparameterization of time, and a $d$-parameter ($\vec
\omega$) vectorial reparameterization of space. Both transformations are
field dependent.

The time transformation is given by an implicit formula involving also the
field~$\theta$~\cite{ref20}, 
\begin{equation}
 t \to T(t, \vec r) = \frac{t}{\cosh c^2\omega} +
\frac{\theta(T, \vec r)}{c^2} \tanh c^2
\omega
\end{equation}  
while the field transforms according to
\begin{equation}
\theta(t, \vec r) \to \theta_\omega (t, \vec r)= \frac{\theta(T,
\vec r)}{\cosh c^2\omega} - c^2 t \tanh c^2 \omega\ .
\label{eq:135}
\end{equation}  
[We record here only the transformation on~$\theta$; how~$\rho$ transforms can
be determined from the (relativistic) Bernoulli equation, obtained by
varying~$\rho$ in \refeq{laglorint}, which expresses~$\rho$ in terms
of~$\theta$. Moreover, \refeq{eq:135} is sufficient for discussing the invariance of
\refeq{B-I}.] The infinitesimal generator, which is time independent by virtue of
the equations of motion, is~\cite{BorHop94}
\begin{align}
 D &= \int \rd{\vec{r}} \Big(c^4 t \rho + \theta
\sqrt{\rho^2c^2+a^2} \sqrt{c^2 + (\grad \theta)^2}
\Big)\nonumber\\
&= \int \rd{\vec{r}} (c^4 t \rho + \theta {\cal H}) \qquad\mbox{(time
reparameterization).}\label{infgener}
\end{align} 

A second class of transformations involving a reparameterization
of the spatial variables is implicitly defined by~\cite{ref20}.
\begin{equation}
\vec r \to \vec R(t, \vec r) = \vec r - \hat{\omega} \theta (t, \vec R) \frac{\tan
c\omega}{c} +
\hat{\omega}(\hat{\omega}\cdot\vec r) \bigg( \frac{1-\cos c\omega}{\cos
c\omega} \bigg)
\end{equation} 
\begin{equation}
\theta (t, \vec r)  \to \theta_{\vec\omega} (t, \vec r) =
\frac{\theta(t, \vec R) - c(\hat{\omega}\cdot \vec r) \sin
c\omega}{\cos c\omega}
\end{equation}  
 $\hat\omega$ is the  unit 
vector $\vec\omega/\omega$ and $\omega
=\sqrt{\vec\omega\cdot\vec\omega}$.    The time-independent generator of the
infinitesimal transformation reads~\cite{BorHop94}
\begin{align}
{\vec G} &=\int \rd{\vec{r}} (c^2 \vec r \rho + \theta \rho
\grad\theta)\nonumber\\
&= 
\int
\rd{\vec{r}} (c^2
\vec r\rho +\theta {\cal P})\qquad\mbox{(space re\pr).}
\label{eq:88}
\end{align} 
Of course the Born-Infeld action \refeq{laglorint} or \refeq{B-I} is invariant
against these transformations, whose infinitesimal form is generated by the
constants.

With the addition of $D$ and $\vec G$ to the previous
generators, the Poincar\'e algebra in $(d+1,1)$ dimension is
reconstructed, and $(t, \vec r,\theta)$ transforms linearly as a
$(d+2)$-dimensional Lorentz vector (in Cartesian
components)~\cite{Baz}.  Note that this symmetry also
holds in the free, $a=0$, theory.

It is easy to exhibit solutions of the relativistic equation \refeq{followsfrom},
which reduce to solutions of the nonrelativistic, Chaplygin gas equation
\refeq{interstren} [after
$-c^2t$ has been removed, as in \refeq{thetaRNR}].   For example 
\numeq{nonrelChap}{
\theta(t,\vec r) = -c \sqrt{c^2t^2 + \frac{r^2}{d-1}}
}
solves \refeq{followsfrom} and reduces to \refeq{disclat}.  The relativistic analog
of the lineal solution \refeq{intersol} is 
\numeq{relanal}{
\theta(t,\vec r) = \Theta(\hat n\cdot\vec r) + \vec n\cdot\vec r - ct
\sqrt{c^2 + u^2 -(\hat n \cdot \vec u)^2} .
}
 [Note that the above profiles continue to solve \refeq{followsfrom}, even when
the sign of the square root is reversed, but then they no longer possess a 
nonrelativistic limit.] 

Additionally there exists an essentially relativistic solution, describing massless
propagation in one direction: according to \refeq{followsfrom}, $\theta$ can satisfy
the wave equation
$\sqcap\llap{$\sqcup$} \theta = 0$, provided
$(\partial_\mu \theta)^2 = $~constant, as for example with plane waves 
\numeq{withplanew}{
\theta(t,\vec r) = f(\hat n\cdot\vec r \pm ct)
}
 where $(\partial_\mu \theta)^2$ vanishes.  Then $\rho$ reads, from
\refeq{followsfrom2},
\numeq{eq:143}{
\rho = \mp \frac a{c^2} f' \ . 
}
Other solutions are given in ref.~\cite{Gib}.

\subsection{Some remarks on relativistic fluid mechanics}

The Born-Infeld model reduces in the nonrelativistic limit to the Chaplygin gas.  
Equations governing the latter belong to fluid mechanics, but the Born-Infeld
equations do not readily expose their fluid mechanical structure.  Nevertheless
they do in fact describe a relativistic fluid.  In order to demonstrate this, we give
a pr\'ecis of relativistic fluid mechanics.  

Usually the dynamics of a relativistic fluid is presented in terms of the
energy-momentum tensor, $\theta^{\mu\nu}$, and the equations of motion are just
the conservation equations $\partial_\mu \theta^{\mu\nu}=0$~\cite{Wein}. [We
denote the relativistic energy-momentum tensor by $\theta^{\mu\nu}$, to
distinguish it from the nonrelativistic $T^{\mu\nu}$ introduced in \refeq{conteqs}
and \refeq{conteqsII}. The limiting relation between the two is given below.] But
we prefer to begin with a Lagrange density 
\numeq{Lagdens}{
{\cal L} = -j^\mu a_\mu - f(\sqrt{j^\mu j_\mu}).
}
Here $j^\mu$ is the current Lorentz vector $j^\mu= (c\rho,\vec j)$. The $a_\mu$
comprise a set of auxilliary variables; in the relativistic analog of irrotational
fluids we take $a_\mu=\partial_\mu\theta$, more generally 
\numeq{relanalirr}{
a_\mu = \partial_\mu\theta + \alpha \partial_\mu\beta 
}
so that the \CS\ density of $a_i$ is a total derivative 
[compare \refeq{notethat}, \refeq{curr4vec}]. The function~$f$ depends on the
Lorentz invariant $j^\mu j_\mu = c^2 \rho^2 -\vec j^2$ and encodes the specific
dyanamics (equation of state).

The energy momentum tensor for  
$\cal L$ is 
\numeq{enmomten}{
\theta_{\mu\nu} = -g_{\mu\nu} {\cal L} + \frac{j_\mu j_\nu}{\sqrt{j^\alpha
j_\alpha}} f' (\sqrt{j^\alpha j_\alpha}). 
}
 [One way to derive \refeq{enmomten} from \refeq{Lagdens} is to
embed that expression in an external metric tensor $ g_{\mu\nu}$, which is then
varied; in the variation $j^\mu$ and $a_\mu$ are taken to be metric-independent
and
$j_\mu= g_{\mu\nu} j^\nu$.] 
Furthermore,  varying $j^\mu$ in \refeq{Lagdens} shows that 
\numeq{varyingjmu}{
a_\mu = -\frac{j_\mu}{\sqrt{j^\alpha j_\alpha}} f'(\sqrt{j^\alpha j_\alpha})
}
so that \refeq{enmomten}  becomes  
\numeq{enmombec}{
\theta_{\mu\nu} = -g_{\mu\nu} [nf'(n) - f(n)] + u_\mu u_\nu n f'(n) 
}
where we have introduced the proper velocity $u_\mu$ by
\numeq{intrumu}{
j_\mu= nu_\mu\qquad u^\mu u_\mu=1
}
so that $n$ is proportional to the proper density and $1/n$  is proportional to the
specific volume.  Eq.~\refeq{enmombec} identifies the proper energy density~$e$
and the pressure~$P$ (which coincides with $\cal L$) through the conventional
formula~\cite{Wein}. 
\numeq{convform}{
\theta_{\mu\nu} = -g_{\mu\nu} P + u_\mu u_v (P + e) 
}
Therefore, in our case
\begin{align}
e &= f(n) \label{efn}\\
P &= n f'(n) - f(n)\ . 
\label{pnfn}
\end{align}
The thermodynamic relation involving entropy~$S$ reads
\numeq{thermts}{
P \rd{\Bigl(\frac1n\Bigr)} + \rd{\Bigl(\frac en\Bigr)} \propto \rd S 
}
where the proportionality constant is determined by the temperature.
With \refeq{efn} and \refeq{pnfn}  the left side of \refeq{thermts} vanishes
and we verify that entropy is constant, that is, we are dealing with an
isentropic system, as has been stated in the very beginning.

For the free system, the pressure vanishes, so we choose $f(n) = cn$
\numeq{pressvan}{
{\cal L}_0 = -j^\mu a_\mu - c\sqrt{j^\mu j_\mu}.
}
In the ``irrotational'' case, $a_\mu=\partial_\mu \theta$,  and with $\vec j =
\rho\vec v$ this Lagrange density  produces the free 
$\bar L_0^{\mathrm{Lorentz}}$ of \refeq{laglor}, \refeq{Lorcov} (apart from a 
total time derivative).

For the Born-Infeld theory, we  present 
the pressure~$P$ in \refeq{pnfn} by choosing $f(n)\! = c\sqrt{a^2\! +\! n^2}$, 
 which corresponds to the pressure $P =
-a^2 c/ \sqrt{a^2 + n^2}$.  When the current is written as 
\begin{subequations}
\numeq{curwrit}{
j_\mu = \frac{a\partial_\mu \theta}{\sqrt{c^2 - (\partial_\mu \theta)^2 }}
}
so that 
\numeq{curwrit2}{
n = a \sqrt{\frac{(\partial_\mu \theta)^2}{c^2 - (\partial_\mu \theta)^2}}
}
the Lagrange density
\numeq{curwrit3}{
{\cal L} = P = -\frac{a^2 c}{\sqrt{a^2+n^2}}
}
 coincides with that for the Born-Infeld 
model in eq.~\refeq{B-I}.

\end{subequations}

Other forms for~$f$ give rise to relativistic fluid mechanics with other equations of
state.  

It is interesting to see how the nonrelativistic limit of
$\theta^{\mu\nu}$ in \refeq{enmombec} produces $T^{\mu\nu}$ of 
\refeq{conteqs}--\refeq{conteqsII}. It is especially intriguing to notice that
$\theta^{\mu\nu}$ is symmetric but $T^{\mu\nu}$ is not. To make the connection 
we recall that $u^\mu = 1/\sqrt{1-v^2/c^2} (1, \vec v/c)$, we observe that
$n=\sqrt{\rho^2 c^2 -\vec j^2}$, set $\vec j = \rho\vec v$ and
conclude that
$n=\linebreak[4]\rho c \sqrt{1-v^2/c^2} \sim \rho c - (\rho v^2/2c)$. Also $f(n)$ is
chosen to be
$cn + V(n/c)$, and thus $P=n f'(n) - f(n) = (n/c) V' (n/c) - V(n/c)$. It follows
that 
\begin{align}
\theta^{oo} &= \frac{nc - ({v^2 n}/{c^3})V'}{1-v^2/c^2} + V \approx 
\frac{\rho c^2 -   \rho v^2/2}{1-v^2/c^2} + V(\rho) \nonumber\\
&\approx \rho c^2 + \frac{\rho v^2}2 + V(\rho) = \rho c^2 + T^{oo}\ .
\label{itfollowsthat}
\end{align}
Thus, apart from the relativistic ``rest energy'' $\rho c^2$, $\theta^{oo}$ passes to
$T^{oo}$. The relativistic energy flux is $c\theta^{jo}$ (because
$\frac\partial{\partial x^\mu} \theta^{\mu o} = \frac1c \dot \theta^{oo} + 
\partial_j\theta^{jo}$)
\begin{align}
c \theta^{jo} &= \frac{v^j}{1-v^2/c^2} \Bigl(nc + \frac nc V'\Bigr) \approx 
v^j \frac{\rho c^2 -   \rho v^2/2 + \rho V'(\rho)}{1-v^2/c^2}  \nonumber\\
&\approx j^j c^2 +  \rho v^j\bigl(v^2/2 +V'(\rho)\bigr) = j^j c   + T^{jo}\ .
\label{relenflux}
\end{align}
Again, apart from the $O(c^2)$ current, associated with the $O(c^2)$ rest energy in
$\theta^{oo}$, $T^{jo}$ is obtained in the limit. The momentum density is
$\theta^{oi}/c$ (because $\theta^{\mu\nu}$ has dimension of energy density). Thus 
\numeq{dimenerdens}{
\theta^{oi}/c = \frac{v^i/c^2}{1-v^2/c^2} \Bigl(nc + \frac nc V'\Bigr) \approx \rho v^i
= {\cal P}^i\ . 
 }
Finally, the momentum flux is obtained directly from $\theta^{ij}$. 
\begin{align}
\theta^{ij} &= \delta^{ij} \Bigl(\frac nc V' - V\Bigr) + \frac{v^i v^j}{c^2 - v^2} \Bigl(nc
+ \frac nc V'\Bigr)\nonumber\\
&\approx \delta^{ij} \bigl( \rho V'(\rho) - V(\rho)\bigr) + v^i v^j \rho = T^{ij}  
\label{finmomflux} 
\end{align}

From the limiting formula $n\sim\rho c$ we also see that the pressure in
\refeq{curwrit3} tends to the Chaplygin expression~$-a/\rho$.

\newpage
\section{Common Ancestry:  The Nambu-Goto Action}

The ``hidden'' symmetries and the associated transformation laws for the
Chaplygin and Born-Infeld models may be given a coherent setting by considering
the Nambu-Goto action for a d-brane in $(d+1)$ spatial dimensions, moving on
$(d+1,1)$-dimensional space-time.  In our context, a~d-brane is simply a
$d$-dimensional extended object: a $1$-brane is a string, a $2$-brane is a
membrane and so on.  A~d-brane in $(d+1)$ space divides that space in two.

The Nambu-Goto action reads
\begin{align}
I_{\rm NG} &= \int\rd{ \phi^0} \rd\phi {\cal L}_{\rm NG} = 
\int \rd{ \phi^0} \rd{
\phi^1}
\cdots \rd{
\phi^d}
\sqrt{G}\label{N-Gr1}\\
 G &= (-1)^d \det \frac{\partial X^\mu}{\partial \phi^\alpha}
\frac{\partial X_\mu}{\partial \phi^\beta}
\end{align}
Here $X^\mu$ is a $(d+1,1)$ target space-time (d-brane) variable, with $\mu$
extending over the range $\mu=0,1,\dots,d,d+1$.  The $\phi^\alpha$ are
``world-volume'' variables describing the extended object with $\alpha$ ranging
$\alpha=0,1,\dots,d$; $\phi^\alpha$,
$\alpha=1, \dots, d$, parameterizes the $d$-dimensional d-brane that evolves
in $\phi^0$.

The Nambu-Goto action is parameterization invariant, and we shall show that
two different choices of parameterization (``light-cone'' and ``Cartesian'') lead to
the Chaplygin gas and Born-Infeld actions, respectively.  For both
parameterizations we choose
$(X^1, \dots, X^d)$ to coincide with $(\phi^1,
\dots, \phi^d)$, renaming them as $\vec r$ (a $d$-dimensional vector). This is
usually called the ``static \pr''. (The ability to carry out this parameterization
globally  presupposes that the extended object is topologically trivial; in the
contrary situation, singularities will appear, which are spurious in the sense that
they disappear in different parameterizations, and parameterization-invariant
quantities are singularity-free.)

\subsection{Light-cone \pr}\label{l-cpar}
For the light-cone parameterization we define $X^\pm$ as $\frac{1}{\sqrt{2}}
(X^0 \pm X^{d+1})$.  $X^+$ is renamed $t$ and identified with $\sqrt{2\lambda}\, \phi^0$.  This completes the fixing of the parameterization and the
remaining variable is $X^-$, which is a function of $\phi^0$ and $\vec\phi$,
or after redefinitions, of $t$ and $\vec r$.  $X^-$ is renamed as
 $\theta(t,\vec r)$ and then the Nambu-Goto action in this
parameterization coincides with the Chaplygin gas action 
$I^{\mathrm{Chaplygin}}_\lambda$ in \refeq{rindependent}~\cite{Gold}.

\subsection{Cartesian \pr}\label{c-cpar}
For the second, Cartesian parameterization $X^0$ is
renamed $ct$ and identified with $c \phi^0$.  The remaining target space
variable $X^{d+1}$, a function of $\phi^0$ and $\vec\phi$, equivalently of $t$
and $\vec r$, is renamed
$\theta(t,\vec r)/c$.  Then the Nambu-Goto action reduces
to the Born-Infeld action $\int \rd t L_a^{\BI}$, (\ref{B-I})~\cite{Gold}.

\subsection{Hodographic transformation}\label{h-trans}
There is another derivation of the Chaplygin gas from the Nambu-Goto action
that makes use of a hodographic transformation, in which independent and
dependent variables are interchanged.  Although the derivation is more
involved than the light-cone/static \pr\ used in Section~\ref{l-cpar} above,
the hodographic approach is instructive in that it gives a natural definition for
the density~$\rho$, which in the above static \pr\ approach is determined
from~$\theta$ by the Bernoulli equation~\refeq{conseqBern}.

We again use light-cone combinations: $\frac{1}{\sqrt{2}}
(X^0  + X^{d+1})$ is called $\tau$ and is identified with $\phi^0$, while 
$\frac{1}{\sqrt{2}} (X^0 - X^{d+1})$ is renamed~$\theta$. At this stage the
dependent, target-space variables are~$\theta$ and the transverse coordinates
$\vec X\colon X^i$, $(i=1,\ldots,d)$, and all are functions of the
world-volume parameters $\phi^0=\tau$ and
$\vec\phi\colon \phi^r$, $(r=1,\ldots,d)$; $\partial_\tau$ indicates differentiation
with respect to $\tau=\phi^0$, while $\partial_r$ denotes derivatives with
respect to $\phi^r$. The induced metric
$G_{\alpha\beta} = \frac{\partial X^\mu}{\partial\phi^\alpha} \frac{\partial
X_\mu}{\partial\phi^\beta}$ takes the form
\numeq{inducmetr}{
G_{\alpha\beta} = \begin{pmatrix}
G_{oo} & G_{os}\\
G_{ro} & -g_{rs}
\end{pmatrix} = 
\begin{pmatrix}
2\partial_\tau\theta - (\partial_\tau \vec X)^2& \partial_s\theta -
\partial_\tau \vec X \cdot \partial_s \vec X\\ 
\partial_r\theta -
\partial_r \vec X \cdot \partial_\tau \vec X&
-\partial_r \vec X \cdot \partial_s \vec X
\end{pmatrix}
}
The Nambu-Goto Lagrangian now leads to the canonical momenta
\begin{subequations}\label{NGL}
\begin{align}
\frac{\partial {\cal L}_{\mathrm NG}}{\partial {\partial_\tau} \vec X} &= \vec
p\label{NGLa}\\[1ex]
\frac{\partial {\cal L}_{\mathrm NG}}{\partial {\partial_\tau}  \theta} &=
\Pi\label{NGLb}
\end{align}
\end{subequations}
and can be presented in first-order form as 
\numeq{foform}{
 {\cal L}_{\mathrm NG} = \vec p\cdot \partial_\tau \vec X +
\Pi\partial_\tau\theta + \frac1{2\Pi} (p^2+g) +
u^r (\vec p \cdot \partial_r \vec X + \Pi\partial_r \theta)
}
where $g=\det g_{rs}$ and 
\numeq{Lagmult}{
u_r\equiv \partial_\tau \vec X \cdot \partial_r \vec X - \partial_r \theta 
}
acts as a Lagrange multiplier. Evidently the equations of motion are
\begin{subequations}\label{eveqmot}
\begin{align}
\partial_\tau \vec X &= - \frac1\Pi \vec p - u^r \partial_r 
\vec X\label{eveqmota}\\
\partial_\tau\theta &= \frac1{2\Pi^2} (p^2+g) - u^r
\partial_r\theta\label{eveqmotb}\\
\partial_\tau\vec p &= -\partial_r \Bigl(\frac1\Pi g g^{rs}\partial_s\vec X\Bigr)
- \partial_r(u^r\vec p)\label{eveqmotc}\\
\partial_\tau \Pi &= -\partial_r ( u^r \Pi)\label{eveqmotd}
\end{align}
\end{subequations}
Also there is the constraint
\numeq{alsoconstr}{
\vec p\cdot\partial_r \vec X  + \Pi \partial_r \theta = 0
}
[That $u^r$ is still given by \refeq{Lagmult} is a consequence of
\refeq{eveqmota} and \refeq{alsoconstr}.] Here $g^{rs}$ is inverse to~$g_{rs}$, and
the two metrics are used to move the $(r,s)$ indices.  The theory still possesses an
invariance against redefining the spatial parameters with a $\tau$-dependent
function of the parameters; infinitesimally: $\delta\phi^r= -f^r(\tau, \vec\phi)$,
$\delta\theta = f^r\partial_r \theta$, $\delta X^i = f^r \partial_r X^i$. This freedom
may be used to set $u^r$ to zero and $\Pi$ to~$-1$. 

Next the hodographic
transformation is performed: Rather than viewing the dependent variables
$\vec p$, $\theta$, and $\vec X$ as functions of $\tau$ and $\vec \phi$,
$\vec X(\tau,\vec \phi)$ is inverted so that $\vec \phi$ becomes a function of
$\tau$ and $\vec X$ (renamed $t$ and $\vec r$, respectively), and $\vec p$ and
$\theta$ also become functions of~$t$  and~$\vec r$. It then
follows from the chain rule that the constraint~\refeq{alsoconstr} (at $\Pi=-1$)
becomes
\numeq{constbec}{
0=\frac{\partial X^i}{\partial \phi^r} \Bigl(p^i -\frac\partial{\partial
X^i}\theta\Bigr)
 }
and is solved by 
\numeq{constbecsolv}{
\vec p = \grad \theta\ .
}
Moreover, according to the chain rule and the implicit function theorem, the partial
derivative with respect to~$\tau$ at fixed $\vec\phi$ [this derivative is present in
\refeq{foform}] is related to the partial derivative with respect to~$\tau$ at
fixed $\vec X=\vec r$ by
\numeq{relpartder}{
\partial_\tau = \frac\partial{\partial t} + \grad\theta\cdot\grad
}
where we have used the new name ``$t$'' on the right. Thus the Nambu-Goto
Lagrangian -- the $\phi$ integral of the Lagrange density \refeq{foform}
(at $u^r=0$, $\Pi=-1$) -- reads
\begin{subequations}\label{NGLII}
\numeq{NGLIIa}{
L_{\mathrm NG} = \int \rd\phi \bigl\{
\vec p\cdot\grad\theta - \dot\theta - \grad\theta\cdot\grad \theta -
\fract12 (p^2 + g)
\bigr\}.
}
But use of \refeq{constbecsolv} and of the Jacobian relation $\rd\phi =\rd r
\det \frac{\partial\phi^s}{\partial X^i} = \frac{\rd r}{\sqrt g}$ shows that
\numeq{NGLIIb}{
L_{\mathrm NG} = \int \rd r \bigl\{
-\frac1{\sqrt g}\dot\theta - \frac1{2\sqrt g}(\grad \theta)^2 - \fract12
\sqrt g
\bigr\}.
}
With the definition
\numeq{withdef}{
\sqrt g = \sqrt{2\lambda}/\rho
}
$L_{\mathrm NG}$ becomes, apart from a total time derivative
\numeq{TotTimder}{
L_{\mathrm NG} =\fract1{\sqrt{2\lambda}} \int \rd r \bigl\{
\theta \dot\rho - \fract12 \rho(\grad \theta)^2 - \frac\lambda\rho
\bigr\}.
}
Up to an overall factor, this is just the Chaplygin gas Lagrangian in 
\refeq{chapgasact}.

The present derivation has the advantage of relating the density~$\rho$ to the
Jacobian of the $\vec X ^i \to \vec \phi$ transformation: $\rho = \sqrt{2\lambda}
\det \frac{\partial\phi^s}{\partial X^i}$. (This in turn shows that the
hodographic transformation is just exactly the passage from Lagrangian to
Eulerian fluid variables -- a remark aimed at those who are acquainted with
the Lagrange formulation of fluid motion~\cite{Salm}.)

Finally, let me observe that the expansion of the Galileo symmetry in $(d,1)$
space-time to a Poincar\'e symmetry in $(d+1,1)$ space-time can be understood
from a Kaluza-Klein--type framework~\cite{HasHor2}.

\end{subequations}

\subsection{Interrelations}\label{sec:Interrelations}
The relation to the Nambu-Goto action explains the origin of the
hidden $(d+1,1)$ Poincar\'e group in our two nonlinear models on
$(d,1)$ space-time: Poincar\'e invariance is what remains of the
reparameterization invariance of the Nambu-Goto action after choosing
either the light-cone or Cartesian parameterizations.  Also the
nonlinear, field dependent form of the transformation laws leading to
the additional symmetries is understood: it arises from the
identification of some of the dependent variables ($X^\mu$) with the
independent variables $(\phi^\alpha)$.

The complete integrability of the $d=1$ Chaplygin gas and Born-Infeld
model is a consequence of the fact that both descend from a string in
2-space; the associated Nambu-Goto theory being completely
integrable. We shall discuss this in Section~\ref{sec:6}.  

We observe that in addition to the nonrelativistic descent from the
Born-Infeld theory to the Chaplygin gas, there exists a mapping of one
system on another, and between solutions of one system and the other,
because both have the same d-brane ancestor.  The mapping is achieved
by passing from the light-cone parameterization to the Cartesian, or
vice-versa.  Specifically this is accomplished as follows:

\paragraph{Chaplygin gas \boldmath$\to$ Born-Infeld:}
  
Given $\theta_{NR} (t,\vec r)$, a
nonrelativistic solution, determine $T(t,\vec r)$ from the equation
\numeq{eq:172}{
T+\frac{1}{c^2} \theta_{NR} (T,\vec r) = \sqrt{2} \, t
} 
Then the relativistic solution is
\numeq{relatsol}{
\theta_R(t,\vec r)=\frac{1}{\sqrt{2}} c^2 T - \frac{1}{\sqrt{2}}
\theta_{NR} (T,\vec r) = c^2 (\sqrt2T - t)
} 

\paragraph{Born-Infeld \boldmath$\to$ Chaplygin gas:}

Given $\theta_R(t,\vec r)$, a relativistic solution, find $T(t,\vec r)$ from
\begin{equation}
T+\frac{1}{c^2} \theta_{R} (T,\vec r) = \sqrt{2} \, t
\end{equation} 
Then the nonrelativistic solution is
\numeq{nonrelsol}{
\theta_{NR}(t,\vec r)=\frac{1}{\sqrt{2}} c^2 T - \frac{1}{\sqrt{2}}
\theta_{R} (T,\vec r) = c^2 (\sqrt2T - t)
} 

\begin{figure}
\begin{center}
\BoxedEPSF{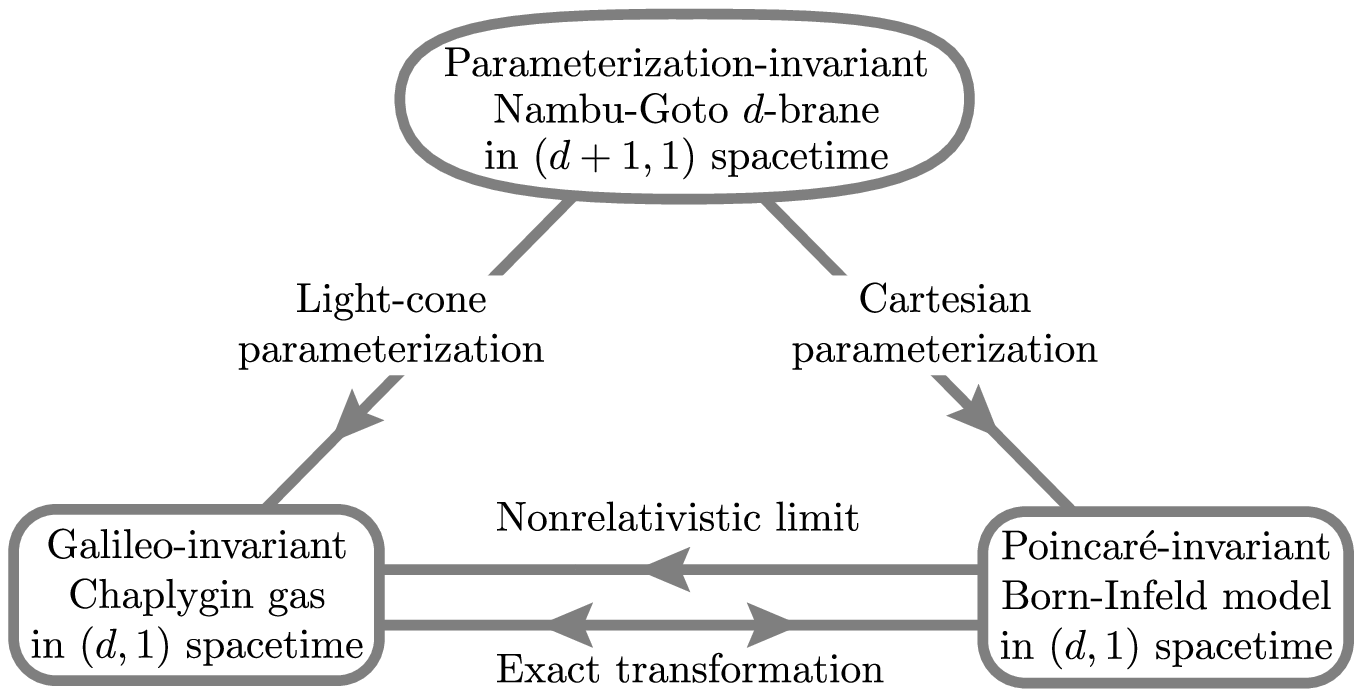 scaled 1000}
 \end{center}
\centerline{Dualities and other relations between nonlinear
equations.}
\end{figure}

The relation between the different models is depicted in the figure below. 

As a final comment, I recall that the elimination of~$\rho$, both in the
nonrelativistic (Chaplygin) and relativistic (Born-Infeld) models is possible only in
the presence of interactions. Nevertheless, the $\theta$-dependent
($\rho$-independent) resultant Lagrangians contain the interaction strengths
only as overall factors; see \refeq{rindependent} and \refeq{B-I}. It is these
$\theta$-dependent Lagrangians that correspond to the Nambu-Goto action in
various parameterizations. Let us further recall the the Nambu-Goto action also
carries an overall multiplicative factor: the d-brane ``tension'', which has been
suppressed in~\refeq{N-Gr1}. Correspondingly, for a ``tensionless'' d-brane, the
Nambu-Goto expression vanishes, and cannot generate dynamics. This suggests
that an action for ``tensionless'' d-branes could be the noninteracting fluid
mechanical expressions \refeq{chapgasact}, \refeq{laglorint}, with vanishing
coupling strengths
$\lambda$ and $a$, respectively. Furthermore, we recall that the noninteracting
models retain the higher, dynamical symmetries, appropriate to a d-brane in
one higher dimension.

\newpage
\section{Supersymmetric Generalization}

Once proven the fact that a bosonic Nambu-Goto theory gives rise to and links
together the  Chaplygin gas and Born-Infeld models, which are irrotational in that
the velocity of the former and the momentum of the latter are given by a gradient
of a potential, one can ask whether there is a d-brane that produces a fluid
model with nonvanishing vorticity.

We shall show that this indeed can be achieved if one starts with a
super--d-brane, and moreover the resulting fluid model possesses
supersymmetry. However, since theories of extended ``super'' objects cannot
be formulated in arbitrary dimensions, we shall consider the fluid in two
spatial dimensions, namely, on the plane~\cite{JacPol2}. 

\subsection{Chaplygin gas with Grassmann variables}
We begin by  positing the fluid model. The Chaplygin gas Lagrangian in 
\refeq{chapgasact}
is supplemented
by Grassmann variables $\psi_a$ that are Majorana 
spinors [real, two-component: $\psi_a^* = \psi_a$, $a=1,2$, $(\psi_1\psi_2)^* =
\psi_1^*\psi_2^*$].

The associated Lagrange density reads
\begin{equation}
{\cal L} = -\rho (\dot\theta - \fract12 \psi \dot\psi) - \fract12 \rho
(\grad\theta - \fract12 \psi \grad \psi)^2 -
\frac{\lambda}{\rho}  -\frac{\sqrt{2\lambda}}{2} \, \psi
\vec\alpha \cdot \grad \psi\ .\label{lagsusy}
\end{equation} 
Here $\alpha^i$ are two ($i=1,2$), $2 \times 2$, real
symmetric Dirac ``alpha'' matrices; in terms of Pauli matrices
we can take $\alpha^1=\sigma^1$, $\alpha^2=\sigma^3$. 
Note that the matrices satisfy the following relations, which
are needed to verify subsequent formulas
\begin{align}
\epsilon_{ab} \alpha^i_{bc} &= \epsilon^{ij} \alpha^j_{ac}
\nonumber \\
\alpha^i_{ab} \alpha^j_{bc} &= \delta^{ij} \delta_{ac}
-\epsilon^{ij} \epsilon_{ac}
\nonumber \\
\alpha^i_{ab} \alpha^i_{cd} &= \delta_{ac} \delta_{bd}
- \delta_{ab} \delta_{cd} + \delta_{ad} \delta_{bc};
\end{align}
$\epsilon_{ab}$ is the $2 \times 2$ antisymmetric matrix
$\epsilon \equiv i \sigma^2$.  In equation (\ref{lagsusy}) $\lambda$ is
a coupling strength which is assumed to be positive.  
  The Grassmann term enters with coupling
$\sqrt{2\lambda}$, which is correlated with the strength of the Chaplygin
potential $V(\rho)={\lambda}/{\rho}$  in order to ensure supersymmetry, as we
shall show below.  It is evident that the velocity should be defined as 
\begin{equation}
\vec v = \grad \theta - \fract12 \psi \grad \psi\ .\label{clebfer}
\end{equation} 

The Grassmann variables directly give rise to a Clebsch
formula for $\vec v$, and provide the Gauss potentials.  The
two-dimensional vorticity reads
$\omega=\epsilon^{ij}\partial_i v^j = -\fract12
\epsilon^{ij}\partial_i \psi \partial_j \psi = -\fract12
\grad \psi \times \grad \psi$. The variables 
$\{\theta, \rho\}$ remain a canonical pair, while the
canonical 1-form in (\ref{lagsusy}) indicates that the
canonically independent Grassmann variables are
$\sqrt{\rho} \, \psi$ so that the antibracket of the $\psi$'s is
\begin{equation}
\{ \psi_a (\vec r), \psi_b (\vec r')\} = -\frac{\delta_{ab}}{\rho(\vec r)}
\delta(\vec r- \vec r')\ .\label{eq:2.5}
\end{equation} 
One verifies that the algebra (\ref{algebra}) or (\ref{algebrap})
is satisfied, and further, one has
\begin{align}
\{ \theta (\vec r), \psi(\vec r)\} &= -\frac{1}{2\rho(\vec r)}
\psi (\vec r) \delta (\vec r-\vec r') \label{eq:2.5a} \\[1ex]
\{ \vec v (\vec r), \psi (\vec r')\}
&= 
-\frac{\grad \psi(\vec r)}{\rho(\vec r)}
 \delta(\vec r-\vec r') \label{eq:2.5b} \\[2ex]
\{ \vec{\cal P} (\vec r), \psi(\vec r') \} &= -\grad \psi(\vec r)
\delta(\vec r-\vec r')\ .
\label{eq:2.5c}
\end{align}
The momentum density $\vec{\cal P}$ is given by the bosonic formula $\vec{\cal
P} =\rho\vec v$, but the Grassmann variables are hidden in $\vec v$, by virtue
of~\refeq{clebfer}.

The equations of motion read
\begin{align}
\dot\rho + \grad \cdot (\rho\vec v) &= 0 \label{eq:2.6a}
\\
\dot\theta + \vec v \cdot \grad \theta
&= \fract12 v^2 + \frac{\lambda}{\rho^2}+
\frac{\sqrt{2\lambda}}{2 \rho}\, \psi \vec\alpha \cdot
\grad \psi \label{eq:2.6b} \\[1ex] 
\dot\psi + \vec v \cdot \grad \psi 
&= \frac{\sqrt{2\lambda}}{\rho} \, \vec\alpha
\cdot \grad \psi \label{eq:2.6c}
\end{align}
and together with (\ref{clebfer}) they imply
\begin{equation}
\dot{\vec v} + \vec v \cdot \grad \vec v = 
\grad \frac{\lambda}{\rho^2} +
\frac{\sqrt{2\lambda}}{\rho}\, (\grad \psi) \vec\alpha \cdot
\grad \psi\ .
\label{eq:2.6d}
\end{equation} 

All these equations may be obtained by bracketing with the
Hamiltonian
\begin{equation}
H=\int \rd{\null^2 r}   \Bigl(\fract12 \rho v^2 + \frac{\lambda}{\rho}
+\frac{\sqrt{2\lambda}}{2}\, \psi \vec\alpha \cdot \grad
\psi \Bigr) = \int \rd{\null^2 r} {\cal H}
\label{eq:2.7}
\end{equation} 
when (\ref{algebra}), (\ref{algebrap}) as well as  \refeq{eq:2.5}--\refeq{eq:2.5b}
are used.

We record the components of the energy-momentum ``tensor'',
and the continuity equations they satisfy.  The energy
density ${\cal E}=T^{oo}$, given by
\begin{equation}
{\cal E} = \fract12 \rho v^2 +
\frac{\lambda}{\rho}+\frac{\sqrt{2\lambda}}{2}\, \psi
\vec\alpha \cdot  \grad \psi  = T^{oo}
\label{eq:2.8}
\end{equation} 
satisfies a continuity equation with the energy flux $T^{jo}$.
\begin{equation}
T^{jo}
=   \rho  v^j   \bigl(\fract12  v^2 - \frac\lambda{\rho^2}\bigr) + 
\frac{\sqrt{2\lambda}}{2}\, \psi \alpha^j \vec v \cdot 
\grad \psi - \frac{\lambda}{\rho} \psi \partial_j \psi
+  \frac{\lambda}{\rho} \epsilon^{jk} \psi \epsilon
\partial_k \psi 
\label{eq:2.9b} 
\end{equation} 
\begin{equation}
\dot T^{oo} + \partial_j T^{jo}= 0
\label{eq:2.9a} 
\end{equation} 
This ensures that the total energy, that is, the
Hamiltonian, is time-indepen\-dent.  Conservation of the total
momentum
\begin{equation}
\vec P = \int \rd{\null^2 r}   \vec{\cal P}
\label{eq:2.10}
\end{equation} 
follows from the continuity equation satisfied by the
momentum density ${\cal P}^i=T^{oi}$ and the momentum flux, that is, the stress
tensor $T^{ij}$.
\begin{gather}
T^{ji}= \rho v^i v^j - \delta^{ij} \Bigl(\frac{2\lambda}{\rho} 
+ \frac{\sqrt{2\lambda}}{2}\,  \psi \vec\alpha \cdot \grad
\psi \Bigr) +  \frac{\sqrt{2\lambda}}{2}\, \psi \alpha^j
\partial_i \psi 
\\
\dot T^{oi} + \partial_j T^{ji}= 0 
\end{gather} 
But $T^{ij}$  is not
symmetric in its spatial indices, owing to the presence of spin
in the problem.  However, rotational symmetry makes it
possible to effect an ``improvement'', which modifies the
momentum density by a total derivative term, leaving the
integrated total momentum unchanged (provided surface
terms can be ignored) and rendering the stress tensor
symmetric.  The improved quantities are
\begin{gather}
\hskip-1.5in {\cal P}^i_I = T^{oi}_I  = \rho v^i + \fract18
\epsilon^{ij}
\partial_j (\rho \psi \epsilon \psi)
\label{imprquant}\\
 T^{ij}_I
= \rho v^i v^j -\delta^{ij} \Bigl(\frac{2 \lambda}{\rho} + 
\frac{\sqrt{2\lambda}}{2} \, \psi \vec\alpha \cdot \grad
\psi \Bigr) + \frac{\sqrt{2\lambda}}{4} \, \Big( \psi 
\alpha^i \partial_j \psi + \psi \alpha^j \partial_i \psi  \Big)
\nonumber \\ 
{}-
\fract18 \partial_k \Big[ (\epsilon^{ki} v^j + \epsilon^{kj}
v^i) \rho \psi \epsilon\psi\Big] \qquad\qquad\qquad\\ 
\dot T^{oi}_I + \partial_j T^{ij}_I = 0\ .
\end{gather}
It immediately follows from the symmetry of $T^{ij}_I$ that the angular
momentum
\begin{equation}
M=\int \rd{\null^2 r}   \epsilon^{ij} r^i {\cal P}^j_I
= \int \rd{\null^2 r}  \rho \epsilon^{ij} r^i v^j + \fract14\int \rd{\null^2 r}
\rho\psi\epsilon\psi 
\end{equation} 
is conserved. The first term is clearly the orbital part
(which still receives a Grassmann contribution through $\vec v$),
whereas the second, coming from the improvement, is the spin
part. Indeed, since
$\frac{i}{2}\epsilon=\fract12 \sigma^2 \equiv \Sigma$, we
recognize this as the spin matrix in (2+1)~dimensions. The
extra term in the improved momentum density~\refeq{imprquant},  $\fract18
\epsilon^{ij}\partial_j (\rho\psi\epsilon\psi)$, can then be
readily interpreted as an additional localized momentum
density, generated by the nonhomogeneity of the spin density.
This is analogous to the magnetostatics formula giving the
localized current density~$\vec j_m$ in a magnet in terms of its
magnetization $\vec m$: $\vec j_m = \grad\times\vec m$. All in
all, we are describing a fluid with spin.

Also the total number 
\begin{equation}
N=\int \rd{\null^2 r}  \rho
\label{eq:2.15new} 
\end{equation} 
is conserved
by virtue of the continuity equation (\ref{eq:2.6a}) satisfied
by $\rho$.  Finally, the theory is Galileo invariant, as is seen
from the conservation of the Galileo boost,
\begin{equation}
\vec B = t \vec P-\int \rd{\null^2 r}   \vec r\rho
\label{eq:2.16} 
\end{equation} 
which follows from (\ref{eq:2.6a}) and (\ref{eq:2.10}).  The
generators $H, \vec P, M, \vec B$ and $N$ close on the (extended)
Galileo group.  [The theory is not Lorentz invariant in
$(2+1)$-dimensional space-time, hence the energy flux
$T^{jo}$ does not coincide with the momentum density,
improved or not.]

We observe that $\rho$ can be eliminated from
(\ref{lagsusy}) so that ${\cal L}$ involves only $\theta$ and
$\psi$.  From (\ref{eq:2.6b}) and (\ref{eq:2.6c}) it follows
that
\begin{equation}
\rho=\sqrt\lambda \bigl(\dot\theta - \fract12 \psi \dot\psi + \fract12
v^2\bigr)^{- 1/2}\ .
\label{eq:2.17} 
\end{equation} 
Substituting into (\ref{lagsusy}) produces the supersymmetric generalization of
the Chaplygin gas Lagrange density in \refeq{rindependent}.
\begin{equation}
{\cal L} = - 2\sqrt\lambda \, \Big\{\sqrt{ 2\dot\theta -  \psi
\dot\psi +  (\grad \theta - \fract12 \psi \grad \psi)^2}
+ \fract12 \psi \vec\alpha \cdot \grad \psi \Big\} 
\label{eq:2.18} 
\end{equation} 
Note that the coupling strength has disappeared from the
dynamical equations, remaining only as a normalization factor
for the Lagrangian.  Consequently the above elimination of
$\rho$ cannot be carried out in the free case, $\lambda=0$.

\subsection{Supersymmetry}
As we said earlier, this theory possesses 
supersymmetry.  This
can be established, first of all, by verifying that the
following two-component supercharges are time-independent 
Grassmann quantities.
\begin{equation}
Q_a =  \int \rd{\null^2 r}   \Big[ \rho \vec v \cdot  (\vec\alpha_{ab}
\psi_b) +  \sqrt{2\lambda} \psi_a \Big]\ .
\label{eq:2.19} 
\end{equation} 
Taking a time derivative and using the evolution equations
(\ref{eq:2.6a})--(\ref{eq:2.6d}) establishes that $\dot{Q}_a=0$.

Next, the supersymmetric transformation rule for the dynamical variables is
found by constructing a bosonic symmetry generator~$Q$, obtained by contracting 
the Grassmann charge with a constant Grassmann parameter $\eta^a$,
 $Q=\eta^aQ_a$, and commuting with the dynamical variables.  Using the canonical
brackets one verifies the following field transformation rules:
\begin{align}
\delta \rho = \{ Q, \rho\} &= -\grad \cdot
\rho( \eta\vec\alpha\psi)
\label{eq:2.20a} \\[1ex]
 \delta \theta = \{ Q, \theta\} &=
-\fract12( \eta\vec\alpha \psi) \cdot \grad \theta - \fract14
 ( \eta\vec\alpha\psi) \cdot \psi \grad\psi +
\frac{\sqrt{2\lambda}}{2\rho}\,  \eta  \psi
\label{eq:2.20b} \\[1ex] 
 \delta \psi = \{ Q, \psi\} &=
-(\eta\vec\alpha \psi) \cdot \grad \psi -
\vec v \cdot \vec\alpha  \eta - \frac{\sqrt{2\lambda}}{\rho}\, \eta
\label{eq:2.20c} \\[1ex] 
 \delta \vec v = \{ Q, \vec v\} &=
-(\eta\vec\alpha \psi) \cdot \grad \vec v +
\frac{\sqrt{2\lambda}}{\rho}\,
 \eta \grad \psi\ .
\end{align}
Supersymmetry is
reestablished by determining the variation of the action $\int
\rd t \rd{\null^2 r}  {\cal L}$  consequent to the above field variations:  the
action is invariant.  One then reconstructs the supercharges
(\ref{eq:2.19}) by Noether's theorem. 
Finally, upon computing the bracket of two supercharges,
one finds
\begin{equation}
\{ \eta^a_1 Q_a, \eta^b_2 Q_b\} = 2  (\eta_1 \eta_2) H
\end{equation} 
which again confirms that the charges are time-independent:
\begin{equation}
\{ H, Q_a\} = 0\ .
\end{equation} 

Additionally a further, kinematical, supersymmetry can be
identified.  According to the equations of motion the following
two supercharges are also time-independent:
\begin{equation}
\bar{Q}_a = \int \rd{\null^2 r}  \rho \psi_a\ .\label{susyq}
\end{equation} 
$\bar{Q}=\bar{\eta}^a \bar{Q}_a$ effects a shift of the
Grassmann field:
\begin{align}
\bar{\delta} \rho = \{ \bar{Q}, \rho\} &= 0
\\
\bar{\delta} \theta = \{ \bar{Q}, \theta\} &=
-\fract12(  \bar{\eta}\psi) \\ 
\bar{\delta} \psi = \{ \bar{Q}, \psi\} &=
- \bar{\eta}\\
\bar{\delta} \vec v = \{ \bar{Q}, \vec v\} &= 0\ .
\end{align}
This transformation leaves the Lagrangian invariant, and
Noether's theorem reproduces (\ref{susyq}).  The algebra
of these charges closes on the total number $N$.
\begin{equation}
\{ \bar{\eta}_1^a \bar{Q}_a, \bar{\eta}_2^b \bar{Q}_b \} 
=  (\bar{\eta}_1 \bar{\eta}_2) N
\label{eq:2.25} 
\end{equation} 
while the algebra with the generators (\ref{eq:2.19}), closes
on the total momentum, together with a central extension,
proportional to volume of space $\Omega = \int \rd{\null^2 r}$
\begin{equation}
\{ \bar{\eta}^a \bar{Q}_a, \eta^b Q_b \} 
=    (\bar{\eta}\vec\alpha \eta) \cdot \vec P +   \sqrt{2\lambda}\,
(\bar{\eta}\epsilon \eta) \Omega\ .
\label{eq:2.26} 
\end{equation} 
The supercharges $Q_a, \bar Q_a$, together with the Galileo
generators ($H$, $\vec P$, $M$, and $\vec B$),  with
$N$ form a superextended Galileo algebra. The
additional, nonvanishing brackets are
\begin{align}
 \{ M, Q_a\} &= \fract12 \epsilon^{ab}Q_b\\
 \{ M, \bar Q_a\} &= \fract12 \epsilon^{ab}\bar Q_b\\
 \{\vec B,  Q_a\} &= \vec\alpha_{ab}\bar Q_b\ .
\end{align}

\subsection{Supermembrane Connection}

The equations for the supersymmetric Chaplygin fluid devolve
from a supermembrane Lagrangian, $L_M$.  We shall
give two different derivations of this result, which make use
of two different parameterizations for the
parameter\-ization-invariant membrane action and give rise,
respectively, to \refeq{lagsusy} and \refeq{eq:2.18}. The two
derivations follow what has been done in the bosonic case in
Sections~\ref{l-cpar} and \ref{h-trans}.

We work in a light-cone gauge-fixed theory:  The supermembrane
in 4-dimensional space-time is described by coordinates
$X^\mu$ $(\mu=0,1,2,3)$, which are decomposed into
light-cone components $X^\pm=\frac{1}{\sqrt{2}} (X^0 \pm
X^3)$ and transverse components $X^i$ $\{i=1,2\}$.  These
depend on an evolution parameter $\phi^0\equiv\tau$ and two
space-like parameters $\phi^r$ $\{r=1,2\}$.  Additionally
there are two-component, real Grassmann spinors $\psi$,
which also depend on $\tau$ and $\phi^r$.  In the light-cone
gauge, $X^+$ is identified with $\tau$, $X^-$ is renamed
$\theta$, and the supermembrane Lagrangian is~\cite{ref:3}
\begin{equation}
L_M=\int \rd{\null^2 \phi}  {\cal L}_M = -\int \rd{\null^2 \phi}  \, 
\{\sqrt{G} -
\fract12
\epsilon^{rs}
\partial_r \psi \balpha \partial_s \psi \cdot \vec X \}
\label{eq:3.1} 
\end{equation}
where $G=\det G_{\alpha\beta}$;
\begin{align}
G_{\alpha\beta} &=  
\begin{pmatrix}
G_{oo} &\quad  G_{os} \\ 
G_{ro} & -g_{rs}
\end{pmatrix}
=  
\begin{pmatrix}
2 \partial_\tau \theta-(\partial_\tau \vec X)^2
- \psi \partial_\tau \psi & \quad u_s \\ 
u_r & -g_{rs}
\end{pmatrix}\label{eq:3.2}
\\[2ex]
G &= g\Gamma
\nonumber \\[1ex]
\Gamma &\equiv  2\partial_\tau \theta-(\partial_\tau \vec X)^2
 -  \psi \partial_\tau \psi
+g^{rs} u_r u_s \nonumber  \\[1ex] 
g_{rs} &\equiv \partial_r \vec X \cdot \partial_s \vec X
\ ,  \quad g=\det g_{rs} \nonumber  \\[1ex] 
u_s &\equiv \partial_s \theta - \fract12 \psi
\partial_s\psi - \partial_\tau \vec X \cdot \partial_s \vec X \ .
\label{eq:3.3}
\end{align}
Here $\partial_\tau$ signifies differentiation with respect to
the evolution parameter $\tau$, while $\partial_r$ differentiates with respect to
the space-like parameters $\phi^r$; $g^{rs}$  is the inverse of
$g_{rs}$, and the two are used to move the $(r,s)$ indices.  Note that
the dimensionality of the transverse coordinates $X^i$ is the
same as of the parameters $\phi^r$, namely two.

\subsection{Hodographic transformation}

To give our first derivation following the procedure in Section~\ref{h-trans}, we
rewrite the Lagrangian in canonical, first-order form, with the help of bosonic
canonical momenta defined by
\begin{subequations}\label{eq:3.4}
\begin{align}
\frac{\partial {\cal L}_M}{\partial \partial_\tau \vec X} &=
\vec p = -\Pi  \partial_\tau \vec X - \Pi u^r \partial_r \vec X
\label{eq:3.4a} \\[1ex]
\frac{\partial {\cal L}_M}{\partial \partial_\tau \theta} &=
\Pi = \sqrt{{g}/{\Gamma}}\ .
\label{eq:3.4b}
\end{align}
\end{subequations}%
(The Grassmann variables already enter with first-order derivatives.) The
supersymmetric extension of \refeq{foform} then reads
\begin{align}
{\cal L}_M&=\vec p \cdot \partial_\tau \vec X + \Pi \partial_\tau
\theta - \fract12\Pi \psi \partial_\tau\psi + \frac{1}{2\Pi}
(p^2+g) +
\fract12 \epsilon^{rs}
\partial_r \psi \balpha \partial_s \psi \cdot \vec X 
\nonumber\\[1ex] 
&\quad {}+ u^r \Big(\vec p \cdot\partial_r \vec X  + \Pi \partial_r
\theta -
\fract12 \Pi \psi \partial_r \psi \Big)\ .
\label{eq:3.5} 
\end{align}
In (\ref{eq:3.5}) $u^r$ serves as a Lagrange multiplier
enforcing a subsidiary condition on the canonical variables, and $g=\det g_{rs}$. 
The equations that follow from (\ref{eq:3.5}) coincide with
the Euler-Lagrange equations for
\refeq{eq:3.1}.  The theory still possesses an
invariance against redefining the spatial parameters with a
$\tau$-dependent function of the parameters.  This freedom
may be used to set $u_\tau$ to zero and fix $\Pi$ at $-1$.  Next
we introduce the hodographic transformation, as in Section~\ref{h-trans}, 
whereby independent-dependent variables are
interchanged, namely we view the $\phi^r$ to be
functions of $X^i$.  It then follows that the constraint on
(\ref{eq:3.5}), which with
$\Pi=-1$ reads
\begin{equation}
 \vec p \cdot\partial_r \vec X - \partial_r \theta + \fract12 \psi
\partial_r \psi =0
\label{eq:3.6a}%
\end{equation}
becomes
\begin{equation}
\partial_r \vec X \cdot \Big(\vec p -\grad \theta +
\fract12 \psi\grad \psi \Big) =0\ .
\label{eq:3.6b}%
\end{equation}
Here $\vec p$, $\theta$ and $\psi$ are viewed as functions of
$\vec X$, renamed $\vec r$, with respect to which acts the gradient
$\grad$.  Also we rename $\vec p$ as $\vec v$, which
according to (\ref{eq:3.6b}) is
\begin{equation}
\vec v = \grad \theta -
\fract12 \psi\grad \psi\ .
\label{eq:3.6c}%
\end{equation}

As in Section~\ref{h-trans}, from the chain rule and the implicit function theorem 
it follows that
\begin{equation}
\partial_\tau = \partial_t + \partial_\tau \vec X \cdot
\grad
\label{eq:3.7}
\end{equation}
and according to (\ref{eq:3.4a}) (at $\Pi=-1$, $u^r=0$)
$\partial_\tau \vec X = \vec p = \vec v$.  Finally, the measure 
transforms according to $\rd{\null^2 \phi}  \to
\rd{\null^2 r}  \frac{1}{\sqrt{g}}$.  Thus the Lagrangian
for (\ref{eq:3.5}) becomes, after setting $u^r$ to zero and
$\Pi$ to $-1$,
\begin{subequations}
\begin{gather}
L_M=\!\int \!\frac{\rd{\null^2 r}}{\sqrt{g}} \Bigl( v^2 -\dot\theta - \vec v
\!\cdot\!
\grad \theta + \fract12 \psi(\dot\psi + \vec v \cdot \grad
\psi) -\fract12(v^2+g)\nonumber\\
\hbox to 2in{\hfill} -\fract12 \epsilon^{rs}\, \psi
\alpha^i\, \partial_j \psi \, \partial_s x^j\,
\partial_r x^i
\Bigr)\ .
\label{eq:3.8}
\end{gather}
But $\epsilon^{rs} \partial_s x^j \partial_r x^i = \epsilon^{ij}
\det \partial_r x^i = \epsilon^{ij} \sqrt{g}$.  After
$\sqrt{g}$ is renamed $\sqrt{2\lambda}/\rho$,
(\ref{eq:3.8}) finally reads
\begin{equation}
L_M= \frac1{\sqrt{2\lambda}}  \int \rd{\null^2 r}\,  \Big(
{-\rho} (\dot\theta -
\fract12 \psi\dot\psi) -
\fract12 \rho (\grad \theta - \fract12 \psi \grad
\psi)^2 - \frac{\lambda}{\rho} - \frac{\sqrt{2\lambda}}{2} 
\psi \balpha \times \grad \psi \Big)\ .
\label{eq:3.9}
\end{equation}
\end{subequations}
Upon replacing $\psi$ by $\frac{1}{\sqrt{2}} (1-\epsilon)
\psi$, this is seen to reproduce the Lagrange density
\refeq{lagsusy}, apart from an overall factor.

\subsection{Light-cone parameterization}

For our second derivation, we return to
(\ref{eq:3.1})--(\ref{eq:3.3}) and use the remaining
reparameterization freedom to equate the two $X^i$
variables with the two $\phi^r$ variables, renaming both as
$r^i$.  Also $\tau$ is renamed as $t$. This parallels the method in
Section~\ref{l-cpar}.  Now in (\ref{eq:3.1})--(\ref{eq:3.3}) $g_{rs} = \delta_{rs}$,
and
$\partial_\tau \vec X=0$, so that (\ref{eq:3.3}) becomes simply
\begin{align}
G=\Gamma &= 2\dot\theta -  \psi \dot\psi + u^2
\label{eq:3.10} \\[1ex]
 \vec u &= \grad \theta - \fract12 \psi \grad \psi\ .
\label{eq:3.11}
\end{align}
Therefore the supermembrane Lagrangian (\ref{eq:3.1}) reads
\begin{equation}
L_M=-\int \rd{\null^2 r} \biggl\{ \sqrt{2 \dot\theta -  \psi \dot\psi +
\bigl(\grad \theta -\fract12 \psi \grad \psi\bigr)^2}
+ \fract12 \psi \balpha \times \grad \psi \biggr\}\ .
\label{eq:3.12}
\end{equation}
Again a replacement of $\psi$ by $\frac{1}{\sqrt{2}}
(1-\epsilon) \psi$ demonstrates that the integrand
coincides with the Lagrange density in (\ref{eq:2.18}) (apart
from a normalization factor).

\subsection{Further consequences of the supermembrane
connection}

Supermembrane dynamics is Poincar\'e invariant in
(3+1)-dimensional space-time.  This invariance is hidden
by the choice of light-cone parameterization: only the
light-cone subgroup of the Poincar\'e group is left as a
manifest invariance.  This is just the $(2+1)$ Galileo group
generated by $H$, $\vec P$, $M$, $\vec B$, and $N$.  (The
light-cone subgroup of the Poincar\'e group is isomorphic to
the Galileo group in one lower dimension~\cite{Sus}.)  The Poincar\'e
generators not included in the above list correspond to
Lorentz transformations in the ``$-$'' direction.  We expect
therefore that these generators are ``dynamical'', that is,
hidden and unexpected conserved quantities of our
supersymmetric Chaplygin gas, similar to the situation with
the purely bosonic model.

One verifies that the following quantities
\begin{align}
D &= tH-\int \rd{\null^2 r}\,  \rho \theta 
\label{eq:3.13} \\[1ex]
\vec G &= \int \rd{\null^2 r}  (\vec r {\cal H} - \theta \bcP_I - \fract18
\psi \balpha \balpha \cdot \bcP_I \psi) \nonumber \\
&= \int \rd{\null^2 r} (\vec r {\cal H}  - \theta \bcP - \fract14
\psi \balpha \balpha \cdot \bcP \psi) 
\label{eq:3.15}
\end{align}
are time-independent by virtue of the equations of motion
(\ref{eq:2.6a})--(\ref{eq:2.6d}), and they supplement the Galileo
generators to form the full $(3+1)$ Poincar\'e algebra, which becomes
the super-Poincar\'e algebra once the supersymmetry is taken
into account. Evidently \refeq{eq:3.13}, \refeq{eq:3.15} are the supersymmetric
generalizations of \refeq{eq:105}, \refeq{eq:106}.

We see that fluid dynamics can be extended to
include Grassmann variables, which also enter in a
supersymmetry-preserving interaction.  Since our
construction is based on a supermembrane in
(3+1)-dimensional space-time, the fluid model is
necessarily a planar Chaplygin gas.  It remains for the future to show
how this construction could be generalized to arbitrary
dimensions and to different interactions. Note that Grassmann
Gauss potentials~$\psi$ can be used even in the absence of
supersymmetry. For example, our theory \refeq{lagsusy}, with
the last term omitted, posseses a conventional, bosonic
Hamiltonian without supersymmetry, while the Grassmann
variables are hidden in $\vec v$ and occur only in the canonical
1-form. 

\newpage
\section{One-dimensional Case}\label{sec:6}

In this section, I shall discuss the nonrelativistic/relativistic models in one
spatial dimension. Complete integrability has  been established for both the
Chaplygin gas \cite{Nutku} and the Born-Infeld theory \cite{BarChe}. We can now
understand this to be a consequence of the complete integrability of the
Nambu-Goto 1-brane (string) moving on 2-space (plane), which is the antecedent
of both models. [Therefore, it suffices to discuss only the Chaplygin gas since
solutions of the Born-Infeld model can then be obtained by the mapping
\refeq{eq:172}--\refeq{relatsol}.] 

As remarked previously, in one dimension there is no vorticity, and the
nonrelativistic velocity~$v$ can be presented as a derivative with respect to the
single spatial variable of a potential~$\theta$. 
Similarly, the relativistic momentum $p=v/\sqrt{1-v^2/c^2}$ is a derivative of
a potential~$\theta$.   In both cases the potential is canonically conjugate to the
density
$\rho $ governed by the canonical 1-form  $\int \rd x \theta\dot\rho $.
Moreover, it is evident that at the expense of a spatial nonlocality, one may
replace $\theta$ by its antiderivative, which is $p$ both nonrelativistically and
relativistically (nonrelativistically $p = v $), so that in both cases the Lagrangian
reads 
\numeq{repltheta}{
L = -\fract12 \int \rd x \rd y \rho(x) \eps(x-y)\dot{p}(y) - H\ .
} 
For the Chaplygin gas and the Born-Infeld models, $H$ is given respectively by 
\begin{align}
H^{\mathrm{Chaplygin}} &= \int \rd x \Bigl( \fract12 \rho {p}^2 +
\frac\lambda\rho\Bigr) \label{ChapH}\\
H^{\BI} &= \int \rd x \bigl( \sqrt{\rho^2c^2+ a^2} \sqrt{c^2+p^2}\bigr)\ .
\label{ChapBI}
\end{align}
The equations of motion are, respectively
\begin{align}
&\mbox{Chaplygin gas:} & \dot\rho + \frac\partial{\partial x}(p\rho)
&=0\label{eqmot234}\\
& & \dot p + \frac\partial{\partial x}\Bigl( \frac{p^2}2 -
\frac\lambda{\rho^2}\Bigr) &=0\label{eqmot235}\\
& &\mbox{or}\quad  \frac\partial{\partial t} \frac1{\sqrt{\dot\theta +
\fract{p^2}2}} +
\frac\partial{\partial x} \frac p{\sqrt{\dot\theta +
\fract{p^2}2}} &=0 \label{eqmot236} \\[3ex]
&\mbox{Born-Infeld model:} &  \dot\rho + \frac\partial{\partial
x}\biggl(p \sqrt{\frac{\rho^2c^2 + a^2}{c^2+p^2}}\biggr) &=0\label{eqmot237}\\
& & \dot p + \frac\partial{\partial x}\biggl(\rho c^2 \sqrt{\frac{c^2 + p^2}{\rho^2
c^2 + a^2}}\biggr) &=0\label{eqmot238}\\
& &\llap{$\displaystyle\mbox{or}\quad  \frac\partial{c^2\partial t} \biggl(
\frac{\dot\theta}{\sqrt{c^2-\fract1{c^2} \dot\theta^2 + p^2}}\biggr)$} 
- \frac\partial{\partial x} \biggl(\frac{p}{\sqrt{c^2-\fract1{c^2}
\dot\theta^2 + p^2}}\biggr) &=0 \label{eqmot239}
\end{align}
In the above, eqs.~\refeq{eqmot236} and \refeq{eqmot239} result by determining
$\rho$ in terms of $\theta$ ($p=\frac\partial{\partial x} \theta$) from
\refeq{eqmot235} and \refeq{eqmot238}, and using that expression for $\rho$ in
\refeq{eqmot234} and \refeq{eqmot237}.

\subsection{Specific solutions for the Chaplygin gas on a line}
Classes of solutions for a Chaplygin gas in one dimension can be given in closed
form. For example, to obtain general, time-rescaling--invariant solutions, we make
the \emph{Ansatz} that $\theta\propto 1/t$. Then \refeq{interstren} or
\refeq{eqmot236} leads to a second-order nonlinear differential equation for the
$x$-dependence of~$\theta$. Therefore solutions involve two arbitrary constants,
one of which fixes the origin of~$x$ (we suppress it);  the other we call $k$, and
take it to be real. The solutions then read
\numeq{solthenread}{
\theta(t,x) = - \frac1{2k^2 t} \cosh^2 kx\ .
}
[Other solutions can be obtained by relaxing the reality condition on $k$ and/or
shifting the argument $kx$ by a complex number. In this way one  finds that
$\theta$ can also be $\frac1{2k^2 t} \sinh^2 kx$, $\frac1{2k^2 t} \sin^2 kx$,
$\frac1{2k^2 t} \cos^2 kx$; but these lead to singular or unphysical forms
for~$\rho$.] The density corresponding to \refeq{solthenread} is found from
\refeq{conseqBern} or \refeq{eqmot235} to be 
\numeq{densecorresp}{
\rho(t,x) = \sqrt{2\lambda} \frac{k\left|t\right|}{\cosh^2 kx}\ . 
}
The velocity/momentum $v=p=\frac\partial{\partial x}\theta$ is 
\numeq{velmom}{
v(t,x) = p(t,x) = -\frac1{kt} \sinh kx \cosh kx
}
while the sound speed
\numeq{sounspee}{
s(t,x) = \frac{\cosh^2 k x}{k\left|t\right|}
} 
is always larger than $\left| v\right|$. Finally, the current $j=  \rho
\frac{\partial\theta}{\partial x}$ exhibits a kink profile, 
\numeq{kinkprof}{
 j(t,x) = -\eps(t) \sqrt{2\lambda} \tanh kx
}
which is suggestive of complete integrability. 

Another particular  solution is the Galileo boost of the static profiles 
\refeq{intersol}, \refeq{timeindep}:
\begin{align}
p(t,x) &= p(x-ut)\label{galboost1}\\
\rho(t,x) &= \frac{\sqrt{2\lambda}}{\left|p-u\right|}\ .
\end{align}
Here $u$ is the boosting velocity and $p(x-ut)$ is an arbitrary function of its
argument (provided $p\neq u$). Clearly this is a constant profile
solution, in linear motion with velocity~$u$. 

Further evidence  for complete integrability is found by identifying an infinite
number of constants of motion. One verifies that the following quantities
\numeq{infnumconsmo}{
I_n^{\pm} = \int \rd x \rho \Bigl(p \pm \frac{\sqrt{2\lambda}}\rho \Bigr)^n\ ,
\quad n=0, \pm1, \ldots
 }
are conserved. 

The combinations $p \pm \frac{\sqrt{2\lambda}}\rho $ are just the velocity
$(\pm)$ the sound speed, and they are known as Riemann coordinates.
\numeq{riemcoor}{
R_{\pm} = p \pm \frac{\sqrt{2\lambda}}\rho  
}
The equations of motion for this system [continuity \refeq{eqmot234} and Euler 
 \refeq{eqmot235}] can be succinctly presented in terms
of $R_{\pm}$: 
\numeq{succpres}{
\dot R_{\pm} = - R_{\mp}\frac\partial{\partial x} R_{\pm}\ .
}

\subsection{Aside on the integrability of the cubic potential in one dimension}

Although it does not belong to the models that we have discussed, the cubic
potential for 1-dimensional motion, $V(\rho) = \ell \rho^3/3$, is especially
interesting because it is secretly free -- a fact that is exposed when Riemann
coordinates are employed. For this problem these read $R_{\pm} =
p\pm\sqrt{2\ell} \rho$ and again they are just the velocity $(\pm)$ the sound
speed. In contrast to \refeq{succpres} the Euler and continuity equations for this
system decouple: $\dot R_{\pm} = - R_{\pm} \frac\partial{\partial x} R_{\pm}$.
Indeed, it is seen that
$ R_{\pm}$ satisfy essentially the free Euler equation [compare with
\refeq{freeEuler} and identify $R_{\pm}$ with~$v$]. Consequently, the solution
\refeq{initdata}--\refeq{initdata3} works here as well.

Recall the previous remark in Section~\ref{sec:3.1} on the Schr\"odinger group
[Galileo
$\oplus$ SO(2,1)]: in one dimension the cubic potential is invariant against this
group of transformations, and in all dimensions the free theory is
invariant~\cite{HasHor}, \cite{Jac72}. Therefore a natural speculation is that the
secretly noninteracting nature of the cubic potential in one dimension is a
consequence of Schr\"odinger group invariance. 

Another interesting fact about a one-dimensional nonrelativistic fluid with cubic
potential is that it also arises in a collective, semiclassical description of
nonrelativistic free fermions in one dimension, where the cubic potential
reproduces fermion repulsion \cite{JevSak}. In spite of the nonlinearity of the
fluid model's equations of motion, there is no interaction in the underlying
fermion dynamics. Thus,  the presence of the Schr\"odinger group  and the 
equivalence to free  equations for this fluid system is an understandable
consequence. 

\subsection{General solution for the Chaplygin gas on a line}\label{sec:6.3}

The general solution to the Chaplygin gas can be found by linearizing the
governing equations (continuity and Euler) with the help of a Legendre
transform, which also effects a hodographic transformation that exchanges the
independent variables $(t, x)$ with the dependent ones $(\rho, \theta)$;
actually instead of
$\rho$ we use the sound speed $s = \sqrt{2\lambda}/\rho$ and instead of
$\theta$ we use the momentum $p = \frac\partial{\partial x}\theta$.  

Define
\numeq{eq:245}{
\psi(p,s) = \theta(t,x) - t\dot\theta(t,x) - x\frac\partial{\partial x}\theta(t,x)\ . 
}
From the Bernoulli equation we know that 
\numeq{eq:246}{
\dot\theta = -\fract12 p^2 + \fract12 s^2\ .
}
Thus 
\numeq{eq:247}{
\psi(p,s) = \theta(t,x) + \frac t2 (p^2-s^2) - xp
}
and the usual Legendre transform rules govern the derivatives.
\begin{subequations}\label{eq:248}
\begin{align}
\frac{\partial\psi}{\partial p} &= tp -x \label{eq:248a}\\
\frac{\partial\psi}{\partial s} &= -ts \label{eq:248b}
\end{align}
\end{subequations}
 It remains to incorporate the continuity equation
\refeq{eqmot234} whose content must be recast by the
hodographic transformation. This is achieved by rewriting  equation
\refeq{eqmot234} in terms of $s = \sqrt{2\lambda}/\rho$:
\numeq{eq:249}{
\frac{\partial s}{\partial t} + p \frac{\partial s}{\partial x} - s
\frac{\partial p}{\partial x} = 0\ .
}
Next \refeq{eq:249} is presented as a relation between Jacobians:
\begin{subequations}\label{eq:250}
\numeq{eq:250a}{
\frac{\partial (s,x)}{\partial (t,x)} + p \frac{\partial (t,s)}{\partial (t,x)} - s
\frac{\partial (t,p)}{\partial (t,x)} = 0
}
which is true because here $\partial x/\partial t= \partial t/\partial x=0$.
Eq.~\refeq{eq:250a} implies, after multiplication by $\partial (t,x)/\partial
(s,p)$
\begin{align}
0 &= \frac{\partial (s,x)}{\partial (s,p)} + p \frac{\partial (t,s)}{\partial (s,p)} - s
\frac{\partial (t,p)}{\partial (s,p)}\nonumber\\
&= \frac{\partial x}{\partial p}  - p \frac{\partial t}{\partial p}  -
s\frac{\partial t}{\partial s}\ .\label{eq:250b}
\end{align}
The second equality holds because now we take $\partial s/\partial p= \partial
p/\partial s=0$. Finally, from \refeq{eq:247}, \refeq{eq:248} it follows that
\refeq{eq:250b} is equivalent to 
\numeq{eq:250c}{
\frac{\partial^2 \psi}{\partial p^2}  - \frac{\partial^2 \psi}{\partial s^2}  +
 \frac2s \frac{\partial \psi}{\partial s} = 0\ .
}
\end{subequations}
This linear equation is solved by two arbitrary functions of $p\pm s$ ($p\pm
s$ being just the Riemann coordinates)
\numeq{eq:251}{
\psi(p,s) = F(p+s) - sF'(p+s) + G(p-s) + sG'(p-s)\ . 
}

In summary, to solve the Chaplygin gas equations, we choose two functions $F$
and $G$, construct $\psi$ as in \refeq{eq:251}, and regain $s$
($=\sqrt{2\lambda}/\rho$), $p$ ($=\frac\partial{\partial x}\theta$), and $\theta$
from
\refeq{eq:247},
\refeq{eq:248}. In particular, the solution \refeq{solthenread},
\refeq{densecorresp} corresponds to
\numeq{eq:252}{
F(z) = G(-z) = \pm \frac z{2k} \ln z
}
where the sign is correlated with the sign of~$t$. 

\subsection{Born-Infeld model on a line}

Since the Born-Infeld system is related to Chaplygin gas by the transformation
described in Section~\ref{sec:Interrelations}, there is no need to discuss
separately Born-Infeld solutions. Nevertheless, the formulation in terms of
Riemann coordinates is especially succinct and gives another view on the
Chaplygin/Born-Infeld relation. 

The Riemann coordinates $R_{\pm}$ for the Born-Infeld model are contructed
by first defining
\begin{align}
\frac1c \frac\partial{\partial x}\theta  = p/c &= \tan \phi_p\nonumber\\
a/\rho c &= \tan \phi_\rho \label{eq:253}\\
\intertext{and}
R_{\pm}  &= \phi_p \pm \phi_\rho\ . \label{eq:254}
\end{align}
The 1-dimensional version of the equations of motion
\refeq{followsfrom1}, \refeq{followsfrom2}, that is, \refeq{eqmot237},
\refeq{eqmot238} can be presented as 
\numeq{eq:255}{
\dot R_{\pm} = -c(\sin R_{\mp}) \frac\partial{\partial x}R_{\pm}\ . 
}

The relation to the Riemann description of the Chaplygin gas can now be seen 
in two ways:  a nonrelativistic limit and an exact transformation. For the former,
we note that at large~$c$, $\phi_p \approx p/c$, $\phi_\rho\approx a/\rho c$ so
that
\numeq{eq:256}{
R_{\pm}^{\BI} \approx \frac1c \Bigl(p \pm \frac a\rho\Bigr) = \frac1c
R_{\pm}^{\mathrm{Chaplygin}} \Bigr|_{\lambda=a^2/2}\ .
}
Moreover, the equation \refeq{eq:255} becomes, in view of \refeq{eq:256},
\numeq{eq:257}{
\frac1c \dot R_{\pm}^{\mathrm{Chaplygin}} = - R_{\mp}^{\mathrm{Chaplygin}}
\frac1c\frac\partial{\partial x} R_{\pm}^{\mathrm{Chaplygin}}
 }
so that  \refeq{succpres} is regained. On the other hand, for the exact
transformation we define new Riemann coordinates in the relativistic, Born-Infeld 
case by
\numeq{eq:258}{
{\cal R}_{\pm} = c \sin R_{\pm}\ .
}
Evidently \refeq{eq:255} implies that ${\cal R}_{\pm}$ satisfies the
nonrelativistic equations \refeq{succpres}, \refeq{eq:257} when
$R_{\pm}$ solves the relativistic equation \refeq{eq:255}. Expressing
${\cal R}_{\pm}$ and  $R_{\pm}$ in terms of the corresponding nonrelativistic
and relativistic variables produces a mapping between the two sets. Calling
$p_{\mathrm{NR}}$, $\rho_{\mathrm{NR}}$ and $p_{\mathrm{R}}$,
$\rho_{\mathrm{R}}$ the  momentum and density of the nonrelativistic
and of the relativistic theory, respectively, the mapping implied by
\refeq{eq:258} is 
\begin{align}
p_{\mathrm{NR}} &=
\frac{c^2 \rho_{\mathrm{R}}
p_{\mathrm{R}}}{\sqrt{(p^2_{\mathrm{R}} +
c^2)(\rho^2_{\mathrm{R}}c^2 + a^2)}}\nonumber\\[2ex]
\rho_{\mathrm{NR}} &= \frac1{c^2} 
\sqrt{(p^2_{\mathrm{R}} + c^2)(\rho^2_{\mathrm{R}}c^2 + a^2)}\ .
\label{eq:259}
\end{align}
As can be checked, this maps the Chaplygin equations into the Born-Infeld
equations. But the mapping is not canonical.

We record the infinite number of constants of motion, which put into evidence
the (by now obvious) complete integrability of the Born-Infeld equations on a
line. The following quantities are time-independent:
\numeq{eq:260}{
I_n^{\pm} = ac^{n-1} \int \rd x \frac{(\phi_p \pm
\phi_\rho)^n}{\sin\phi_\rho\cos \phi_p}\ , \quad
n=0,\pm1,\ldots
 }
The nonrelativistic limit takes the above into \refeq{infnumconsmo}, while
expressing $I_n^{\pm}$ in terms of ${\cal R}_{\pm}$ according to
\refeq{eq:258} shows that the integrals in \refeq{eq:260} are expressible as
series in terms of the integrals in \refeq{infnumconsmo}.

In the relativistic model $\rho$ need not be constrained to be positive
(negative $\rho$ could be interpreted as antiparticle density). The
transformation $p \to -p$, $\rho \to -\rho$ is a symmetry and can be
interpreted as charge conjuguation. Further, $p$ and $\rho$ appear in 
an equivalent way. As a result, this theory enjoys a 
duality transformation:
\numeq{eq:261}{
\rho \to \pm\frac{a}{c^2} p \qquad  p \to \pm\frac{c^2}{a}
\rho\ .
}
Under the above, both the canonical structure and the
Hamiltonian remain invariant. Solutions are mapped in
general to new solutions. Note that the nonrelativistic limit
is mapped to the ultra-relativistic one under the above duality.
Self-dual solutions, with $\rho = \pm\frac{a}{c^2} p$, satisfy
\numeq{eq:262}{
\dot{\rho} = \mp c \frac\partial{\partial x} \rho 
}
and are, therefore, the chiral relativistic solutions that
were presented at the end of Section~\ref{L-irm}.  In the self-dual
case, when $p$ is eliminated from the canonical 1-form and from
the Hamiltonian with the help of \refeq{eq:261}, one arrives at an action for
$\rho$, which coincides (apart from irrelevant constants) with the
self-dual action, constructed some time ago~\cite{FloJac}
\begin{align}
\biggl\{
\fract12 &\displaystyle\int \rd t \rd x \rd y  \dot\rho(x)\epsilon (x-y)
p(y)  - \int  \rd t \rd x \sqrt{\rho^2c^2+a^2} \sqrt{c^2+p^2}  \rd t
\bigg\}\biggr|_{p=\frac{c^2}a \rho} \nonumber\\
= \frac{2c^2}a \biggl\{ \fract14  &\displaystyle\int \rd t \rd x \rd y
\dot\rho(x)\epsilon (x-y)
\rho(y)  
- \frac c2 \int \rd t \rd x\Bigl( \rho^2(x) + \frac{a^2}{c^2}\Bigr)
\biggr\}       \label{eq:263}
\end{align}

\subsection{General solution of the Nambu-Goto  theory
 for a
(d=1)-brane (string) in two spatial dimensions  (on a plane)}

The complete integrability of the Chaplygin gas and of the Born-Infeld theory,
as well as the relationships between the two, derives from the fact that the
different models descend  by
fixing in different ways the parameterization invariance of the Nambu-Goto 
theory for string on a plane.  At the same time, the equations governing  the
planar motion of a string can be solved completely. Therefore it is instructive
to see how the string solution produces this Chaplygin solution~\cite{Baz}. 

We follow the development in Section~\ref{h-trans}. The Nambu-Goto action
reads 
\begin{subequations}\label{eq:264}
\begin{align}
I_{\textrm{NG}} &= \int \rd{\phi^0} L_{\textrm{NG}}\label{eq:264a}\\
 L_{\textrm{NG}} &= \int \rd {\phi^1} {\cal L}_{\textrm{NG}}\label{eq:264b}\\
{\cal L}_{\textrm{NG}}&= \Bigl[ -\det \frac{\partial X^\mu}{\partial\phi^\alpha}
\frac{\partial X_\mu}{\partial\phi^\beta}\Bigr]^{1/2}\ .\label{eq:264c}
\end{align}
\end{subequations}
Here $X^\mu$, $\mu=0,1,2$, are string variables and $(\phi^0, \phi^1)$ are its
parameters. As in Section~\ref{h-trans}, we define light-cone combinations
$X^{\pm} = \frac1{\sqrt2}(X^0\pm X^2)$, rename $X^-$ as $\theta$, and choose
the parameterization $X^+=\phi^0\equiv \tau$. After suppressing the
superscripts on $\phi^0$ and $X^1$, we construct the Nambu-Goto Lagrange
density as
\begin{align}
{\cal L}_{\textrm{NG}}&= \det\nolimits^{1/2} \begin{pmatrix}
2\partial_\tau \theta -(\partial_\tau X)^2 & u \\
u & -(\partial_\phi X)^2
\end{pmatrix}\label{eq:265}\\
u &= \partial_\phi \theta - \partial_\tau X \partial_\phi X\label{eq:266}
\end{align}
Equations of motion are presented in Hamiltonian form:
\numeq{eq:267}{
p \equiv \frac{\partial{\cal L}_{\textrm{NG}}}{\partial\partial_\tau X} \qquad
\Pi\equiv \frac{\partial{\cal L}_{\textrm{NG}}}{\partial\partial_\tau \theta}
}
\begin{subequations}\label{eq:268}
\begin{align}
\partial_\tau X &= -\frac1\Pi p - u \partial_\phi X\label{eq:268a}\\
\partial_\tau \theta &= \frac1{2\Pi^2} \bigl(
          p^2 + (\partial_\phi X)^2  \bigr) - u \partial_\phi \theta\label{eq:268b}\\
\partial_\tau p &= -\partial_\phi \Bigl(\frac1\Pi \partial_\phi
X \Bigr) - \partial_\phi (up)\label{eq:268c}\\
\partial_\tau \Pi &= - \partial_\phi (u\Pi)\label{eq:268d}
\end{align}
\end{subequations}
and there is the constraint
\numeq{eq:269}{
p\partial_\phi X + \Pi \partial_\phi\theta = 0\ .
}
There still remains the reparameterization freedom of replacing $\phi$ by an
arbitrary function of $\tau$ and $\phi$; this freedom may be used to set
$u=0$, $\Pi=-1$. Consequently, in the fully parameterized equations of
motion  Eq.~\refeq{eq:268d} disappears; instead of \refeq{eq:268a} and
\refeq{eq:268c}, we have $\partial_\tau X = p$, $\partial_\tau p =
\partial^2_\phi X$, which imply
\begin{subequations}\label{eq:270}
\numeq{eq:270a}{
(\partial_\tau^2 -\partial_\phi^2)X = 0 
}
\refeq{eq:268b} reduces to
\numeq{eq:270b}{
\partial_\tau \theta = \fract12 \bigl[ (\partial_\tau X)^2 + (\partial_\phi
X)^2\bigr] 
 }
and the constraint \refeq{eq:269} requires
\numeq{eq:270c}{
\partial_\phi \theta = \partial_\tau X \partial_\phi X\ .
}
\end{subequations}

Solution to \refeq{eq:270a} is immediate in terms of two functions $F_{\pm}$, 
\numeq{eq:271}{
x(\tau,\phi) = F_+ (\tau +\phi) + F_- (\tau-\phi)
}
and then \refeq{eq:270b}, \refeq{eq:270c} fix $\theta$:
\numeq{eq:272}{
\theta(\tau,\phi) = \int^{\tau+\phi} \rd z \bigl[F'_+(z)\bigr]^2 + 
 \int^{\tau-\phi} \rd z \bigl[F'_- (z)\bigr]^2\ .
}
This completes the description of a string moving on a plane.
But we need to convert this information into a solution of the Chaplygin gas,
and we know from Section~\ref{h-trans} that this can be accomplished by a
hodographic transformation: instead of $X$ and $\theta$ as a function of
$\tau$ and $\phi$, we seek $\phi$ as a function of $\tau$ and $X$, and this
renders $\theta$ to be a function of $\tau$ and $X$ as well. The density $\rho$
is determined by the Jacobian $\left|\partial X/\partial \phi\right|$. 

Replace $\tau$ by $t$ and $X$ by $x$ and define $\phi$ to be $f(t,x)$. Then from
\refeq{eq:271} it follows that
\numeq{eq:273}{
x = F_+ \bigl(t + f(t,x)\bigr) + F_- \bigl(t - f(t,x)\bigr)\ .
}
This equation may be differentiated with respect to $t$ and $x$, whereupon
one finds
\begin{subequations}\label{eq:274}
\begin{align}
\frac{\partial f}{\partial t} &= -\frac{F'_+(t+f) + F'_-(t-f)}{F'_+(t+f) - F'_-(t-f)}
      \label{eq:274a}\\
\frac{\partial f}{\partial x} &=  \frac1{F'_+(t+f) - F'_-(t-f)}\ .
      \label{eq:274b}
\end{align}
\end{subequations}

Thus the procedure for constructing a Chaplygin gas solution is to choose two
functions $F_{\pm}$, solve the differential equations \refeq{eq:274} for $f$, and
then the fluid variables are
\begin{align}
\theta(t,x) &= \int^{t+f(t,x)} \bigl[F'_+(z)\bigr]^2 \rd z 
                       + \int^{t-f(t,x)} \bigl[F'_-(z)\bigr]^2 \rd z \label{eq:275}\\
\frac{\sqrt{2\lambda}}\rho &= \left| F'_+(t+\phi) - F'_-(t-\phi)\right|\ .
                              \label{eq:276}
\end{align}

One may verify directly that \refeq{eq:275} and \refeq{eq:276} satisfy the
required equations: Upon differentiating \refeq{eq:275} with respect to $t$ and
$x$, we find
\begin{subequations}\label{eq:277}
\begin{align}
\frac{\partial \theta}{\partial t}&= (F'_+)^2\Bigl(1+
\frac{\partial f}{\partial t}\Bigr)  + (F'_-)^2 \Bigl(1-
\frac{\partial f}{\partial t}\Bigr)\nonumber\\
        &= -2F'_+ F'_-    \label{eq:277a}\\
\frac{\partial \theta}{\partial x}&= (F'_+)^2\Bigl( 
\frac{\partial f}{\partial x}\Bigr)  - (F'_-)^2 \Bigl( 
\frac{\partial f}{\partial x}\Bigr)\nonumber\\
        &= F'_+ + F'_-     \label{eq:277b}
\end{align}
\end{subequations}
The second equalities follow with the help of \refeq{eq:274}. From
\refeq{eq:277} one sees that
\numeq{eq:278}{
\frac{\partial \theta}{\partial t} + \fract12 \Bigl(\frac{\partial \theta}{\partial
x}\Bigr)^2 = 
\fract12 ( F'_+ - F'_- )^2 = \frac\lambda{\rho^2} 
 }
the last equality being the definition \refeq{eq:276}. Thus the Bernoulli (Euler)
equation holds. For the continuity equation, we first find from \refeq{eq:276}
and \refeq{eq:277}
\begin{subequations}\label{eq:279}
\begin{align}
\frac{\partial\rho}{\partial t}&= 
\pm \frac{\partial}{\partial t} \frac{\sqrt{2\lambda}}{F'_+ - F'_-} \nonumber\\
     &= \mp \frac{\sqrt{2\lambda}}{(F'_+ - F'_-)^2} \Bigl[
F''_+ \Bigl(1 +\frac{\partial f}{\partial t} \Bigr) -
F''_- \Bigl(1 -\frac{\partial f}{\partial t} \Bigr)\Bigr] \nonumber\\
        &= \pm \frac{2\sqrt{2\lambda}}{(F'_+ - F'_-)^3}  
              \bigl( F''_+ F'_- + F''_- F'_+  \bigr) \label{eq:279a}\\[2ex]
\intertext{\pagebreak[1]}
\frac{\partial }{\partial x}\Bigl(\rho \frac{\partial \theta}{\partial x}\Bigr)&=
     \frac{\partial }{\partial x}\Bigl(\pm \sqrt{2\lambda}
\frac{F'_+  + F'_- }{ F'_+ - F'_-}\Bigr)  \nonumber\\
 &= \mp \frac{\sqrt{2\lambda}}{(F'_+ - F'_-)^2}  
              \bigl( F''_+ F'_- + F''_- F'_+  \bigr) \frac{\partial f}{\partial
x}\nonumber\\
&= \mp \frac{2\sqrt{2\lambda}}{(F'_+ - F'_-)^3}  
              \bigl( F''_+ F'_- + F''_- F'_+  \bigr) 
\label{eq:279b}
\end{align}
\end{subequations}
The last equalities follow from \refeq{eq:274};  since \refeq{eq:279a} and 
\refeq{eq:279b} sum to zero, the continuity equation holds.

We observe that the differentiated functions $F'_{\pm}$ are just the Riemann
coordinates: from \refeq{eq:277b} and \refeq{eq:276} [with the absolute value
ignored] we have 
\numeq{eq:280}{
p \pm \frac{\sqrt{2\lambda}}\rho \equiv R_{\pm} = 2F'_{\pm}\ .
}
Also it is seen with the help of \refeq{eq:274} that the Riemann formulation
\refeq{succpres} of the Chaplygin equations is satisfied by $2F'_{\pm}$. 

The constants of motion \refeq{infnumconsmo} become proportional to 
\begin{align}
I^{\pm}_n &\propto \int \rd x \frac1{F'_+ - F'_-} \bigl[  F'_{\pm}\bigr]^n
\nonumber\\
      &= \int \rd x \frac{\partial f}{\partial x} \bigl[  F'_{\pm}(t\pm f)\bigr]^n
\nonumber\\ 
      &\propto \int \rd z  \bigl[  F'_{\pm}(z) \bigr]^n\ .\label{eq:281}
\end{align}

Finally we remark that the solution 
\refeq{solthenread}, \refeq{densecorresp} corresponds to 
\numeq{eq:282}{
F_+(z) = - F_- (z) = \pm  \frac{\ln z}{2k}\ .
}
There exists a relation between the two functions~$F$ and~$G$  in \refeq{eq:251},
which encode the Chaplygin gas solution in the linearization approach of
Section~\ref{sec:6.3}, and the above two functions $F_{\pm}$, which do the same
job in the Nambu-Goto approach. The relation is that $2F'_+$ is inverse to $2F''$
and $2F'_-$ is inverse to $2G''$, that is,
\begin{align}
2F'' [2F'_+ (z)] &=z \nonumber\\
2G'' [2F'_- (z)] &=z \label{eq:282b}
\end{align}

\centerline{\hfill\hbox to 2in{\hrulefill} \hfill}

\begin{problem}\label{prob:7}
\emph{
Derive \refeq{eq:282b}. Verify this relation with \refeq{eq:252}
and~\refeq{eq:282}. }
\end{problem}
\centerline{\hfill\hbox to 2in{\hrulefill} \hfill}

\newpage
\section{Towards a Non-Abelian Fluid Mechanics}\label{sec:7}

Fluid mechanics and fluid magnetohydrodynamics may very well describe the
long-wavelength degrees of freedom in a quark-gluon plasma. Moreover it is
plausible that the group (color) degrees of freedom remain distinct in that
regime, so that one should incorporate them in the fluid approximation. In this
way one is led to think about constructing  non-Abelian fluid mechanics and
(color) magnetohydrodynamics. In this section we describe an approach to this
task~\cite{JNP}. In the course of development,  we encounter and solve an
interesting mathematical problem: how to para\-meter\-ize a non-Abelian vector
potential so that the non-Abelian \CS\  density becomes a total derivative, and the
volume-integrated \CS\  term is given by a surface integral. Obviously this is
 the non-Abelian generalization of the similar Abelian problem, which
is solved by presenting the Abelian vector potential in Clebsch form. So we shall
determine the non-Abelian version of the \Cpr. 

\subsection{Proposal for non-Abelian fluid mechanics}\label{sec:7.1}

We review our Lagrange density for relativistic Abelian fluid mechanics
\refeq{Lagdens}:  
\numeq{eq:283}{
{\cal L} = -j^\mu a_\mu - f(\sqrt{j^\mu j_\mu})\ .
}
The equation of state is encoded in the function~$f$. For free fluid motion
$f(\sqrt{j^\mu j_\mu}) = c \sqrt{j^\mu j_\mu}$. Here $j^\mu$ is the matter
current and $a_\mu$ is an auxilliary 4-vector, which is presented in the form
\numeq{eq:284}{
a_\mu = \partial_\mu \theta + \alpha\partial_\mu\beta\ .
}
The time component $a_0$, involving time derivatives, determines the
canonical 1-form; the spatial components $\vec a$ are in the \Cpr, as
is needed for overcoming the obstacle created by a Casimir invariant of the fluid
in the algebra 
\refeq{algebra}, \refeq{algebrap}. 
Another way of characterizing the \pr\ of the vector $\vec a$ is that it casts the
\CS\ density of $\vec a$, namely, $\vec a \cdot \grad \times \vec  a$ into  
total derivative form: $\grad\theta \cdot (\grad\alpha\times\grad\beta) =
\grad \cdot (\theta \grad\times\vec  a)$.

For a \nA\ generalization, it is plausible to suppose that the current 4-vector
acquires an internal symmetry index: $J^\mu_a$; correspondingly, the auxiliary
4-vector must also acquire an internal symmetry index: $A^a_\mu$. It remains
to give a rule for parameterizing $A^a_\mu$, which generalizes the Abelian
rule~\refeq{eq:284}.

Our proposal -- and it is a speculative one, since at this stage we have no
derivation from microscopic considerations -- is that $A^a_\mu$ should be written
in a form so that its \nA\ \CS\ density, $\textrm{CS}(A) = A^a \rd {A^a} + \fract13 
f^{abc}\linebreak[2] A^a A^b A^c$, is a total derivative ($f^{abc}$ are the structure
constants of the group). This leads us to the purely mathematical problem of
constructing a
\pr\ for a \nA\ vector that ensures this property.

\subsection{Non-Abelian \Cpr}\label{sec:7.2}
\subsubsection*{(or, casting the \nA\ \CS\ density into total
derivative form)}

We enquire whether it is possible to parameterize the \nA\ 1-form, $A^a$, such
that the
\CS\ 3-form is a total derivative (is exact):
\numeq{eq:285}{
\textrm{CS}(A) = A^a \rd {A^a} + \fract13  f^{abc}
A^a A^b A^c = \rd\Omega \ .
}
That this should be possible follows from the observation that the left side of
\refeq{eq:285} is a 3-form on 3-space; hence it is closed, because a 4-form 
does not exist in 3-space. [Of course on a 4-dimensional space the exterior
derivative of \refeq{eq:285} is proportional to the \nA\ anomaly
(Chern-Pontryagin density)~\cite{TrJaZuWi}.] But a closed form is also exact, at
least locally; this justifies the right side of \refeq{eq:285}. 

How this works in the Abelian case has already been explored in
Sections~\ref{sec:2.4} and \ref{sec:2.5}: the \Cpr\ \refeq{threescalar},
\refeq{1-form} for $A$ leads to the desired result. But the generalization of 
\refeq{threescalar}, \refeq{1-form} for a \nA\ 1-form is not evident. However,
at the end of Section~\ref{sec:2.5}, an alternative approach is presented, wherein
the Abelian 1-form is  projected from a \nA\ pure gauge 1-form. This
construction can be generalized to the \nA\ case and yields the sought-for \pr.

The mathematical problem can therefore be formulated in the following way:
{\em For a given group H, how can one construct a potential $A^a_\mu =
(A^a_0,  A^a_i)$ such that the non-Abelian Chern-Simons integrand
$CS(A)$ is a total derivative?}  Here we shall only sketch the solution to the
problem, referring those interested to Ref.~\cite{JNP} for a detailed
discussion.

In the solution that we present, the ``total derivative''
form for the Chern-Simons density of
$A^a$ is
achieved in two steps.  The \pr, which we find, 
directly leads to an Abelian form of the Chern-Simons
density:
\numeq{eq:286}{
A^a \rd{A^a} + \fract13 f^{abc}
A^a A^b A^c  =
\gamma \rd \gamma
}
for some $\gamma$.  Then Darboux's theorem~\cite{Darb} (or usual fluid
dynamical theory \cite{Lam}) ensures that $\gamma$ can be presented
in Clebsch form, so that $\gamma \rd \gamma$ is
explicitly a total derivative.

We begin with a pure gauge $g^{-1}\rd g$ in
some non-Abelian group $G$ (called the Ur-group) 
whose Chern-Simons integral coincides
with the winding number of $g$.
\numeq{eq:287}{
W(g) = \frac1{16\pi^2} \int \rd{^3r} \textrm{CS}(g^{-1}\rd g) 
= \frac1{24\pi^2} \int \tr (g^{-1}\rd g)^3 
}
We consider a normal subgroup $H\subset G$, with
generators $I^a$, and construct a non-Abelian gauge potential for $H$ by
projection:
\numeq{eq:288}{
A^a \propto ~\tr (I^a g^{-1} \rd g )\ .
}
Within $H$, this is not a pure gauge.
We determine the group structure that    ensures
 the Chern-Simons 3-form  of $A^a$
to be proportional to 
$\tr (g^{-1} \rd g)^3$. Consequently, the  
constructed non-Abelian gauge fields, belonging to the group
$H$, carry quantized Chern-Simons number.  Moreover, we
describe the properties of the Ur-group $G$ 
that guarantee that the
projected potential $A^a$ enjoys sufficient generality to
represent an arbitrary potential in $H$.

Since $\tr (g^{-1} \rd g)^3$ is a total derivative for an
arbitrary group (although this fact cannot in general be
expressed in finite terms~\cite{CroMic}) our construction ensures that the
form of $A^a$, which is achieved through the projection \refeq{eq:288},
produces a ``total derivative'' expression (in the limited sense indicated above) for
its Chern-Simons density. 

Conditions on the Ur-group
$G$, which we take to be compact and semi-simple, are
the following. First of all $G$ has to
be so chosen that it has sufficient number of parameters to
make  $\tr (I^a g^{-1} \rd g)$ a generic potential for
$H$.  Since we are in three dimensions, an $H$-potential
$A^a_i$ has $3\times {\rm dim} H$ independent functions; so a
minimal requirement will be
\numeq{eq:289}{
{\rm dim~} G \ge 3 {\rm ~dim~} H \ \ .
}
Secondly we require that the $H$-Chern-Simons form for $A^a$
should coincide with that of $g^{-1}\rd g$.
As we shall show in a moment, this is achieved if $G/H$ is a
symmetric space.  In this case, if we split the Lie algebra of
$G$ into the $H$-subalgebra spanned by $I^a$,
$a=1,\dots, {\rm dim~} H$, and the orthogonal
complement spanned by $S^A$, $A=1,\dots, ({\rm
dim~}G-{\rm dim~}H)$, the commutation rules are of the
form
\begin{subequations} 
\begin{align}
\relax [I^a, I^b{]} &= f^{abc}
I^c
\label{eq:2.2a} \\
\relax[ I^a, S^A{]} &= h^{a AB} S^B
\label{eq:2.2b} \\
\relax[S^A, S^B{]} &= N ~h^{a AB} I^a \ .
\label{eq:2.2c}
\end{align}
\end{subequations} 
$(h^a)^{AB}$ form a (possibly reducible)
representation of the
$H$-generators $I^a$. The constant  $N$
depends on 
normalizations. More explicitly, if the structure
constants for the Ur-group $G$ are named ${\bar f}^{abc},~
a,b,c =1,\dots,{\rm dim}G$, then the conditions (\ref{eq:2.2a}--c) require
that ${\bar f}^{abc}$ vanishes whenever an odd number of indices belongs to
the orthogonal complement labeled by $A,B,..$. Moreover, $f^{abc}$ 
are taken to be the conventional structure constants for $H$ and this may
render them proportional to (rather than equal to) ${\bar f}^{abc}$.

We define the
traces of the generators by
\begin{align}
\tr (I^a I^b) &= -N_1 \delta^{ab} \ \ ,
\quad \tr (S^A S^B) = -N_2 \delta^{AB}
\nonumber  \\
\tr (I^a S^A) &=0 \ \ .
\label{eq:2.3}
\end{align}
We can evaluate the quantity $\tr[S^A,S^B]I^a =
\tr S^A[S^B,I^a]$ using the commutation rules.  This immediately gives the
relation $N_1N=N_2$. 

Expanding $g^{-1} \rd g$ in terms of generators, we write
\begin{equation}
g^{-1}  \rd g= (I^a A^a + S^A\alpha^A)
\label{eq:2.4}
\end{equation}
which defines the $H$-potential $A^a$. 
Equivalently
\begin{equation}
A^a =-\frac{1}{N_1} \tr(I^a g^{-1}  \rd g)
\label{eq:2.5n}
\end{equation}
From $\rd{(g^{-1} \rd g)} =-g^{-1} \rd g  g^{-1} \rd g$, we get the
Maurer-Cartan relations
\begin{align}
F^\alpha\equiv \rd{A^a} + \fract12
f^{abc}A^b A^c &= -\frac{N}{2}
h^{a A B} \alpha^A \alpha^B  \nonumber \\
\rd{\alpha^A} + h^{\alpha B A} A^a \alpha^B &= 0 \ \ .
\label{eq:2.6}
\end{align}
Using these results, the following chain of equations shows that
the
Chern-Simons 3-form for the $H$-gauge group is
proportional to $\tr (g^{-1} \rd g)^3$:
\begin{align}
\frac {1}{16\pi^2} (A^a
\rd{ A^a}+\fract13 f^{abc}A^a
A^b A^c) &=\frac{1}{48\pi^2} ( A^a
\rd {A^a}+2~ A^a F^a) \nonumber \\
&=\frac {1}{48\pi^2} ( A^a \rd{ A^a} - N
h^{a AB}A^a \alpha^A\alpha^B) \nonumber \\
&=\frac {1}{48\pi^2} ( A^a
 \rd {A^a}+ N \rd{ \alpha^A} \alpha^A)   \nonumber\\
&=-\frac
{1}{48\pi^2}\Bigl[\frac1{N_1} \tr(A \rd
A)+\frac{N}{N_2}\tr(\rd\alpha\alpha)\Bigr]
 \nonumber\\
&=-\frac {1}{48\pi^2 N_1} \tr(A\rd A + \alpha \rd  \alpha)
\nonumber\\ 
&=-\frac {1}{48\pi^2N_1} \tr g^{-1} \rd g  ~\rd{(g^{-1}
\rd g)}\nonumber \\ 
&=\frac {1}{48\pi^2N_1} \tr(g^{-1} \rd g)^3\ .\label{eq:2.7n}
\end{align}
In the
above sequence of manipulations, 
we have used the Maurer-Cartan relations
(\ref{eq:2.6}), which rely on the symmetric space structure of
(\ref{eq:2.2a}--c), and the trace relations (\ref{eq:2.3}), along with $N_1N=N_2$.

We thus see that $\int \textrm{CS}(A)$ is indeed the winding number of
the configuration $g\in G$.  Since $\tr(g^{-1} \rd g)^3$ is a
total derivative locally on $G$, the potential (\ref{eq:2.5n}),
with the symmetric space structure of (\ref{eq:2.2a}--c), does indeed fulfill
the requirement of making $\textrm{CS}(A)$ a total derivative. 
It is therefore appropriate to call our construction (\ref{eq:2.5n})
a ``non-Abelian Clebsch parameterization".

In explicit realizations, given a gauge group of interest $H$,
we need to choose a group $G$ 
such that the conditions
(\ref{eq:289}), (\ref{eq:2.2a}--c) hold. In general this is not possible.
However, one can proceed recursively. Let us suppose that the
desired result has been established for a group, which we call
$H_2$. Then we form $H\subset G$ obeying (\ref{eq:2.2a}--c)
as $H=H_1 \times H_2$, where
$H_1$ is the gauge group of interest, satisfying ${\rm dim}G \geq
3~{\rm dim} H_1$. For this choice of $H$, the result (\ref{eq:2.7n})
becomes
\numeq{eq:2.8n}{
\textrm{CS} (H_1) +\textrm{CS}  (H_2) =  \frac {1}{48\pi^2 N_1}\tr(g^{-1}
\rd g)^3
}
But since $\textrm{CS}  (H_2)$ is already known to be a total derivative,
(\ref{eq:2.8n}) shows the desired result: $\textrm{CS} (H_1)$ is a total
derivative.

To see explicitly how this works we work out the representation for a
SU$(2)\approx O(3)$ potential $A^a_i$, which possesses nine independent
functions. 

We take $G=O(5), H=O(3)\times O(2)$.  We consider the
4-dimensional spinorial representation of $O(5)$.
With the generators normalized  by $\tr(t^at^b) = -
\delta^{ab}$, the Lie algebra generators of $O(5)$ are given by
\begin{align}
I^a &=\frac{1}{ 2i} 
\left(
\begin{array}{cc}
\sigma^a & 0 \\
0 &\sigma^a
\end{array}
\right)  \nonumber \\
I^0 &=\frac{1}{ 2i} 
\left(
\begin{array}{cc}
-1 & 0 \\
0 &1
\end{array}
\right)  \label{eq:3.1n}   \\
S^A &=\frac{1}{ i\sqrt{2}}
\left(
\begin{array}{cc}
0 & 0 \\
\sigma^A &0
\end{array}
\right)
\qquad
\tilde{S}^A = \frac{1}{ i\sqrt{2}}
\left(
\begin{array}{cc}
0 & \sigma^A \\
0 &0
\end{array}
\right)
\nonumber
\end{align}
$\sigma$'s are the $2 \times 2$ Pauli matrices. $I^a$ generate
$O(3)$, with the conventional structure constants $\epsilon^{abc}$,
and $I^0$ is the generator of $O(2)$.  $S,\tilde{S}$ are the coset
generators. 

A general group element in $O(5)$ can be written in the form $g=Mhk$
where $h\in O(3)$, $k\in O(2)$, and
\begin{equation}
M = \frac{1}{\sqrt{1+{\bf \bar{w}} \cdot {\bf w} - \fract{1}{ 4}({\bf
w}\times{\bf {\bar w}})^2 }}
\left(
\begin{array}{cc}
1- \frac{i}{ 2}({\bf w} \times {\bf \bar{w}})\cdot {\bf \sigma} & -{\bf w} \cdot
{\bf \sigma}
\\[2ex]
{\bf \bar{w}} \cdot {\bf \sigma} & 1+ \frac{i}{ 2}({\bf w} \times
{\bf\bar{w}})\cdot {\bf \sigma}
\end{array}
\right)
\label{eq:3.2n}
\end{equation}
$w^a$ is a complex 3-dimensional vector, with the bar denoting complex
conjugation. 
${\bf w}\cdot {\bf
\bar{w}} = w^a \bar{w}^a$ and $({\bf w} \times {\bf\bar{w}})^a =
\epsilon^{abc} w^b \bar{w}^c$. 
The general $O(5)$ group element contains ten independent real functions.
These are collected as six from $M$ (in the three complex functions $w^a$), 
three in $h$, and one in~$k$. 

 The $O(3)$ gauge
potential given by $- \tr (I^a g^{-1} \rd  g)$ reads
\begin{align}
A^a &= R^{ab} (h) ~a^b  + (h^{-1} \rd  h)^a
\nonumber \\
a^a &= \frac{1}{1+ {\bf w} \cdot {\bf\bar{w}} - \fract{1}{ 4}({\bf w} \times
{\bf\bar{w}})^2}
\Bigg\{ \frac{w^a \rd{\bf\bar{w}} \cdot ({\bf w} \times
{\bf\bar{w}}) + \bar{w}^a \rd  {\bf w} \cdot ({\bf \bar{w}}\times {\bf w})}{2} \label{eq:3.3n}\\
&~~~~~~~~~~~~~~~~~~~~~~~~~~~~~~~~~~~~~~~~~~~~~~~~{}+
\epsilon^{abc} (\rd { w^b} \bar{w}^c -
w^b \rd {\bar{w}^c}) \Bigg\}
\nonumber
\end{align}
where $R^{ab} (h)$ is defined by $hI^a h^{-1} = R^{ab} h^b$ and $k$ does not
contribute. 
$A^a$ is the $h$-gauge transform of $a^a $,  which depends
on six real  parameters $(w^a)$.  The three gauge
parameters of $h\in O(3)$,
along with the six, give the
nine functions needed to parameterize a general $O(3)$- [or $SU(2)$-] 
potential
in three dimensions.  The Chern-Simons form is 
\begin{align}
\textrm{CS} (A) &= \frac{1}{16\pi^2} (A^a \rd{A^a} + \fract13
\epsilon^{abc} A^a A^b A^c)
\nonumber \\
&=  \frac{1}{16\pi^2} (a^a \rd{ a^a} + \fract13
\epsilon^{abc} a^a a^b a^c) -
\rd{\left[\frac{1}{16\pi^2}  (\rd h h^{-1})^a a^a )\right]} \nonumber\\
&\qquad{}+
\frac{1}{24\pi^2} \tr (h^{-1} \rd  h)^3
\label{eq:3.4n}
\end{align}
The second equality reflects the usual response of the Chern-Simons density
to gauge transformations.
Using the explicit form of
$a^a$ as given in (\ref{eq:3.3n}), we can further reduce this.
Indeed we find that
\begin{equation}
a^a \rd{a^a} + \fract13
\epsilon^{abc} a^a a^b a^c =
(-2) \frac{({\bf \bar{w}} \times \rd  {\bf\bar{w}}) \cdot {\bf\rho} + ({\bf w} \times 
\rd {\bf w}) \cdot {\bf\bar{\rho}}}{[1+ {\bf w}\cdot {\bf \bar{w}} -\fract{1}{ 4}
({\bf w} \times {\bf{\bar w}})^2 ]^2}
\label{eq:3.5n}
\end{equation}
$$
\rho_k = \fract12 \epsilon_{ijk} \rd {\bar{w}^i} \rd { \bar w^j}
$$
Defining an Abelian potential
\begin{equation}
a = \frac{{\bf w} \cdot \rd {\bf\bar{w}} -{\bf \bar{w}} \cdot \rd  {\bf w}}{1+
{\bf w}\cdot {\bf \bar{w}} - \fract{1}{ 4}({\bf w} \times {\bf\bar{w}})^2}
\label{eq:3.6}
\end{equation}
we can easily check that $a \rd  a$ reproduces (\ref{eq:3.5n}).  In other
words
\begin{equation}
\textrm{CS} (A)  = \frac{1}{16\pi^2}  a\rd  a + \rd { \left[ \frac{ (\rd
h h^{-1})^a a^a )}{16 \pi^2} \right] }+ \frac{1}{48\pi^2} \tr(h^{-1} \rd  h)^3
\label{eq:3.7n}
\end{equation}
If desired, the Abelian potential $a$ can now be written in the Clebsch form making
$a \rd  a$ into a total derivative, while the remaining two terms already are total
derivatives, though in a ``hidden'' form for the last expression. This completes our
construction.

\subsection{Proposal for non-Abelian
magnetohydro\-dynamics}\label{sec:7.3}

We return to our construction of a Lagrange density for non-Abelian kinetic
theory. As explained in Section~\ref{sec:7.1}, a plausible non-Abelian
generalization for~\refeq{eq:283} is 
\numeq{eq:300}{
{\cal L} = -J^{\mu a} \frac1{N_1} \tr I^a g^{-1} \partial_\mu g - c\sqrt{J^{\mu a}
J^a{}_\mu} 
}
where for simplicity we have taken the ``free'' form for~$f$. When the desired
group is SU(2), $g$ is an $O(5)$  group element, as detailed in Section~\ref{sec:7.2}. 

Magnetohydrodynamics is achieved by introducing a further interaction with a
dynamical gauge potential ${\cal A}^a_\mu$. This is accomplished by promoting
the derivative of~$g$ to a gauge-covariant derivative, gauged on the right
\numeq{eq:301}{
{\cal L}_{\textrm{magnetohydrodynamics}} = -\frac1{N_1} J^{\mu a}  
\tr (I^a g^{-1} D_\mu g) -  c\sqrt{J^{\mu a} J^a_\mu} -\fract14 {\cal F}^{a\mu\nu}
{\cal F}^a_{\mu\nu}
}
with
\numeq{eq:302}{
D_\mu g = \partial_\mu g + eg {\cal A}_\mu\ . 
}
${\cal A}_\mu = {\cal A}_\mu^a I^a$ are independent, dynamical gauge potentials
(not given by~$g$) leading to the field strengths~${\cal F}^a_{\mu\nu}$. The gauge
transformation properties by the gauge function~$h$ are 
\begin{gather}
g' = gh \qquad {\cal A}' = h^{-1} {\cal A}  h + \frac1e  h^{-1}\rd h\nonumber\\
J_\mu^{\prime\, a} I^a =  h^{-1} J_\mu^a I^a h\ .\label{eq:303}
\end{gather}

We expect that the Lagrangian \refeq{eq:301} will describe non-Abelian
magnetohydrodynamics, namely the dynamics of a fluid with non-Abelian charge
coupled to non-Abelian fields. [The Abelian version of \refeq{eq:301} does indeed
describe ordinary magnetohydrodynamics.]  This gluon hydrodynamics can be
useful for non-Abelian plasmas such as the quark-gluon plasma. Details of
\refeq{eq:301} and possible applications are under further study.

\newpage
\addcontentsline{toc}{section}{\kern1.4em Solutions to Problems}
\subsection*{Solutions to Problems}

\paragraph{Problem \ref{prob:1}} The imaginary part of the Schr\"odinger
equation gives the continuity equation in the form $\dot\rho +
\grad\cdot(\rho\grad
\theta)=0$. This identifies the velocity $\vec v$ as $\grad\theta$, that is, $\vec v$
is irrotational and there is no vorticity. The real part becomes the Bernoulli
equation $\dot\theta + \fract12 (\grad\theta)^2 = \frac{\hbar^2}2 \rho^{-1/2}
\nabla^2\rho^{1/2}$, whose gradient gives the Euler equation and identifies the
force $\vec f$ as $\grad (\frac{\hbar^2}2 \rho^{-1/2}
\nabla^2\rho^{1/2})$. 

\paragraph{Problem \ref{prob:2}} $\vec j = \rho\vec v$ with $\vec v = \grad
\theta$. 

\paragraph{Problem \ref{prob:3}} 
${\cal L}_{\textrm{Schr\"odinger}} = \theta\dot\rho -
\fract12\rho(\nabla\theta)^2 -
\frac{\hbar^2}8\frac{\grad\rho\cdot\grad\rho}\rho$ where the time derivative of
$i\frac{\hbar\rho}2 -\rho\theta$ has been dropped. 

The results in the solutions to Problems~\ref{prob:1} and \ref{prob:3} are called
the M\"adelung formulation of quantum mechanics~\cite{Mad}.

\paragraph{Problem \ref{prob:4}}
$\textrm{CS}(A) = \eps^{ijk} \partial_i \Phi \partial_j \cos\Theta
\partial_k h(r)$ 
\begin{itemize}
\item[(a)] Extracting the first derivative leaves 
$\textrm{CS}(A) =   \partial_i V^i_a$, $V^i_a = \eps^{ijk} \Phi \partial_j
\cos\Theta\linebreak[3]
\partial_k h(r)$. (This is true because $\eps^{ijk}   \partial_i \partial_j\cos\Theta =
0 = \eps^{ijk} \partial_i \partial_k h(r)$, since $\cos\Theta$ and $h(r)$ are
nonsingular.) Note that $V^i_a = \eps^{i\Theta r} \Phi(-\frac{\sin\Theta}r) h'(r) =
\delta^{i\Phi}\Phi \bigl(\fract1r\sin\Theta\bigr) h(r)$. Since $V^r_a=0$, the
surface integral does not contribute. However, since $\Phi$ is multivalued, there is
a contribution from the $\Phi$ integral:
$
\int\rd{^3r} \textrm{CS}(A) = \int_0^R r^2\rd r \int_0^\pi \sin\Theta \rd\Theta
\int_0^{2\pi} \rd\Phi \bigl(\frac1{r \sin\Theta} \frac\partial{\partial\Phi}
\Phi\bigr)\times\linebreak[3]
\bigl(\frac1r \sin\Theta\bigr) h'(r) = 4\pi \bigl[h(R) - h(0)\bigr]$.

\item[(b)] Extracting the second derivative leaves $\textrm{CS}(A) =
\partial_j V^j_b - \eps^{ijk}
(\partial_j \partial_i \Phi) \cos\Theta\linebreak[3] \partial_k h(r)$. The last term
is present, owing to the singularity of $\Phi$ at the origin, which gives rise to
$\eps^{kij}\partial_i \partial_j \Phi = \delta^{k3} 2\pi \delta(x)\delta(y)$.
(See~\cite{JacPi2}.) Also we have $V^i_b =  \eps^{ijk} \partial_i \Phi \cos\Theta
\partial_k h(r) = 
\eps^{\Phi jr} \bigl(\frac1{r \sin\theta}\bigr) \cos\Theta h'(r) = -\frac1r
\delta^{j\theta} \cot\Theta\linebreak[3] h'(r)$. Again there is no $r$-component to
contribute to the surface integral, but the second, singular term leaves 
\begin{align*}
\int\rd{^3r} \textrm{CS}(A) &= \int\rd{^3r} (2\pi) \delta(x) \delta(y) \cos\Theta
\frac\partial{\partial z} h(r)\\
 &= 2\pi \int_{-R}^R \rd z \frac z{|z|}
\frac\partial{\partial z} h(|z|)\\
 &= 4\pi\int_0^R \rd z \frac\partial{\partial z} h(z) = 4\pi \bigl[h(R) - h(0)\bigr]\ .
\end{align*}

\item[(c)] Extracting the last derivative leaves
\begin{align*}
\textrm{CS}(A) &= \partial_k V^k_c - \eps^{ijk}
(\partial_k \partial_i \Phi) \partial_j\cos\Theta h(r)\ ,\\
V^k_c &=  \eps^{ijk} \partial_i \Phi\partial_j \cos\Theta h(r)\\
&= \eps^{\Phi\theta k}\Bigl(\frac1{r\sin\Theta}\Bigr)
\Bigl(-\frac1r \sin\Theta\Bigr) h(r) \\
&= \delta^{kr} \frac{h(r)}{r^2}\ .
\end{align*}
Here the surface integral contributes $4\pi h(R)$. The singular term is 
\begin{align*}
-\delta^{j3} (2\pi)\delta(x)\delta(y) ( \partial_j \cos\Theta) h(r) 
&= - (2\pi)\delta(x)\delta(y) \Bigl(\frac\partial{\partial z} \frac z{|z|}\Bigr) h(|z|)\\
&= -(4\pi) \delta^3 (\vec r) h(0)\ .
\end{align*}
Hence this contribution to the spatial integral is $-4\pi h(0)$, for a total of
$4\pi\bigl[ h(R)-h(0)\bigr]$.
\end{itemize}

\paragraph{Problem \ref{prob:5}}
In the \Cpr, $\vec B = \grad\alpha\times\grad\beta$, and $\delta \vec A =
\grad\delta\gamma + \delta\alpha\grad\beta + \alpha\grad\delta\beta$.
Therefore
\begin{align*}
\vec B\cdot\delta\vec A &= \vec B \cdot \grad\delta\gamma + 
      \vec B\cdot\alpha\grad\delta\beta + (\vec B
\cdot \grad \beta)\delta\alpha\\ 
&= \grad\cdot (\vec
B\delta\gamma) + \grad\cdot(\vec B\alpha\delta\beta) -
    (\vec B\cdot\grad\alpha)\delta\beta
 + (\vec B \cdot  \grad \beta)\delta\alpha \ .
\end{align*}
The last two terms vanish, so $\int\rd{^3r} \vec A \cdot \vec B$ is the surface
term
$\int\rd{\vec S}\cdot \vec B(\delta\gamma + \alpha\delta\beta)$ with no
contribution from the bulk (finite $\vec r$ space). This of course is consistent with
the \CS\ integral being a surface term, since $\delta \fract12 \int\rd{^3r} \vec
A\cdot\vec B =  \int\rd{^3r} \vec B\cdot\delta\vec A$. When demanding the
variation of $\fract12 \int\rd{^3r} B^2$ to vanish, we first find $\int\rd{^3r}
(\grad\times\vec B)\cdot \delta\vec A = 0$. When $\delta\vec A$ is arbitrary,
this condition implies the vanishing of $\grad\times\vec B$. However, in the \Cpr,
all we can conclude is that $\grad\times\vec B$ is proportional to $\vec B$, with a
position-dependent proportionality factor $\grad\times\vec B = \mu\vec B$.
Taking the divergence shows that $\vec B\cdot\grad\mu = 0$, that is, $\mu$ can
be a function of the magnetic surfaces; see \refeq{follows}, \refeq{follows2}.

\paragraph{Problem \ref{prob:6}}
$\theta(t,\vec r) = r^2/2t$. Time rescaling: $\theta_\omega(t,\vec r) = e^\omega
\theta(T,\vec r)$, $T=e^\omega t$; $e^\omega\theta(T,\vec r) = e^\omega
r^2/2e^\omega t = r^2/2t = \theta(t,\vec r)$. 

Space-time mixing: $\theta_{\vec\omega} (t,\vec r) = \theta(T,\vec R)$, 
$T= t +\vec\omega\cdot\vec r +\fract12\omega^2 \theta(T,\vec R) = t
+\vec\omega\cdot\vec r + \omega^2 R^2/4T$, $\vec R = \vec r
+\vec\omega\theta(T,\vec R) = \vec r +\vec\omega R^2/2T$. Squaring the second
equation gives
$R^2 = r^2 + \vec\omega\cdot\vec r R^2/T + \omega^2  R^4/4T^2$.
Multiplying  the first equation by $R^2/T$ gives 
$R^2 = t R^2/T  + \vec\omega\cdot\vec r R^2/T + \omega^2
 R^4/4T^2$. Comparing the two shows that $R^2/T = r^2/t$ or 
$\theta(T,\vec R) = \theta(t,\vec r)$.

\paragraph{Problem \ref{prob:7}}
From \refeq{eq:248} and \refeq{eq:251} we
learn that $t=F'' + G''$, $x= (p+s) F''-F' + (p-s) G'' - G'$, where $F$ is a function of
$p+s = R_+$ and $G$ is  a function of 
$p-s=R_-$. Differentiating these equations with respect to $t$ and $x$, it follows
that $1=F''' \dot R_+ + G''' \dot R_-$, $0=F''' \frac{\partial R_+}{\partial x} + G'''
\frac{\partial R_-}{\partial x}$, $0= R_+F''' \dot R_+ + R_- G''' \dot R_-$, $1= R_+ F'''
\frac{\partial R_+}{\partial x} + R_- G'''
\frac{\partial R_-}{\partial x}$, which in turn imply
$\frac{\partial R_+}{\partial x} = 1/(R_+-R_-)F'''$, $\frac{\partial R_-}{\partial x} =
-1/(R_+-R_-)G'''$, and $\dot R_{\pm} = - R_{\mp} \frac{\partial R_{\pm}}{\partial
x}$ [the Riemann equation~\refeq{succpres} again].

On the other hand, the functions $F_{\pm} (t\pm f)$ describing the Chaplygin gas
solution from the Nambu-Goto equation are related to $R_{\pm}$ by
\refeq{eq:280}: $R_{\pm} = 2F'_{\pm}$. Hence $\frac{\partial R_{\pm}}{\partial
x} = \pm 2 F''_{\pm} \frac{\partial f}{\partial x} = \pm 2 F''_{\pm}/(F'_+ - F'_-) =
\pm 4 F''_{\pm}/(R_+ - R_-)$, where \refeq{eq:274b} is used. It follows that $4
F''_+(z) F''' \bigl(2 F'_+(z)\bigr)  = \frac{\rd{}}{\rd z}\Bigl( 2F'' \bigl(2
F'_+ (z)\bigr)\Bigr) = 1$ and
$4 F''_-(z) G'''\bigl(2F'_-(z)\bigr) =
\frac{\rd{}}{\rd z}\Bigl( 2G'' \bigl(2 F'_-(z)\bigr)\Bigr) = 1$, or 
$2F''\bigl(F'_+(z)\bigr) = z$ and 
$2G''\bigl(2F'_-(z)\bigr) = z$. When $F_+(z) = \ln z/2k$, $2F'_+(z) = 1/zk$; with
$F(z) =
\frac z{2k} \ln z$, $2F''(z) = 1/kz$ and $2F'' \bigl(2F'_+(z)\bigr) = z$; similarly for
$F_-(z)$ and~$G(z)$. 

\def\Journal#1#2#3#4{{#1} {\bf #2}, #3 (#4)}
\def\add#1#2#3{{\bf #1}, #2 (#3)}
\def\Book#1#2#3#4{{\em #1}  (#2, #3 #4)}
\def\Bookeds#1#2#3#4#5{{\em #1}, #2  (#3, #4 #5)}

\def\NPB{{\em Nucl. Phys.}} 
\def\PLA{{\em Phys. Lett.}} 
\def\PLB{{\em Phys. Lett.}} 
\def\PRL{{\em Phys. Rev. Lett.}}
\def\PRD{{\em Phys. Rev. D}}
\def\PR{{\em Phys. Rev.}}
\def\ZPC{{\em Z. Phys.} C}
\def\SJNP{{\em Sov. J. Nucl. Phys.}}
\def\AnnP{{\em Ann. Phys.}\ (NY)}
\def\JETPL{{\em JETP Lett.}}
\def\LMP{{\em Lett. Math. Phys.}}
\def\CMP{{\em Comm. Math. Phys.}}
\def\PTP{{\em Prog. Theor. Phys.}}
\def\PNAS{{\em Proc. Nat. Acad. Sci.}}


\begin{thebibliography}{99}
\bibitem{Darb}
A constructive discussion of the Darboux theorem can be found it the second
work of Ref.~\cite{FadJac}.
\bibitem{AnDet} 
Analytic details of the implementation of the space-time mixing
transformation on~\refeq{disclat} are presented by Bazeia and
Jackiw~\cite{BazJac}.
\bibitem{ref20} 
The additional transformation rules were derived in Ref.~\cite{JacPol}, on the basis
of constants of motion identified previously in the
$d=2$ case; see Ref.~\cite{BorHop94}.
\end{thebibliography}
\end{document}